\pdfoutput=1
\documentclass[11pt]{article}
\usepackage[margin=1in]{geometry}
\usepackage{amsmath,amssymb,amsthm,booktabs,array}
\usepackage[shortlabels]{enumitem}
\usepackage[hidelinks]{hyperref}
\usepackage{algorithm}
\usepackage[noend]{algpseudocode}
\usepackage{tikz}
\usetikzlibrary{arrows.meta,backgrounds,calc}
\tikzset{
  v/.style={circle,draw,fill=white,inner sep=0pt,minimum size=6pt,line width=0.6pt},
  d/.style={circle,fill=black,inner sep=0pt,minimum size=4.5pt},
  e/.style={line width=0.6pt},
  faint/.style={line width=0.5pt,black!30},
  arc/.style={line width=0.6pt,-{Stealth[length=4.5pt]}},
  lab/.style={font=\scriptsize},
  fcap/.style={font=\footnotesize,align=center},
}

\setlist{itemsep=1pt,topsep=3pt}

\newtheorem{theorem}{Theorem}[section]

\newtheorem{corollary}[theorem]{Corollary}

\newtheorem{claim}{Claim}
\theoremstyle{definition}

\newtheorem{example}[theorem]{Example}
\theoremstyle{remark}
\newtheorem{remark}[theorem]{Remark}

\theoremstyle{plain}
\newtheorem*{namedinner}{\namedtext}
\newenvironment{named}[1]{\def\namedtext{#1}\phantomsection\namedinner}{\endnamedinner}

\newcommand{\ThmSSSP}{\hyperref[thm:A]{Exact SSSP Theorem}}
\newcommand{\ThmNeg}{\hyperref[thm:B]{Negative Weights Theorem}}
\newcommand{\ThmFlow}{\hyperref[thm:C]{Maximum Flow Theorem}}
\newcommand{\ThmFace}{\hyperref[thm:Cp]{Common-Face Flow Theorem}}
\newcommand{\ThmCut}{\hyperref[thm:D]{Minimum Cut Theorem}}
\newcommand{\HypL}{\hyperref[prop:L]{Label Hypothesis}}
\newcommand{\LemVS}{\hyperref[lem:V]{Virtual Source Lemma}}
\newcommand{\LemBoost}{\hyperref[lem:boost]{Discretized Boosting Lemma}}
\newcommand{\LemDD}{\hyperref[lem:D]{Dual Distance Lemma}}
\newcommand{\CorDual}{\hyperref[cor:dual]{Dual SSSP Corollary}}
\newcommand{\LemTrees}{\hyperref[lem:exact41]{Exact Trees Lemma}}
\newcommand{\LemAnn}{\hyperref[lem:N]{Annulus Lemma}}
\newcommand{\LemRep}{\hyperref[lem:R]{Repaired Graph Lemma}}
\newcommand{\LemWit}{\hyperref[lem:S1]{Witness Interval Lemma}}
\newcommand{\LemInc}{\hyperref[lem:O]{Incision Lemma}}
\newcommand{\LemRec}{\hyperref[lem:S2]{Repaired Recursion Lemma}}
\newcommand{\Exact}{\textsc{ExactSSSP}}
\newcommand{\NegSSSP}{\textsc{NegativeSSSP}}
\newcommand{\DualFlow}{\textsc{DualFlow}}
\newcommand{\FaceFlow}{\textsc{CommonFaceFlow}}
\newcommand{\MinCut}{\textsc{ExactMinCut}}

\newcommand{\Ot}{\tilde O}
\newcommand{\dist}{\mathrm{dist}}
\newcommand{\rev}{\mathrm{rev}}
\newcommand{\eps}{\varepsilon}
\newcommand{\lft}{\mathrm{left}}
\newcommand{\rgt}{\mathrm{right}}
\newcommand{\cin}{\mathrm{in}}
\newcommand{\cout}{\mathrm{out}}

\title{Exact Shortest Paths and Maximum Flow\\ in Planar Networks in the CONGEST Model}
\hypersetup{pdftitle={Exact Shortest Paths and Maximum Flow in Planar Networks in the CONGEST Model},pdfauthor={Sebastian Senge}}
\author{Sebastian Senge\\ Independent Researcher}
\date{}

\begin{document}
\maketitle

\begin{abstract}
In the CONGEST model, exact single-source shortest paths (SSSP) in directed planar networks with
nonnegative weights and exact maximum $st$-flow in directed planar networks have both been solved
in $\Ot(D^2)$ rounds, where $D$ is the hop diameter, while the lower bound is $\Omega(D)$. Given the
approximate distance labels of a recent preprint~\cite{LTY26}, which we use as a black box, we give
an algorithm for exact directed SSSP in $\Ot(D)$ rounds. It rests on a lemma that answers queries
from a virtual source with the labels, and on a corrected discretization of the reduction of exact
to approximate distances of~\cite{RHMGZ23}. All further bounds of the form $\Ot(D)$ and $D\cdot n^{o(1)}$ use these
labels as well. Through the reduction of~\cite{A24}, SSSP with negative
weights, or the detection of a negative cycle, takes $D\cdot n^{o(1)}$ rounds with high probability.
We weight the face-disjoint graph of~\cite{ADPW25} so that its distances are dual distances, which
turns distance problems in the dual into problems in a planar network of hop diameter $O(D)$. This
gives exact maximum $st$-flow in directed planar networks in $\min\{\Ot(D^2),D\cdot n^{o(1)}\}$
rounds with high probability, and in $\Ot(D)$ rounds when $s$ and $t$ lie on a common face. In
undirected planar networks, the value of a minimum $st$-cut and a minimum cut take $\Ot(D)$ rounds.
For this we repair the auxiliary graphs of~\cite{ADW26b} at $s$ and $t$, and we show that its
recursion, as we read it, can miss the minimum cut, also with exact shortest paths: on a network with $11$ vertices
it reports $37$ for a minimum cut of $15$. We prove that two more shortest-path computations per
level make it exact. The algorithms not stated with high probability add no randomness to the labels
and the planar embedding. The virtual-source lemma, the weighting of the face-disjoint graph
and the corrected discretization are machine-checked in Lean~4 with Mathlib; the CONGEST algorithms,
their round counts and all other arguments are proved on paper only. Numerical checks support the
constructions but are not used in any proof.
\end{abstract}

\section{Introduction}

\subsection{Shortest paths and flows in planar networks}

In the CONGEST model the input graph is the network itself. Every vertex is a processor that
initially knows its incident edges and their weights. Computation proceeds in synchronous rounds,
in every round each vertex can send a message of $O(\log n)$ bits over each incident edge, and
local computation is free. The complexity of an algorithm is its number of rounds, as a function of
the number $n$ of vertices and of the hop diameter $D$, the diameter of the underlying undirected
unweighted graph.

We study three problems. In \emph{single-source shortest paths} (SSSP), every edge has an integer
weight of polynomial size, and every vertex $v$ has to learn $\dist(s,v)$ from a given source $s$
and its parent in a shortest-path tree. In the directed version the two directions of an edge have
their own weights, and an edge may be usable in one direction only. With negative weights the
algorithm also has to detect negative cycles. In \emph{maximum $st$-flow}, every edge has an
integer capacity of polynomial size in each direction, and the algorithm has to compute the value of
a maximum flow from $s$ to $t$ and the flow on every edge. The value of a maximum $st$-flow is the
capacity of a \emph{minimum $st$-cut}, and in undirected networks we also ask for this value alone,
together with a minimum cut. Each of these problems needs $\Omega(D)$ rounds on planar networks,
since information has to cross the network.

On general graphs the known bounds for these problems are polynomial in $n$. Exact directed SSSP
takes $\Ot(\sqrt n+D+n^{2/5+o(1)}D^{2/5})$ rounds~\cite{CF23,RHMGZ23} or
$\Ot(\sqrt n\,D^{1/4}+D)$ rounds~\cite{CM20}, for constant $\eps$ a $(1+\eps)$-approximation of
directed SSSP takes $\Ot(\sqrt n+D+n^{2/5+o(1)}D^{2/5})$ rounds~\cite{CFR21}, SSSP with
negative weights takes $(\sqrt n+D+n^{2/5+o(1)}D^{2/5})\cdot n^{o(1)}$ rounds~\cite{A24}, and exact
maximum flow takes $n^{1/2+o(1)}(\sqrt n+D)$ rounds~\cite{dV23}. The bounds for exact SSSP
of~\cite{CF23,RHMGZ23,CM20} and for negative weights of~\cite{A24} are randomized.

\subsection{Previous work on planar networks}

Planar graphs have balanced cycle separators, and the bounded diameter decomposition
of~\cite{LP19} makes them usable in CONGEST: it is a recursive decomposition of height $O(\log n)$
whose bags have spanning trees of small diameter. With it, exact SSSP was solved in $\Ot(D^2)$ rounds
with high probability, for undirected networks in~\cite[Thm.~1.6]{LP19} and for directed networks
in~\cite[Lemma~4]{P20}. The same bound was obtained in~\cite{ADPW25} for exact maximum
$st$-flow in directed planar networks, through SSSP with negative lengths in the dual graph; for
large $D$, the bound $n^{1/2+o(1)}D$ of~\cite{dV23} for excluded-minor graphs is smaller. In a
recent preprint~\cite{LTY26}, $(1+\eps)$-approximate directed SSSP is solved in $\Ot(D/\eps)$
rounds, with distance labels that are built bottom up over the decomposition. In~\cite{ADW26b}, the value of a maximum $st$-flow and a minimum $st$-cut in undirected planar
networks are approximated within a factor $1-o(1)$ in $\Ot(D)$ rounds, deterministically; there,
approximate SSSP is used ``for the lack of almost-$D$-rounds exact distributed SSSP''.
Table~\ref{tab:previous} lists the previous bounds.

\begin{table}[t]\centering
\begin{tabular}{>{\raggedright\arraybackslash}p{0.5\textwidth}ll}
\toprule
Problem & Rounds & \\
\midrule
Exact SSSP, directed, $w\ge0$ & $\Ot(D^2)$ & \cite{LP19,P20}\\
$(1+\eps)$-approximate SSSP, directed & $\Ot(D/\eps)$ & \cite{LTY26} (preprint)\\
Exact maximum $st$-flow, directed & $\min\{\Ot(D^2),n^{1/2+o(1)}D\}$ & \cite{ADPW25,dV23}\\
SSSP in the dual graph, negative lengths & $\Ot(D^2)$ & \cite{ADPW25}\\
Maximum $st$-flow, undirected, $s,t$ on a common face, $(1-o(1))$-approximate & $D\cdot n^{o(1)}$ & \cite{ADPW25}\\
Maximum $st$-flow value and minimum $st$-cut, undirected, $(1-o(1))$-approximate & $\Ot(D)$ & \cite{ADW26b}\\
\bottomrule
\end{tabular}
\caption{Previous bounds on planar networks. The bounds of~\cite{LP19,P20,ADPW25} hold with high
probability. The bounds of~\cite{LTY26} and~\cite{ADW26b} are stated as deterministic, the first
with polynomial local time.}
\label{tab:previous}
\end{table}

One open question is stated in~\cite[Sec.~2.4]{ADPW25}: to find better SSSP algorithms in the
primal graph, which is called ``the bottleneck to overcome in order to obtain better distributed
planar algorithms'', and, before attacking maximum flow, to ``first improve the $\Ot(D^2)$ bound of
the (easier) directed SSSP problem (even with non-negative weights)''.

\subsection{Results}

Each of our results is an algorithm. Table~\ref{tab:results} lists them.

\begin{table}[t]\centering\small
\begin{tabular}{>{\raggedright\arraybackslash}p{0.26\textwidth}l>{\raggedright\arraybackslash}p{0.18\textwidth}>{\raggedright\arraybackslash}p{0.25\textwidth}}
\toprule
Problem & Algorithm & Rounds & Previous\\
\midrule
Exact SSSP, directed, $w\ge0$ & \Exact{} & $\Ot(D)$ & $\Ot(D^2)$~\cite{LP19,P20}\\
Exact SSSP, negative weights & \NegSSSP{} & $D\cdot n^{o(1)}$ & $\Ot(D^2)$ as stated in~\cite[Sec.~1.1]{ADW26b}; bounds for general graphs~\cite{A24}\\
Exact maximum $st$-flow, directed & \DualFlow{} & $\min\{\Ot(D^2),D\cdot n^{o(1)}\}$ & $\min\{\Ot(D^2),n^{1/2+o(1)}D\}$ \cite{ADPW25,dV23}\\
The same, $s,t$ on a common face & \FaceFlow{} & $\Ot(D)$ & approximate and undirected only~\cite{ADPW25,ADW26b}\\
Value of a minimum $st$-cut and a minimum cut, undirected & \MinCut{} & $\Ot(D)$ & $\Ot(D^2)$~\cite{ADPW25}; $(1-o(1))$-approximate in $\Ot(D)$~\cite{ADW26b}\\
SSSP in the dual, nonnegative lengths & \Exact{} on $\hat G$ & $\Ot(D)$ & $\Ot(D^2)$~\cite{ADPW25}\\
SSSP in the dual, negative lengths & \NegSSSP{} on $\hat G$ & $D\cdot n^{o(1)}$ & $\Ot(D^2)$~\cite{ADPW25}\\
\bottomrule
\end{tabular}
\caption{Results, which are the \ThmSSSP{}, the \ThmNeg{}, the \ThmFlow{}, the \ThmFace{}, the
\ThmCut{} and the \CorDual{}. All of them except the $\Ot(D^2)$ branch of \DualFlow{} rest on the
\HypL{}, which is modelled on the distance labels of~\cite{LTY26}. The algorithms for nonnegative weights
and for undirected networks add no randomness to the labels and the planar embedding; the others
hold with high probability.}
\label{tab:results}
\end{table}

Given the labels, the \ThmSSSP{} answers the open question of~\cite{ADPW25} on directed SSSP. The new
term in the \ThmFlow{} is smaller than $\Ot(D^2)$ once $D$ exceeds $(\log n)^{O(\sqrt{\log n})}$. The
\ThmCut{} does not compute the flow on every edge; for that, the \ThmFlow{} is the best bound we know.

\subsection{How the algorithms fit together}

\begin{figure}[t]\centering
\begin{tikzpicture}[
  box/.style={draw,line width=0.6pt,rounded corners=1.5pt,inner xsep=4pt,inner ysep=3pt,font=\footnotesize,align=center},
  ext/.style={box,double,double distance=1pt},
  alg/.style={box,fill=black!8,font=\footnotesize\scshape},
  dep/.style={line width=0.6pt,-{Stealth[length=4.5pt]}},
]
\node[ext] (labels) at (-5.2,3.6) {Distance labels of~\cite{LTY26}\\(Label Hypothesis)};
\node[box] (vs)     at (0,3.6)    {Virtual Source Lemma};
\node[ext] (red)    at (5.2,3.6)  {Reduction of~\cite{RHMGZ23} with the\\Discretized Boosting Lemma};
\node[alg] (exact)  at (0,2.1)    {ExactSSSP};
\node[ext] (neg)    at (-5.2,0.7) {Reduction of~\cite{A24}};
\node[box] (dd)     at (0,0.7)    {Dual Distance Lemma\\on the graph of~\cite{ADPW25}};
\node[ext] (rec)    at (5.2,0.7)  {Recursion of~\cite{ADW26b}, repaired by\\the Annulus, Repaired Graph, Witness\\Interval, Incision and Repaired\\Recursion Lemmas};
\node[alg] (nssp)   at (-5.2,-0.8) {NegativeSSSP};
\node[alg] (dual)   at (-1.9,-0.8) {DualFlow};
\node[alg] (face)   at (1.2,-0.8)  {CommonFaceFlow};
\node[alg] (cut)    at (5.2,-0.8)  {ExactMinCut};
\draw[dep] (labels.south) -- (exact.north west);
\draw[dep] (vs) -- (exact);
\draw[dep] (red.south) -- (exact.north east);
\draw[dep] (exact.south west) -- (neg.north);
\draw[dep] (exact) -- (dd);
\draw[dep] (exact.south east) -- (rec.north);
\draw[dep] (neg) -- (nssp);
\draw[dep] (nssp) -- (dual);
\draw[dep] (dd.south) -- (dual.north);
\draw[dep] (dd.south) -- (face.north);
\draw[dep] (rec) -- (cut);
\end{tikzpicture}
\caption{How the algorithms depend on each other. Boxes with a double frame are the results of other
works that we use as black boxes or repair.}
\label{fig:overview}
\end{figure}
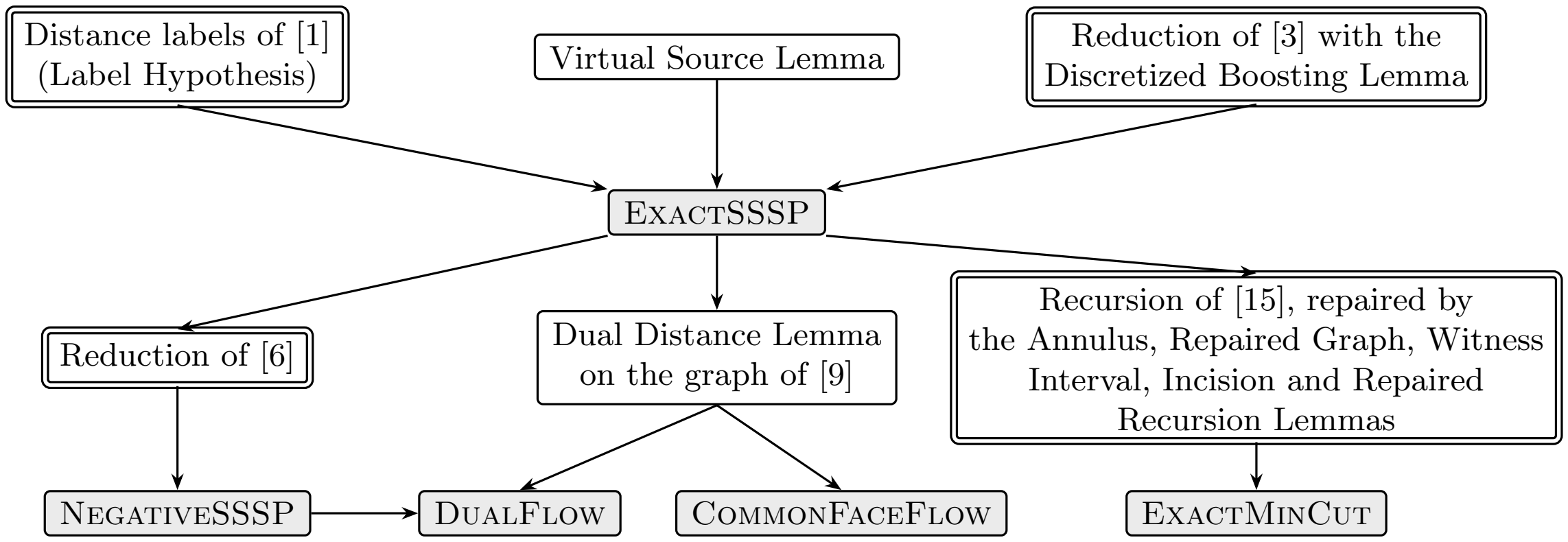

Figure~\ref{fig:overview} shows how the results build on one another. Every algorithm of this paper
calls \Exact{}, directly or through \NegSSSP{}.

\paragraph{\Exact{}.} The reduction of~\cite{RHMGZ23} turns exact SSSP into $O(\log^2n)$ calls to an
approximate oracle, it is noted in~\cite[Sec.~3.1]{RHMGZ23} that this boosting works in distributed
models, and distance labels as in~\cite{LTY26} answer the calls. Two ingredients are needed beyond this
combination. First, the calls ask for distances from a virtual source, which the query
of~\cite{LTY26} does not provide. The \LemVS{} answers such calls with a minimum over all bags on the
chain of a vertex. Second, the discretization of the reduction is only sketched
in~\cite[App.~C.1]{RHMGZ23}; done as written, an update can drive an arc length to $-\eta$. We use
$\ell_i+\eta$ as the base of every update and prove the resulting contraction and exactness in the
\LemBoost{}.

\paragraph{\NegSSSP{}.} This is the reduction of~\cite{A24} with \Exact{} as its oracle. We simulate
its minor-aggregation steps with the low-congestion shortcuts of planar networks.

\paragraph{\DualFlow{} and \FaceFlow{}.} We weight the face-disjoint graph $\hat G$ of~\cite{ADPW25}:
weight $0$ on the cycles around faces, the dual lengths on the edges that cross primal edges, and a
large weight on the star edges. The \LemDD{} shows that $\hat G$ then has a negative cycle exactly
when the dual has one, and that its distances between corners are the dual distances between their
faces. Since $\hat G$ is a planar network of hop diameter at most $3D+2$ that $G$ simulates with a
factor~$2$ in rounds~\cite{ADPW25}, primal algorithms solve dual distance problems. Maximum flow then
follows by the reduction of~\cite{MN95} as used in~\cite[Sec.~6.1]{ADPW25}, and by the method
of~\cite{H81} for $s$ and $t$ on a common face.

\paragraph{\MinCut{}.} This is the technical core of the paper. The algorithm of~\cite{ADW26b}
computes $O(\log n)$ levels of shortest paths in annuli of the dual and approximates the minimum cut
in $\Ot(D)$ rounds, for lack of an exact SSSP algorithm. We run one exact SSSP on the whole auxiliary
graph of a level, with weights of the same kind as in the \LemDD{}, so that the losses of the
approximation disappear. This alone is not enough. As we read~\cite[Remark~3.3]{ADW26b}, the
auxiliary graphs lack, at $s$ and $t$, the properties proved for them in~\cite{ADW26b}; we repair them
and prove these properties in the \LemAnn{} and the \LemRep{}. The recursion
of~\cite[Alg.~3]{ADW26b} drops indices whose witness, by~\cite[Claim~2.4]{ADW26b}, can lie outside the
range of the current annulus. On a network with $11$ vertices it reports $37$ for a minimum cut
of~$15$, also with exact shortest paths. Two more shortest paths per level make it exact, which the
\LemWit{}, the \LemInc{} and the \LemRec{} prove. Their proofs fill the appendices.

\subsection{Verification and status}

Three statements are machine-checked in Lean~4 with Mathlib (Section~\ref{sec:lean}): the \LemVS{},
the \LemDD{} for every graph with the face, dual and star structure of $\hat G$, including its
construction from a combinatorial map, and the discretized boosting step with base $\ell_i+\eta$. In
each case the hypotheses are stated in Section~\ref{sec:lean}. The CONGEST algorithms, their round
counts, every argument that uses planarity and the whole proof of the \ThmCut{} are proved on paper
only. Section~\ref{sec:numerics} reports numerical checks of the sequential meaning of the
statements. They test finite families, not round counts, and no proof relies on them.

The \ThmSSSP{}, the \ThmNeg{}, the \ThmFlow{} and the \ThmFace{} are proved given the \HypL{} and the
other cited results. We completed the proofs of~\cite{LTY26} where this was needed, replacing some
intermediate statements; our public repository holds them, and Remark~\ref{rem:status} gives the details. The \ThmCut{} is
proved given~\cite{ADW26b}, the \ThmSSSP{}, and the five lemmas named above, whose proofs are in
Appendices~\ref{app:R} to~\ref{app:S}. The proofs of these lemmas have had two referee passes by
separate AI referees and an adversarial pass, which found no counterexample; each referee pass was
followed by a revision in answer to it. The proof of the \ThmCut{} rests on our readings of the
recursion of~\cite{ADW26b}, which we state in Section~\ref{sec:thmD}. The proofs were developed and
checked with AI assistance and have not yet been checked by a human expert (see the section on the use
of AI).

Section~\ref{sec:prelim} recalls the model, the reduction of~\cite{RHMGZ23} and the face-disjoint
graph of~\cite{ADPW25}. Section~\ref{sec:A} gives \Exact{}, Section~\ref{sec:B} gives \NegSSSP{},
Section~\ref{sec:flow} gives \DualFlow{} and \FaceFlow{}, and Section~\ref{sec:thmD} gives \MinCut{}.
Sections~\ref{sec:numerics} and~\ref{sec:lean} describe the numerical checks and the formal
verification, and Section~\ref{sec:open} lists open questions. Appendix~\ref{app:R} defines the
repaired graphs and proves the \LemAnn{} and the \LemRep{}, Appendix~\ref{app:O} proves the \LemInc{},
and Appendix~\ref{app:S} proves the \LemWit{} and the \LemRec{}.

\section{Preliminaries}\label{sec:prelim}

\subsection{Model, plane graphs and duals}\label{sec:model}

The network $G=(V,E)$ is connected, with $n\ge2$ vertices and hop diameter $D$. Every vertex has a
distinct identifier of $O(\log n)$ bits. All vertices know a polynomial upper bound on $n$ and on
the largest absolute weight or capacity $W$, so that every weight, every distance and every
identifier fits into $O(1)$ messages. For the randomized results, every vertex has private random bits. In a directed
network every edge $\{u,v\}$ carries the arc $(u,v)$, the arc $(v,u)$, or both, each with its own
integer weight or capacity. For a digraph $H$ with arc lengths, a walk from $a$ to $b$ has as length
the sum of the lengths of its arcs, and $\dist_H(a,b)$ is the infimum of the lengths of these walks,
or $\infty$ if there is none. For a vertex set $X$, $G[X]$ is the induced subgraph. The notation
$\Ot(\cdot)$ hides factors polylogarithmic in $n$, and \emph{with high probability} means with
probability at least $1-n^{-c}$ for an arbitrary fixed constant $c$.

\paragraph{Embeddings.} A combinatorial embedding gives every vertex $v$ a cyclic order
$r_v(0),\dots,r_v(\deg v-1)$ of its incident edges, its \emph{rotation}; indices are taken modulo
$\deg v$. As in~\cite{LTY26,LP19,ADPW25}, every vertex knows its rotation in a fixed planar
embedding of $G$. Such an embedding is computed in $\Ot(D)$ rounds~\cite{GH16}. We also use the
embedding as a drawing of $G$ in the sphere $S^2$. The \emph{corner} $(v,j)$ is the angle at $v$
between $r_v(j-1)$ and $r_v(j)$; it lies in one face of $G$. Every edge $e=uv$ has two \emph{darts},
$u\to v$ and $v\to u$, and $\rev(d)$ is the reverse of a dart $d$. We write $\lft(d)$ and $\rgt(d)$
for the faces on the left and on the right of $d$; they coincide for a bridge. The \emph{dual graph}
$G^*$ has a node for every face of $G$ and, for every dart $d$, a \emph{dual dart} $d^*$ from
$\rgt(d)$ to $\lft(d)$.

\paragraph{Determinism.} Our reductions add no randomness of their own to two steps: the planar
embedding and the labels of the \HypL{}. We call an algorithm \emph{deterministic relative to the
labels} if it is deterministic once these two steps are, for $G$ and for the auxiliary planar networks
that we simulate on $G$. We do not claim that these steps are deterministic. If they succeed only with high probability, a union
bound over the polylogarithmically many calls carries this over. The algorithm
of~\cite[Thm.~4]{LTY26} is stated as deterministic, while the decomposition that its proof cites
from~\cite{LP19} uses a randomized separator and is stated as randomized
in~\cite[Lemma~5.1]{ADPW25}; a deterministic decomposition is stated in~\cite[Cor.~1]{JMR25}, which
we have not checked. The embedding algorithm of~\cite{GH16} is used as deterministic
in~\cite{ADW26a,ADW26b}; we could not read the full text of~\cite{GH16}.

\subsection{Exact distances from approximate ones}\label{sec:boostprelim}

Let $s$ be a source in a digraph with nonnegative arc lengths $\ell$. In~\cite[Def.~1.5,
1.6]{RHMGZ23}, an estimate $\tilde d:V\to\mathbb R_{\ge0}$ is \emph{smoothly $\alpha$-approximate} if
$\tilde d(v)\ge\dist_\ell(s,v)$ for every $v$, $\tilde d(s)=0$, and
$\tilde d(v)-\tilde d(u)\le\alpha\cdot\dist_\ell(u,v)$ for all $u,v$; in particular
$\tilde d(v)-\tilde d(u)\le\alpha\,\ell(u,v)$ for every arc $(u,v)$. Algorithm~1
of~\cite{RHMGZ23} sets $\ell_1:=\ell$ and, for $i=1,\dots,I$ with $I=O(\alpha\log(n\Delta))$, where
$\Delta$ bounds the distances, asks an oracle for a smoothly $\alpha$-approximate estimate
$\tilde d_i$ for $\ell_i$ and sets
\[
  \ell_{i+1}(u,v):=\ell_i(u,v)-\tilde d_i(v)/\alpha+\tilde d_i(u)/\alpha .
\]
It outputs $\sum_i\tilde d_i(v)/\alpha$ rounded to the nearest integer, which equals
$\dist_\ell(s,v)$ for integer lengths~\cite[Lemma~3.1]{RHMGZ23}. By~\cite[Thm.~1.7]{RHMGZ23},
Algorithm~3 of~\cite{RHMGZ23} computes smoothly $(1+\eps)$-approximate estimates with $O(\log n)$
calls to a $(1+O(\eps/\log n))$-approximate distance oracle, on graphs with weights in
$[1,\mathrm{poly}(n)]$. With $\alpha=2$, exact SSSP therefore takes $O(\log n)$ iterations of
Algorithm~1 with $O(\log n)$ oracle calls each, so $O(\log^2n)$ calls to a $(1+\eps_0)$-approximate
oracle with $\eps_0=\Theta(1/\log n)$~\cite[Lemma~3.1, Thm.~1.7, App.~C.1]{RHMGZ23}. The calls are
made on $G$ with new arc lengths and some arcs deleted (the cross-level edges of the level graphs
of~\cite[Def.~5.7]{RHMGZ23}), from $s$ or from a virtual vertex that is joined to every vertex $u$
by an arc of nonnegative length $\varphi(u)$. Outside the oracle every step is local: an arc needs
the estimates at its two ends, which costs one round. Since Algorithm~1 can make arc lengths much
smaller than~$1$, it is sketched in~\cite[App.~C.1]{RHMGZ23} how to add $\eta=1/\mathrm{poly}(n)$ to
every arc when the oracle is called, and then to scale and discretize. It is noted
in~\cite[Sec.~3.1]{RHMGZ23} that the boosting works in sequential, parallel and distributed models.

\subsection{The face-disjoint graph}\label{sec:fdg}

\begin{figure}[t]\centering
\begin{tikzpicture}[x=1cm,y=1cm]
  \def\rr{0.55}
  \coordinate (A) at (0,0); \coordinate (B) at (3,0);
  \coordinate (C) at (3,3); \coordinate (D) at (0,3);
  \coordinate (A1) at ($(A)+(22.5:\rr)$);  \coordinate (A2) at ($(A)+(67.5:\rr)$);
  \coordinate (A0) at ($(A)+(225:\rr)$);   \coordinate (B1) at ($(B)+(135:\rr)$);
  \coordinate (B0) at ($(B)+(315:\rr)$);   \coordinate (C1) at ($(C)+(247.5:\rr)$);
  \coordinate (C2) at ($(C)+(202.5:\rr)$); \coordinate (C0) at ($(C)+(45:\rr)$);
  \coordinate (D2) at ($(D)+(315:\rr)$);   \coordinate (D0) at ($(D)+(135:\rr)$);
  % the plane graph G (not part of the face-disjoint graph)
  \draw[faint,dashed] (A)--(B)--(C)--(D)--(A)--(C);
  % E_R: one cycle per face, weight 0
  \begin{scope}[line width=0.9pt,blue!70!black]
    \draw (A1)--(B1)--(C1)--cycle;
    \draw (A2)--(C2)--(D2)--cycle;
    \draw (A0) to[out=-135,in=-45,looseness=0.75] (B0) to[out=-45,in=45,looseness=0.75] (C0)
               to[out=45,in=135,looseness=0.75] (D0) to[out=135,in=225,looseness=0.75] (A0);
  \end{scope}
  % E_C: one edge per edge of G, between the two corners on its sides at one end
  \begin{scope}[line width=0.9pt,red!75!black]
    \draw (A1) arc[start angle=22.5,end angle=-135,radius=\rr];
    \draw (B1) arc[start angle=135,end angle=-45,radius=\rr];
    \draw (C1) arc[start angle=247.5,end angle=202.5,radius=\rr];
    \draw (C2) arc[start angle=202.5,end angle=45,radius=\rr];
    \draw (D2) arc[start angle=315,end angle=135,radius=\rr];
  \end{scope}
  % E_S: spokes from the star centres, weight M
  \foreach \v/\c in {A/A1,A/A2,A/A0,B/B1,B/B0,C/C1,C/C2,C/C0,D/D2,D/D0} \draw[e] (\v)--(\c);
  \foreach \c in {A1,A2,A0,B1,B0,C1,C2,C0,D2,D0} \node[d,minimum size=4pt] at (\c) {};
  \foreach \v in {A,B,C,D} \node[v,minimum size=7pt] at (\v) {};
  \node[lab] at (2.05,0.85) {$f_1$};
  \node[lab] at (0.95,2.15) {$f_2$};
  \node[lab] at (-0.55,1.5) {$f_0$};
  % legend
  \begin{scope}[shift={(5.3,2.6)}]
    \draw[line width=0.9pt,blue!70!black] (0,0)--(0.6,0); \node[lab,anchor=west] at (0.7,0) {$E_R$: weight $0$};
    \draw[line width=0.9pt,red!75!black] (0,-0.55)--(0.6,-0.55); \node[lab,anchor=west] at (0.7,-0.55) {$E_C$: dual dart lengths};
    \draw[e] (0,-1.1)--(0.6,-1.1); \node[lab,anchor=west] at (0.7,-1.1) {$E_S$: weight $M$};
    \node[v,minimum size=7pt] at (0.3,-1.65) {}; \node[lab,anchor=west] at (0.7,-1.65) {star centre};
    \node[d,minimum size=4pt] at (0.3,-2.2) {}; \node[lab,anchor=west] at (0.7,-2.2) {corner};
    \draw[faint,dashed] (0,-2.75)--(0.6,-2.75); \node[lab,anchor=west] at (0.7,-2.75) {edge of $G$};
  \end{scope}
\end{tikzpicture}
\caption{The face-disjoint graph $\hat G$ of a $4$-cycle with a diagonal. The faces $f_1,f_2$ and
the outer face $f_0$ become vertex-disjoint cycles of $E_R$ edges. Each edge of $G$ is crossed by one
$E_C$ edge between two corners at one of its ends. The \LemDD{} weights $E_R$ with $0$, $E_C$
with the dual dart lengths and $E_S$ with a large $M$.}
\label{fig:fdg}
\end{figure}
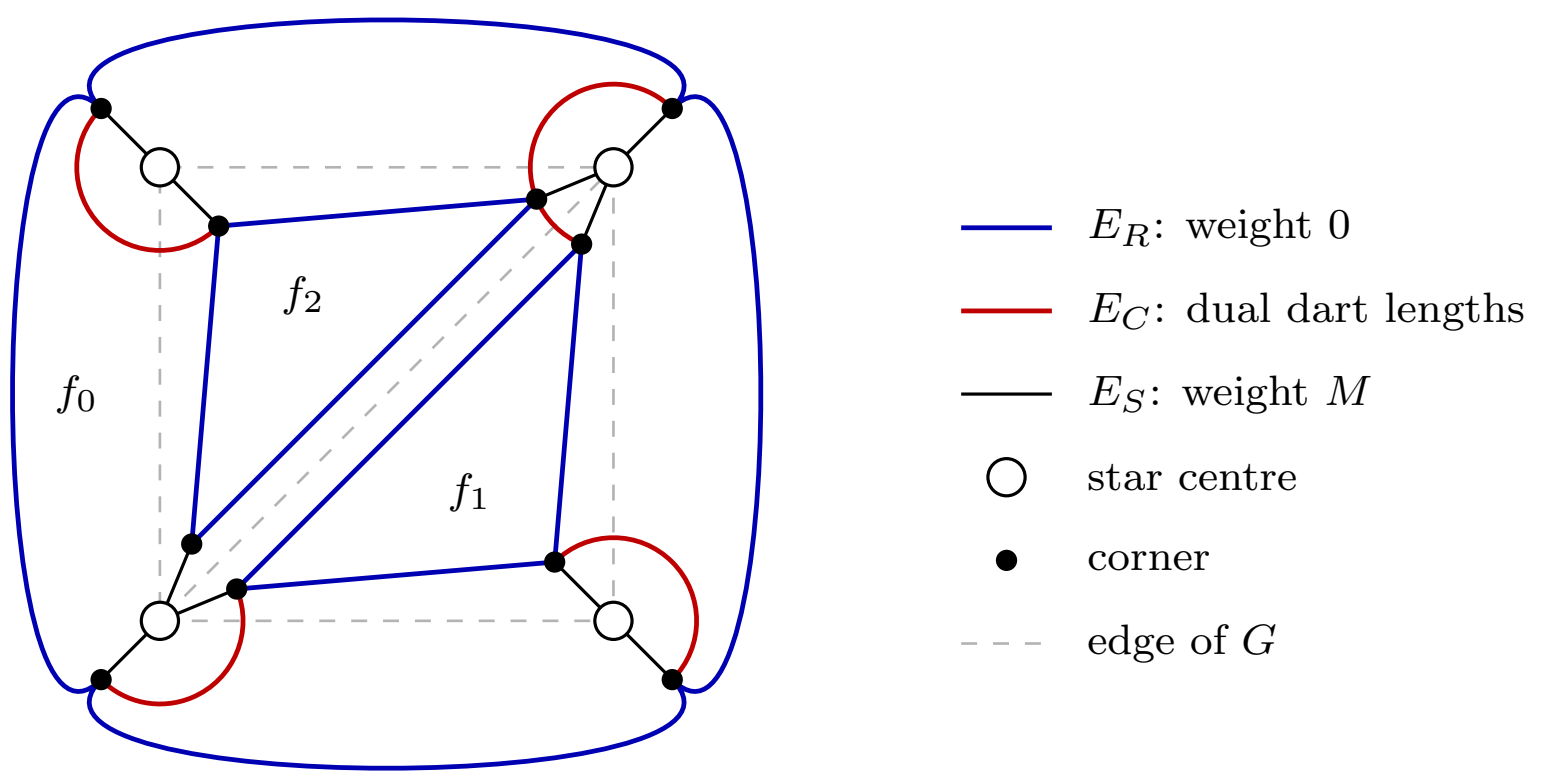

The face-disjoint graph $\hat G$ of~\cite[Sec.~3]{ADPW25} has a \emph{star centre} for every vertex
$v$ of $G$ and a \emph{corner node} $v_i$ for every corner of $v$. Its edges come in three sets
(Figure~\ref{fig:fdg}):
\begin{itemize}
\item the star edges $vv_i$, which form $E_S$;
\item the set $E_R$, with two edges for each edge $uv$ of $G$, one for each side, joining the corner nodes of $u$
  and $v$ that are consecutive along the boundary walk of the face on that side; so every face $f$ of
  $G$ becomes a cycle of $E_R$ edges through the corner nodes of $f$, and these cycles are
  vertex-disjoint;
\item the set $E_C$, with one edge for each edge $e$ of $G$, joining the two corner nodes at one end of $e$ that
  lie on the two sides of $e$.
\end{itemize}
By~\cite[Sec.~3, Properties 1 to 3 and 5]{ADPW25}, $\hat G$ is planar and is built in $O(1)$ rounds,
every vertex hosting its star centre and its corner nodes; its hop diameter is at most $3D$ between
star centres, hence at most $3D+2$; every $r$-round algorithm on $\hat G$ runs in $2r$ rounds on $G$;
and the edges of $E_C$ correspond one to one to the edges of $G^*$, with both ends knowing the weight
and direction of the corresponding dual edge.

\section{Exact SSSP with nonnegative weights}\label{sec:A}

\begin{named}{Exact SSSP Theorem}\label{thm:A}
Let $G$ be a planar digraph with integer weights in $\{0,\dots,n^{O(1)}\}$, hop diameter $D$ and
source $s$. Given the \HypL{}, the algorithm \Exact{} computes, for every vertex $v$, the distance
$\dist(s,v)$ and the parent of $v$ in a shortest-path tree, in $\Ot(D)$ rounds. It is deterministic relative to the
labels, and outside the labels its local computation is polynomial.
\end{named}

The algorithm has three parts: distance labels modelled on~\cite{LTY26}, which we state as the \HypL{}; the
\LemVS{}, which answers distance queries from a virtual source with the labels; and the reduction
of~\cite{RHMGZ23}, with a discretization that we prove in the \LemBoost{}.

\subsection{Bag hierarchies and the labels}\label{sec:labels}

Let $G$ be a digraph with arc lengths in $[0,\infty]$. A \emph{bag hierarchy} is a finite rooted tree
whose nodes $X$, the bags, carry a vertex set $X$, a set $A_X$ of arcs of $G$ with both ends in $X$,
and a set $H_X\subseteq X$ of hubs, such that:
\begin{enumerate}[(V1)]
\item the root bag contains every vertex and every arc;
\item every vertex of $X\setminus H_X$ lies in exactly one child of $X$;
\item every arc of $A_X$ with both ends in $X\setminus H_X$ lies in $A_Y$ for some child $Y$ of $X$.
\end{enumerate}
A leaf has no children, so by (V2) all its vertices are hubs. We write $\dist_X$ for distances in
the digraph $(X,A_X)$. The \emph{chain} of a vertex $v$ starts at the root and, from a bag $X$ with
$v\notin H_X$, continues with the child of $X$ that contains $v$, which is unique by (V2). So the
chain has one bag per depth and ends at the first bag in which $v$ is a hub.

\begin{named}{Label Hypothesis}\label{prop:L}
Let $G$ be a planar network with a known embedding, with integer arc lengths in $[1,W]$ for some
$W=\mathrm{poly}(n)$ and arcs in both directions on every edge, and let $\eps\in(0,1]$ with
$1/\eps\le n^{O(1)}$. A CONGEST algorithm computes, in $\Ot(D/\eps)$ rounds, a bag hierarchy of
height $O(\log n)$ such that:
\begin{enumerate}[(L1)]
\item every bag $X$ has a tree $T_X$ of edges of $G$ that spans $X$ and has diameter $\Ot(D)$;
  $|H_X|=\Ot(D)$ for non-leaf bags, and every leaf has $\Ot(D)$ vertices; for every depth, every edge
  of $G$ lies in at most two of the trees $T_X$ of bags of that depth; every vertex knows the bags of its chain and, for each of its edges, the bags
  whose tree or arc set contains it;
\item for every non-leaf bag $X$, every vertex $v$ with $X$ on its chain and every hub $h$ of $X$, the
  vertex $v$ knows values with
  $\dist_G(v,h)\le\tilde d_X(v,h)\le(1+\eps)\dist_X(v,h)$ and
  $\dist_G(h,v)\le\tilde d_X(h,v)\le(1+\eps)\dist_X(h,v)$; these values have polylogarithmically
  many bits.
\end{enumerate}
\end{named}

The hypothesis is modelled on~\cite{LTY26}. There, the bags are those of~\cite[Lemma~9]{LTY26}, with
$A_X$ the arcs of $G[X]$, $T_X$ the tree $T[X]$, and $H_X$ the set $\partial S^+_X$ of~\cite{LTY26} for a
non-leaf bag and $X$ for a leaf. In this notation (V1) to (V3) correspond to~\cite[Lemma~9 (1) and
(7)]{LTY26}, (L1) to~\cite[Lemma~9 (2), (4), (5), (6) and Lemma~10]{LTY26}, the chain of $v$ to the
sequence $X_0,\dots,X_d$ of~\cite[Def.~4]{LTY26}, and (L2) to~\cite[Thm.~9]{LTY26}. The form that we
obtain has other bags and trees; see Remark~\ref{rem:status}. The bounds in (L2) hold whether the
values approximate distances in $G$, as in~\cite[Def.~4]{LTY26}, or distances in the bag, as in the
construction of~\cite[Lemma~16]{LTY26}, and a two-sided approximation takes this form after scaling. Polynomial local time is stated in~\cite[Thm.~4]{LTY26}, whose proof consists of
these labels and one query.

\begin{remark}[status of the hypothesis]\label{rem:status}
All results of this paper that use $\Ot(D)$-round SSSP are stated given the \HypL{}, since no single
statement of~\cite{LTY26} has this form and~\cite{LTY26} is a preprint. While working through the proofs
of~\cite{LTY26}, we completed them where this was needed, replacing some intermediate statements. The
completed proofs are in the file \texttt{notes/LEMMA15.md} of the repository named in
Section~\ref{sec:lean}. They give the hypothesis from the black boxes listed there, for bags whose $A_X$
are the arcs of a subgraph of $G[X]$, in place of the induced subgraphs of~\cite{LTY26}. All arguments
below use only the properties stated above, so they apply to these bags. The completed proofs find planar
realizations by exhaustive search, which CONGEST does not charge. A polynomial bound on local computation,
as stated in~\cite[Thm.~4]{LTY26}, would need a realization algorithm that respects a prescribed cyclic
order of the terminals, which we have not established. The completed proofs are not part of this paper.
Like the rest of this work, they have been checked by AI referees, adversarial passes and numerical
tests, and not by a human expert.
\end{remark}

\subsection{Queries from a virtual source}

Let $\varphi:V\to[0,\infty]$ be \emph{offsets}. A virtual source $\sigma$ is joined to every vertex $u$
by an arc of length $\varphi(u)$, so
\[
  d(\sigma,v)=\min_{u\in V}\bigl(\varphi(u)+\dist_G(u,v)\bigr).
\]

\begin{named}{Virtual Source Lemma}\label{lem:V}
Let $\eps\in[0,\infty)$, and let a bag hierarchy satisfy (V1) to (V3). Suppose that for every bag
$X$, every vertex $u$ with $X$ on its chain and every $h\in H_X$ there are values with
$\dist_G(u,h)\le\tilde d_X(u,h)\le(1+\eps)\dist_X(u,h)$ and
$\dist_G(h,u)\le\tilde d_X(h,u)\le(1+\eps)\dist_X(h,u)$. Let
\[
  A_X(h):=\min_{u:\,X\text{ on the chain of }u}\bigl(\varphi(u)+\tilde d_X(u,h)\bigr),\quad
  \tilde d(v):=\min_{X\text{ on the chain of }v}\ \min_{h\in H_X}\bigl(A_X(h)+\tilde d_X(h,v)\bigr).
\]
Then $d(\sigma,v)\le\tilde d(v)\le(1+\eps)\,d(\sigma,v)$ for every vertex $v$.
\end{named}

\begin{proof}
\emph{Lower bound.} Let $u$ attain $A_X(h)$. Then
\[
  A_X(h)+\tilde d_X(h,v)\ge\varphi(u)+\dist_G(u,h)+\dist_G(h,v)\ge\varphi(u)+\dist_G(u,v)\ge
  d(\sigma,v).
\]

\emph{Upper bound.} We may assume that $d(\sigma,v)<\infty$. Lengths are nonnegative and the graph
is finite, so some $u^*$ and some walk $P^*$ from $u^*$ to $v$ attain
$d(\sigma,v)=\varphi(u^*)+|P^*|$. Start at the root, which contains all vertices and arcs of $P^*$
by (V1) and lies on the chains of $u^*$ and of $v$. Let $X$ be a bag that contains all vertices of $P^*$,
with all arcs of $P^*$ in $A_X$ and with $X$ on the chains of $u^*$ and of $v$. If $P^*$ has no hub of
$X$, then every vertex of $P^*$ lies in exactly one child by (V2). For an arc $ab$ of $P^*$, (V3) gives
a child $Y$ with $ab\in A_Y$, so $a,b\in Y$, and by uniqueness the child of $a$ is the child of $b$.
Hence all vertices of $P^*$ lie in one child $Y$, by (V3) and uniqueness again all its arcs lie in
$A_Y$, and since $u^*$ and $v$ are no hubs of $X$, the bag $Y$ lies on their chains. So we can move
to $Y$. Since the tree is finite and all vertices of a leaf are hubs, we reach a bag $X$ with these
properties and a hub $h$ of $X$ on $P^*$. Split $P^*$ at $h$ into $P_1$ from $u^*$ to $h$ and $P_2$
from $h$ to $v$. Both lie in $(X,A_X)$, so
\[
  \tilde d(v)\le A_X(h)+\tilde d_X(h,v)\le\varphi(u^*)+(1+\eps)|P_1|+(1+\eps)|P_2|
  \le(1+\eps)\,d(\sigma,v).\qedhere
\]
\end{proof}

The minimum must run over the whole chain: the lowest bag that separates $u$ from $v$ depends on $u$,
and the best $u$ is not known in advance. Only the lower bound by distances in $G$ and the upper
bound by distances in the bag are used, so the lemma covers both kinds of labels above.

\begin{corollary}\label{cor:V}
Given the labels of the \HypL{}, $(1+\eps)$-approximate distances from a virtual source with offsets
$\varphi$ that every vertex knows for itself take $\Ot(D)$ further rounds, and these rounds are
deterministic.
\end{corollary}

\begin{proof}
Every leaf $X$ broadcasts its arcs and their lengths over $T_X$. By (L1) and planarity it has $\Ot(D)$
arcs, so this takes $\Ot(D)$ rounds, and then every vertex of $X$ knows $\dist_X$ on $X$. With these
values for the leaves and (L2) for the other bags, the \LemVS{} applies. For every bag $X$ and hub
$h$, the value $A_X(h)$ is a minimum of values known to vertices of $X$. One convergecast over $T_X$
computes the $|H_X|$ minima of a bag, pipelined, and one broadcast returns them, in
$O(\mathrm{diam}(T_X)+|H_X|)=\Ot(D)$ rounds. The bags of one depth run in parallel, since every edge
lies in at most two of their trees by (L1), and the $O(\log n)$ depths run one after the other. Then
every $v$ computes $\tilde d(v)$ locally.
\end{proof}

\subsection{Exact distances from approximate ones}

Algorithm~1 of~\cite{RHMGZ23} with the perturbation of~\cite[App.~C.1]{RHMGZ23} calls its oracle
on the lengths $\ell_i+\eta$. If the update is applied with base $\ell_i$, as written, a length
can become negative: take one arc $s\to v$ with $\ell_i(s,v)=0$, and the answer $\tilde d(s)=0$,
$\tilde d(v)=\alpha\eta$. It is smooth for $\ell_i+\eta$ and does not underestimate, and the update
gives $\ell_{i+1}(s,v)=0-\alpha\eta/\alpha=-\eta$. We therefore use $\ell_i+\eta$ as the base.

Let $G$ be a finite digraph with $n$ vertices, source $s$, integer lengths $\ell\ge0$, and every
vertex reachable from $s$. Let $\alpha\ge1$ and $\eta\ge0$. A \emph{run} consists of lengths
$L_0=\ell,L_1,L_2,\dots$ and estimates $\tilde d_0,\tilde d_1,\dots$ such that for all $i$:
\begin{enumerate}[(O1)]
\item $\tilde d_i(s)=0$;
\item $\tilde d_i(u)\ge\dist_{L_i+\eta}(s,u)$ for every vertex $u$;
\item $\tilde d_i(b)-\tilde d_i(a)\le\alpha\bigl(L_i(a,b)+\eta\bigr)$ for every arc $(a,b)$;
\end{enumerate}
and $L_{i+1}(a,b)=L_i(a,b)+\eta-\tilde d_i(b)/\alpha+\tilde d_i(a)/\alpha$ for every arc. A smoothly
$\alpha$-approximate estimate for the lengths $L_i+\eta$ satisfies (O1) to (O3). Write
$D_i(u):=\dist_{L_i}(s,u)$ and $q:=1-1/\alpha$.

\begin{named}{Discretized Boosting Lemma}\label{lem:boost}
In every run, for all $i$ and all vertices $u$:
\begin{enumerate}[(a)]
\item the lengths $L_i$ are nonnegative on every arc, so the oracle sees lengths at least $\eta$;
\item the distances contract: $D_{i+1}(u)\le q\,D_i(u)+(n-1)\eta$;
\item hence $D_i(u)\le q^iD_0(u)+\alpha(n-1)\eta$;
\item the estimates telescope: $D_0(u)-D_I(u)\le\sum_{i<I}\tilde d_i(u)/\alpha\le D_0(u)-D_I(u)+I(n-1)\eta$;
\item if $q^ID_0(u)+\alpha(n-1)\eta<1/2$ and $I(n-1)\eta<1/2$, then rounding
  $\sum_{i<I}\tilde d_i(u)/\alpha$ to the nearest integer gives $D_0(u)=\dist_\ell(s,u)$.
\end{enumerate}
\end{named}

For $\eta=0$ this is~\cite[Lemma~3.1]{RHMGZ23}.

\begin{proof}
(a) By induction on $i$. By (O3),
$L_{i+1}(a,b)=L_i(a,b)+\eta-(\tilde d_i(b)-\tilde d_i(a))/\alpha\ge0$.

For a walk $p$ from $a$ to $b$ with $k$ arcs, summing the update along $p$ gives
\[
  L_{i+1}(p)=L_i(p)+k\eta-\tilde d_i(b)/\alpha+\tilde d_i(a)/\alpha. \tag{$*$}
\]
\emph{Lower step.} For a walk $p$ from $s$ to $u$, ($*$) and (O1) give $L_{i+1}(p)\ge
L_i(p)-\tilde d_i(u)/\alpha\ge D_i(u)-\tilde d_i(u)/\alpha$, so $D_{i+1}(u)\ge
D_i(u)-\tilde d_i(u)/\alpha$.

\emph{Upper step.} By (a) the lengths $L_i$ are nonnegative, so some walk $p$ from $s$ to $u$ with
at most $n-1$ arcs has $L_i(p)=D_i(u)$. By ($*$), $D_{i+1}(u)\le L_{i+1}(p)\le
D_i(u)+(n-1)\eta-\tilde d_i(u)/\alpha$.

(b) Every walk is at least as long under $L_i+\eta$ as under $L_i$, so (O2) gives $D_i(u)\le
\tilde d_i(u)$. With the upper step, $D_{i+1}(u)\le D_i(u)-D_i(u)/\alpha+(n-1)\eta$.

(c) By induction from (b), using $(n-1)\eta\sum_{j<i}q^j\le\alpha(n-1)\eta$.

(d) Sum the lower and the upper step over $i<I$.

(e) $D_0(u)$ is an integer $k$, since $\ell$ is integral. By (d), (c) and $D_I(u)\ge0$,
\[
  k-\tfrac12<k-q^Ik-\alpha(n-1)\eta\le k-D_I(u)\le\sum_{i<I}\tilde d_i(u)/\alpha\le
  k+I(n-1)\eta<k+\tfrac12 .\qedhere
\]
\end{proof}

The \LemBoost{} needs (O3) only on the arcs, and its bounds hold at every vertex, so no sum
over the vertices enters.

\subsection{The algorithm}\label{sec:proofA}

We state \Exact{} for a virtual source $\sigma$ joined to a set $S$ of vertices by arcs of
integer lengths $x(u)\in\{0,\dots,n^{O(1)}\}$, $u\in S$, as in the oracle of~\cite[Def.~24]{A24};
the source $s$ is the case $S=\{s\}$, $x(s)=0$. \NegSSSP{} and \MinCut{} use this form.

\begin{algorithm}[t]
\caption{\Exact{}: exact distances from a virtual source}\label{alg:exact}
\begin{algorithmic}[1]
\Require planar network $G$, integer weights $w\ge0$, integer offsets $x$ on a set $S$, all of polynomial size
\Ensure every vertex $v$ learns $d(\sigma,v)$ and a parent in a shortest-path tree
\State $w\gets n^2w+1$ and $x\gets n^2x$ on $S$ \Comment{shortest paths for the new $w$ stay shortest for the old}
\State Offsets of a source $\sigma_0$: $\varphi\gets x$ on $S$ and $\varphi\gets M_1$ elsewhere; $L_0\gets w$, with these offsets as arcs from $\sigma_0$
\For{$i=0,\dots,I-1$} \Comment{$I=O(\log n)$, $\eta=\Theta(1/(In))$}
  \State $\hat\ell_i\gets L_i+\eta$ rounded up to multiples of $\eta/n$ and divided by $\eta/n$ \Comment{integers, at least $n$}
  \State Run Algorithm~3 of~\cite{RHMGZ23} on $\hat\ell_i$ with accuracy $1/3$; at each of its $O(\log n)$ oracle calls:
  \State \hspace{\algorithmicindent}merge the sources into offsets on $G$, and give deleted arcs and missing directions a large length
  \State \hspace{\algorithmicindent}compute the labels of the \HypL{} with accuracy $\eps_0=\Theta(1/\log n)$
  \State \hspace{\algorithmicindent}compute $A_X(h)$ by a convergecast for all bags $X$ and hubs $h$, and $\tilde d(v)$ at every $v$
  \State $\tilde d_i\gets$ the estimate of Algorithm~3, multiplied by $\eta/n$
  \State $L_{i+1}(a,b)\gets L_i(a,b)+\eta-\tilde d_i(b)/2+\tilde d_i(a)/2$ for every arc, the arcs from $\sigma_0$ included
\EndFor
\State Every $v$ rounds $\sum_{i<I}\tilde d_i(v)/2$ to the integer $D(v)$; $D(v)=M_1$ means unreachable, and otherwise $d(\sigma,v)=\lfloor D(v)/n^2\rfloor$
\State Every reachable $v$ picks as parent $\sigma$, if $v\in S$ and $x(v)=D(v)$, or an in-neighbour $u$ with $D(u)+w(u,v)=D(v)$
\end{algorithmic}
\end{algorithm}

\begin{proof}[Proof of the \ThmSSSP{}]
\emph{Tree.} Replace every weight $w$ by $w'=n^2w+1$ and every offset $x$ by $n^2x$. A shortest
path $P$ for $w'$ from $\sigma$ has $w'(P)=n^2w(P)+k$, where $k\le n-1<n^2$ is its number of arcs
in $G$, so it is a shortest path for $w$, and $d_w(\sigma,v)=\lfloor d_{w'}(\sigma,v)/n^2\rfloor$.
Once every vertex knows $d_{w'}(\sigma,\cdot)$, a vertex $v$ that is reachable picks as parent
either $\sigma$, if $v\in S$ and $n^2x(v)=d_{w'}(\sigma,v)$, or an in-neighbour $u$ with
$d_{w'}(\sigma,u)+w'(u,v)=d_{w'}(\sigma,v)$, in one round. Since $w'>0$, the distance strictly
decreases along parent pointers between vertices of $G$, so they form a shortest-path tree. From
now on all weights and offsets are the new ones; let $W_{\max}$ be the largest weight and $x_{\max}$
the largest offset.

\emph{Offsets.} Fix an integer $M_1>x_{\max}+nW_{\max}$ and replace $\sigma$ by a virtual source $\sigma_0$
with $\varphi(u)=x(u)$ for $u\in S$ and $\varphi(u)=M_1$ for $u\notin S$. Every vertex is then
reachable, which the \LemBoost{} needs, and $d(\sigma_0,u)=\min\{d(\sigma,u),M_1\}$. Every
finite $d(\sigma,u)$ is less than $M_1$, so the value $M_1$ means that $u$ is unreachable from
$\sigma$; since the output is exact, every vertex can tell.

\emph{Iterations.} \Exact{} runs Algorithm~1 of~\cite{RHMGZ23} with base $\ell_i+\eta$ on $G$ with the
source $\sigma_0$, where the virtual arcs are updated like all other arcs, so that each vertex $u$
keeps its current offset $L_i(\sigma_0,u)$. Take $\alpha=2$, $I=\lceil2\ln(3M_1)\rceil+2=O(\log n)$ and
$\eta=1/(6I(n+1))$, so that the hypotheses of part~(e) of the \LemBoost{} hold, with $n+1$ vertices,
$D_0\le M_1$ and $I\ge\alpha$. In iteration $i$ the estimate $\tilde d_i$ is computed as follows.
\begin{itemize}
\item[(W1)] Round every length $L_i(a,b)+\eta\ge\eta$ up to a multiple of $\eta/n$ and divide by
  $\eta/n$. The result $\hat\ell_i(a,b)$ is an integer, at least $n$. The rounding never shortens
  a walk and stretches it by a factor of at most $1+1/n$.
\item Run Algorithm~3 of~\cite{RHMGZ23} with accuracy $1/3$ on $\hat\ell_i$, from $\sigma_0$. By
  \cite[Thm.~1.7]{RHMGZ23} it returns a smoothly $(4/3)$-approximate estimate for $\hat\ell_i$ after
  $O(\log n)$ calls to a $(1+\eps_0)$-approximate oracle. The first call asks for accuracy~$1$, and
  every later call~\cite[Alg.~4, line~3]{RHMGZ23} for accuracy $\eps_0=\eps/100$ with
  $\eps=(1/3)/(10t)$ and $t=1+\log_2(3n\Delta)=O(\log n)$~\cite[Alg.~3, lines~2 and~3]{RHMGZ23}, where
  $\Delta$ bounds the distances; so $\eps_0=\Theta(1/\log n)$. Multiplied by $\eta/n$, this
  estimate satisfies (O1) to (O3) for $L_i+\eta$ with the factor $(4/3)(1+1/n)\le2=\alpha$.
\item[(W2)] Every oracle call of Algorithm~3 is on the network $G$ with some arcs deleted and with
  lengths that are a common factor $c\ge1$ times $\hat\ell_i$, from a virtual source $\sigma$. In
  the first call $\sigma=\sigma_0$ and $c=1$~\cite[Alg.~3, line~4]{RHMGZ23}. In the level graphs
  of~\cite[Def.~5.7]{RHMGZ23} a new virtual vertex $\sigma$ is joined to every vertex, $\sigma_0$
  included; the source written $s$ in~\cite[Alg.~4, line~3]{RHMGZ23} is this $\sigma$. Since
  $\sigma_0$ has no entering arc other than the one from $\sigma$, the two merge into the offsets
  $\varphi(u)=\min\{\varphi_\sigma(u),\,\varphi_\sigma(\sigma_0)+c\,\hat\ell_i(\sigma_0,u)\}$ on the
  vertices of $G$, where the second term is left out when the arc $\sigma_0u$ is deleted; the one
  value $\varphi_\sigma(\sigma_0)$ is broadcast in $O(D)$ rounds. Give every deleted arc and every
  missing direction of an edge the length $cM_2$, for an integer $M_2>\max\varphi/c$ that one
  aggregation makes known to all vertices. Every vertex
  $v$ has $d(\sigma,v)\le\varphi(v)<cM_2$, so no walk through such an arc is shorter than
  $d(\sigma,v)$, and the distances from $\sigma$ do not change. Compute the labels of
  the \HypL{} for the integer lengths $\hat\ell_i$ and $M_2$ with accuracy $\eps_0$,
  and multiply them by $c$. Then Corollary~\ref{cor:V} answers the call.
\end{itemize}
The lengths stay polynomial. By (O1) and (O3) along a shortest walk, $\tilde d_i(u)\le
\alpha\dist_{L_i+\eta}(\sigma_0,u)\le2\bigl(D_i(u)+n\eta\bigr)\le2(M_1+2)$, using
part~(c) of the \LemBoost{}. So every iteration raises a length by at most $\eta+M_1+2$, and after
$I=O(\log n)$ iterations all $\hat\ell_i$ and all offsets are bounded by a polynomial in $n$, as
the \HypL{} needs.
After $I$ iterations, every vertex $u$ outputs $\sum_{i<I}\tilde d_i(u)/\alpha$ rounded to an
integer, which is $\dist(\sigma_0,u)$ by part~(e) of the \LemBoost{}.

\emph{Rounds.} Algorithm~1 and Algorithm~3 act locally between oracle calls: an update of an arc
needs the estimates at its two ends, one round. There are $O(\log n)$ iterations with $O(\log n)$
oracle calls each, and every call costs $\Ot(D/\eps_0)=\Ot(D)$ rounds by the \HypL{} and
Corollary~\ref{cor:V}. In total this is $\Ot(D)$. Lengths, offsets and estimates are sums over $O(\log n)$ levels of
rationals with polynomial numerators and denominators, and label values have polylogarithmically many
bits by (L2), so each takes $\Ot(1)$ messages, which $\Ot(\cdot)$ absorbs. Local computation
outside the labels is polynomial: minima, shortest paths in leaves and arithmetic on these numbers.

\emph{Determinism.} The reduction of~\cite{RHMGZ23} is deterministic, and so is
Corollary~\ref{cor:V}, so \Exact{} adds no randomness to the labels.
\end{proof}

\begin{remark}
The \LemBoost{} counts $O(\log n)$ iterations of Algorithm~1, each with one call of
Algorithm~3, which makes $O(\log n)$ oracle calls. So the reduction makes $O(\log^2n)$ oracle calls;
the count ``$O(\log n)$'' in~\cite[App.~C.1]{RHMGZ23} refers to the calls of Algorithm~3.
\end{remark}

\section{Negative weights}\label{sec:B}

\begin{named}{Negative Weights Theorem}\label{thm:B}
Let $G$ be a planar digraph with integer weights of polynomial size, possibly negative, hop
diameter $D$ and source $s$. Given the \HypL{}, the algorithm \NegSSSP{}, which is the reduction
of~\cite{A24} with \Exact{} as its oracle, either computes $\dist(s,v)$ for every vertex $v$ or reports
that $G$ has a negative cycle, in $D\cdot(\log n)^{O(\sqrt{\log n})}=D\cdot n^{o(1)}$ rounds with high
probability.
\end{named}

\begin{proof}
By~\cite[Thm.~2]{A24}, an algorithm for SSSP with nonnegative integer weights in $T(n,D)$ rounds
gives negative-weight SSSP in $O\bigl((T(n,D)+\sqrt n+D)(\log n)^{O(\sqrt{\log n})}\bigr)$ rounds
with high probability. We go through the proof and bound each term on planar networks.

\emph{Oracle calls.} All calls to the nonnegative SSSP oracle are calls to the oracle
of~\cite[Def.~24]{A24}: the network $G$ with new nonnegative integer weights of polynomial size,
some arcs removed by a large polynomial weight, and a virtual source joined to a set of vertices by
arcs with given integer lengths. \Exact{} answers such a call exactly in
$\Ot(D)$ rounds, so $T=\Ot(D)$. In~\cite[Thm.~26]{A24}, the oracle is answered with the
reduction of~\cite{RHMGZ23}, whose own virtual source merges with that of the call into one set
of offsets; \Exact{} merges them in the same way.

\emph{The term $\sqrt n+D$.} It enters only through~\cite[Thm.~14]{A24}, which simulates the steps
of the Distributed Minor-Aggregation Model in $\Ot(\sqrt n+D)$ CONGEST rounds. This simulation is
used in~\cite[Cor.~27, Cor.~33, Cor.~40]{A24} and in the proof of~\cite[Cor.~58]{A24}; every other
step is an oracle call or local computation. On planar networks, part-wise aggregation takes
$\Ot(D)$ rounds with low-congestion shortcuts~\cite[Cor.~4.6]{ADPW25}, so a step of the
minor-aggregation model costs $\Ot(D)$ rounds~\cite[Lemma~4.8]{ADPW25}.

\emph{Negative cycles.} By~\cite[Thm.~60]{A24}, Algorithm~5 of~\cite{A24} outputs ERROR if $G$ has
a negative cycle, and otherwise computes all distances from $s$ with high probability. Its
remaining global steps are a constant number of aggregations, in $O(D)$ rounds each.

Substituting $T=\Ot(D)$ and $\Ot(D)$ for $\sqrt n+D$ gives $D\cdot(\log n)^{O(\sqrt{\log n})}$.
\end{proof}

\section{Dual distances and maximum flow}\label{sec:flow}

\subsection{The weighted face-disjoint graph}

Let every dual dart $d^*$ of $G$ carry an integer length $\mu(d^*)$, of any sign, with
$|\mu|\le\Lambda$. We weight the face-disjoint graph $\hat G$ of Section~\ref{sec:fdg} as follows.
Both arcs of every $E_R$ edge get length $0$. The $E_C$ edge of an edge $e$ joins the corner nodes
of the two faces of $e$ at one end of $e$; for each dart $d$ of $e$, the arc of this edge from the
corner node in $\rgt(d)$ to the corner node in $\lft(d)$ gets length $\mu(d^*)$. Both arcs of every
$E_S$ edge get length $M$, where
\[
  M>|V(\hat G)|\cdot\Lambda .
\]
Every corner node $a$ lies in one face of $G$, which we call the face of $a$. The graph $\hat G$
may have parallel $E_C$ edges, at vertices of degree~$2$, and $E_C$ loops, at vertices of
degree~$1$, where the loop is the dual of a bridge. A negative loop is a negative cycle of both
$\hat G$ and $G^*$; its host detects it locally.

\begin{named}{Dual Distance Lemma}\label{lem:D}
With these weights:
\begin{enumerate}[(i)]
\item the graph $\hat G$ has a negative cycle if and only if $G^*$ has one;
\item if $G^*$ has no negative cycle, then for all corner nodes $a$ and $b$ with faces $g$ and $h$,
  $\dist_{\hat G}(a,b)=\dist_{G^*}(g,h)$ if $\dist_{G^*}(g,h)<\infty$, and
  $\dist_{\hat G}(a,b)\ge M$ otherwise.
\end{enumerate}
Planarity is not used: the statement holds for the face-disjoint graph of a graph embedded in any
surface.
\end{named}

\begin{proof}
\emph{Projection.} A walk in $\hat G$ that avoids the star centres consists of $E_R$ arcs and $E_C$
arcs. An $E_R$ arc joins two corner nodes of the same face, and an $E_C$ arc for $d$ goes from a
corner node of $\rgt(d)$ to one of $\lft(d)$. Replacing every corner node by its face, dropping the
$E_R$ arcs and replacing every $E_C$ arc for $d$ by $d^*$ gives a walk in $G^*$ of the same length,
closed if the walk was closed.

\emph{Lift.} Conversely, let $Q$ be a walk in $G^*$ from $g$ to $h$, and let $a,b$ be corner nodes
of $g$ and $h$. The $E_R$ edges of a face form a cycle through all its corner nodes, of length $0$
in both directions. Replace every dual dart $d^*$ of $Q$ by the $E_C$ arc for $d$ and join
consecutive arcs, and $a$ and $b$ to the ends, by $E_R$ paths inside the faces. The result is a walk
from $a$ to $b$ in $\hat G$ of the same length as $Q$, which avoids the star centres.

\emph{Star centres.} A simple cycle or path of $\hat G$ through a star centre uses two $E_S$ arcs
and at most $|V(\hat G)|-2$ other arcs, each of length at least $-\Lambda$. Its length is at least
$2M-|V(\hat G)|\Lambda>M>0$.

(i) A negative cycle of $G^*$ lifts to a negative closed walk of $\hat G$, which contains a
negative simple cycle. A negative simple cycle of $\hat G$ avoids the star centres, by the previous
paragraph, and projects to a negative closed walk of $G^*$, which contains a negative cycle.

(ii) By (i), $\hat G$ has no negative cycle, so $\dist_{\hat G}(a,b)$ is attained by a simple path,
or is $\infty$. Such a path either avoids the star centres and projects to a walk from $g$ to $h$,
so its length is at least $\dist_{G^*}(g,h)$, or it has length greater than $M$. If
$\dist_{G^*}(g,h)<\infty$, a shortest path from $g$ to $h$ in $G^*$ has fewer than $|V(\hat G)|$
darts, so $\dist_{G^*}(g,h)<|V(\hat G)|\Lambda<M$, and its lift gives
$\dist_{\hat G}(a,b)\le\dist_{G^*}(g,h)$. Hence $\dist_{\hat G}(a,b)=\dist_{G^*}(g,h)$. If
$\dist_{G^*}(g,h)=\infty$, every path from $a$ to $b$ passes a star centre, so
$\dist_{\hat G}(a,b)\ge M$.
\end{proof}

The weights on $E_C$, with their directions, are already given in~\cite[Sec.~3, Property~5]{ADPW25},
where $\hat G$ serves to simulate minor aggregation in the dual. The lengths $0$ on $E_R$ and $M$ on
$E_S$ make the distances of $\hat G$ equal to dual distances, so that primal algorithms run on
$\hat G$ solve dual distance problems. Problems that change when the zero-length $E_R$ cycles are
contracted, such as the girth, are not covered.

\begin{named}{Dual SSSP Corollary}\label{cor:dual}
Let $G$ be a planar network with a known embedding, and let every dual dart carry an integer length
of polynomial size. Given the \HypL{}, for a face $f_0$, given by one of its corners, every vertex learns
$\dist_{G^*}(f_0,g)$ for the face $g$ of each of its corners:
\begin{enumerate}[(a)]
\item in $\Ot(D)$ rounds if the lengths are nonnegative, deterministically relative to the labels;
\item in $D\cdot n^{o(1)}$ rounds with high probability in general, where the algorithm reports a
  negative cycle if $G^*$ has one.
\end{enumerate}
\end{named}

\begin{proof}
Every vertex builds its star centre, its corner nodes and their edges, with the lengths above and
$M=|V(\hat G)|\Lambda+1$, in $O(1)$ rounds~\cite[Sec.~3, Property~1]{ADPW25}. The proof of planarity
in~\cite[App.~A]{ADPW25} gives an embedding of $\hat G$ that every vertex derives from its own
rotation. The graph $\hat G$ has $|V(\hat G)|\le 7n$ nodes and integer lengths of polynomial size,
its hop diameter is at most $3D$ between star centres~\cite[Property~2]{ADPW25} and hence at most
$3D+2$, and an $r$-round algorithm on $\hat G$ runs in $2r$ rounds on $G$~\cite[Property~3]{ADPW25}.
The hosts check for negative loops, drop the other loops and, of parallel edges, keep the lightest arc
in each direction, which changes no distance, no planarity and no hop bound. Run \Exact{}, or \NegSSSP{},
on $\hat G$ from the given corner node of $f_0$. The \LemDD{} turns the result into dual distances:
a negative cycle of $\hat G$ is reported exactly when $G^*$ has one, and otherwise a distance of
at least $M$ means that the face is unreachable. Here \Exact{} uses the embedding of $\hat G$ above. Everything else is local, which gives the
determinism in~(a).
\end{proof}

Part~(a) improves the $\Ot(D^2)$ of~\cite[Lemma~2.2]{ADPW25} for nonnegative lengths, and part~(b)
improves it once $D$ exceeds $(\log n)^{O(\sqrt{\log n})}$.

\subsection{Maximum flow}

A flow is given on darts: $f(\rev d)=-f(d)$, and every dart $d$ has a capacity $c(d)\ge0$, the
capacity of its direction, which is $0$ for a direction that the network does not have. A flow is
feasible if $f(d)\le c(d)$ for every dart, and an $st$-flow of value $\beta$ is a feasible flow whose
net outflow is $\beta$ at $s$, $-\beta$ at $t$ and $0$ elsewhere. We use two standard facts about
a connected plane graph. First, for a function $\psi$ on the faces,
$f(d):=\psi(\lft d)-\psi(\rgt d)$ is a circulation: the darts leaving a vertex, in rotation order,
share faces with their neighbours, and the sum telescopes. Second, the darts whose duals form a
simple directed cycle of $G^*$ are the darts of a cut that leave one side of it.

\begin{named}{Maximum Flow Theorem}\label{thm:C}
Let $G$ be a directed planar network with terminals $s,t$ and integer capacities in
$\{0,\dots,n^{O(1)}\}$. Given the \HypL{}, the algorithm \DualFlow{} computes the value of a maximum
$st$-flow and the flow on every edge. With high probability it takes
$\min\{\Ot(D^2),D\cdot n^{o(1)}\}$ rounds. The algorithm of~\cite[Thm.~1.2]{ADPW25}, which it runs
alongside, gives $\Ot(D^2)$ rounds alone, without the hypothesis.
\end{named}

\begin{algorithm}[t]
\caption{\DualFlow{}: maximum flow by negative-cycle tests in the dual}\label{alg:dualflow}
\begin{algorithmic}[1]
\Require directed planar network $G$, integer capacities $c$, terminals $s,t$
\Ensure the value of a maximum $st$-flow and the flow on every edge
\State Alongside, in alternate rounds, run the algorithm of~\cite[Thm.~1.2]{ADPW25}; stop with the first to finish
\State Compute a path $P$ from $s$ to $t$ in a breadth-first tree
\State Binary search for the largest $\beta\in[0,nU]$ such that $G^*$ with the lengths $c_\beta$ has no negative cycle, each test by the \CorDual{}(b)
\State For this $\beta$, compute the distances $\psi$ in $G^*$ from a fixed face by the \CorDual{}(b)
\State Every dart $d$ gets the flow $\psi(\lft d)-\psi(\rgt d)+\beta[d\in P]-\beta[\rev(d)\in P]$
\end{algorithmic}
\end{algorithm}

\begin{proof}
\DualFlow{} (Algorithm~\ref{alg:dualflow}) is the reduction of~\cite{MN95} in the form
of~\cite[Sec.~6.1]{ADPW25}, with the \CorDual{} as the negative-cycle test.

Compute a path $P$ from $s$ to $t$ in a breadth-first tree, as a sequence of darts, in $O(D)$
rounds. For $\beta\ge0$ let $c_\beta(d)=c(d)-\beta$ for $d\in P$, $c_\beta(d)=c(d)+\beta$ for
$\rev(d)\in P$, and $c_\beta(d)=c(d)$ otherwise, and give every dual dart $d^*$ the length
$c_\beta(d)$. An $st$-flow of value $\beta$ exists if and only if a circulation $f$ with $f\le c_\beta$
exists: subtract or add $\beta$ along $P$. Such a circulation exists if and only if $G^*$ has no
negative cycle. If $\psi$ is a distance function in $G^*$ without negative cycles, the circulation
$f(d)=\psi(\lft d)-\psi(\rgt d)$ satisfies $f(d)\le c_\beta(d)$ by the triangle inequality on $d^*$.
Every dual dart exists, so $G^*$ is strongly connected and $\psi$ is finite. Conversely, a
circulation $f\le c_\beta$ has net flow $0$ across every cut, so every simple dual cycle has length
at least $0$.

The maximum flow value is an integer in $[0,nU]$, where $U$ is the largest capacity, so a binary
search on $\beta$ takes $O(\log n)$ negative-cycle tests. Each is part~(b) of the \CorDual{} in
$D\cdot n^{o(1)}$ rounds. For the largest feasible $\beta$, the same corollary gives $\psi$ at every
corner. The tail $u$ of a dart $d$ holds the corner nodes of $\lft(d)$ and $\rgt(d)$ next to $d$ and
knows whether $d$ or $\rev(d)$ lies on $P$, so it computes the flow
$f(d)+\beta[d\in P]-\beta[\rev(d)\in P]$ locally. The minimum in the round bound comes from the run
alongside the algorithm of~\cite[Thm.~1.2]{ADPW25}: the first to finish tells all vertices over a
breadth-first tree, in $O(D)$ more rounds, and the other stops.
\end{proof}

A minimum $st$-cut follows from the set of vertices reachable from $s$ in the residual graph, which
the \ThmSSSP{} computes in $\Ot(D)$ rounds with weight $0$ on every residual arc.

\begin{named}{Common-Face Flow Theorem}\label{thm:Cp}
If in addition $s$ and $t$ lie on a common face, then, given the \HypL{}, the algorithm \FaceFlow{}
computes a maximum $st$-flow, its value and the flow on every edge in $\Ot(D)$ rounds. It is
deterministic relative to the labels.
\end{named}

\begin{algorithm}[t]
\caption{\FaceFlow{}: maximum flow with $s$ and $t$ on a common face}\label{alg:faceflow}
\begin{algorithmic}[1]
\Require directed planar network $G$, integer capacities $c$, terminals $s,t$ on a common face
\Ensure a maximum $st$-flow and its value
\State Find a common face $f$ and corners $c_s,c_t$ in it: run \Exact{} on $\hat G$ with weight $0$ on $E_R$, weight $\deg(s)+1$ elsewhere, offset $j$ at corner $j$ of $s$, and offset $\deg(s)+1$ at all other nodes
\State Add a chord $e$ from $t$ to $s$ through $c_t$ and $c_s$, with infinite capacity; it splits $f$ into $f_1,f_2$
\State Split the corner nodes of $c_s,c_t$ in $\hat G$ along $e$; run \Exact{} on this graph from a corner node of $f_1$, with the lengths of the \LemDD{} for $\mu=c$, to get dual distances $\psi$
\State Every dart $d$ gets the flow $\psi(\lft d)-\psi(\rgt d)$; the value is $\psi(f_2)$
\end{algorithmic}
\end{algorithm}

\begin{proof}
\FaceFlow{} (Algorithm~\ref{alg:faceflow}) uses the method of~\cite{H81}, which works for darts and so for directed networks, as noted
in~\cite{ADPW25}.

\emph{A common face.} Number the corners of $s$ as $0,\dots,\deg(s)-1$ in rotation order, and let $s$
broadcast $\deg(s)$. Run
the \ThmSSSP{} on $\hat G$ with weight $0$ on both arcs of every $E_R$ edge, weight
$\deg(s)+1$ on all other arcs, and a virtual source with offset $j$ at the corner node of corner
$j$ of $s$ and offset $\deg(s)+1$ at all other nodes. The $E_R$ edges of a face form a cycle through
its corner nodes, so a corner node of $t$ gets distance $j<\deg(s)+1$ exactly when corner $j$ is the
corner of $s$ of smallest number in its face, and distance at least $\deg(s)+1$ when its face has no
corner of $s$. The vertex $t$ picks a corner $c_t$ of smallest distance $j$, which lies in a common
face $f$ since $s$ and $t$ have one, and broadcasts $c_t$ and $j$ in $O(D)$ rounds. Then $s$ knows its
corner $c_s:=j$ in $f$. This takes $\Ot(D)$ rounds.

\emph{The chord.} Add a chord $e$ from $t$ to $s$ inside $f$, through the corners $c_t$ and $c_s$, with
infinite capacity. It splits $f$ into $f_1:=\rgt(e)$ and $f_2:=\lft(e)$. Let $\psi$ be the distance
from $f_1$ in $G^*-\{e^*,\rev(e)^*\}$, where every dual dart $d^*$ has the length $c(d)\ge0$. This
dual graph is the dual of $G+e$ with $e$ contracted, which is connected, so $\psi$ is finite.

\emph{Computing $\psi$.} In $\hat G$, split the corner nodes of $c_s$ and $c_t$ into two each, one
for each side of $e$, hosted by $s$ and $t$. Each new copy keeps the $E_S$ edge of the old node and those of its $E_R$ and $E_C$ edges that
belong to the edge of $G$ on its side of $e$. This cuts the $E_R$ cycle of $f$ into two paths, one
through the corner nodes of $f_1$ and one through those of $f_2$, and $e$ gets no $E_C$ edge. The value
$M$ is taken for the new number of nodes. This split comes before the hosts drop loops and parallel
edges as in the proof of the \CorDual{}, since a loop at a terminal of degree~$1$ can stop being one. The
proof of the \LemDD{} uses of the $E_R$ edges of a face only that they connect its corner nodes
with length~$0$, so it applies to this graph and to $G^*-\{e^*,\rev(e)^*\}$. The modified graph is
still planar, has hop diameter at most $3D+2$, and is simulated by $G$ with a factor $2$, since the
split nodes are hosted locally and no $E_R$ edge of an edge of $G$ is removed. The \ThmSSSP{} on
it, from a corner node of $f_1$, gives $\psi$ in $\Ot(D)$ rounds.

\emph{Flow.} Set $f(d):=\psi(\lft d)-\psi(\rgt d)$ for every dart $d$ of $G$, and
$f(e):=\psi(f_2)-\psi(f_1)=\psi(f_2)$. This is a circulation in $G+e$, and $f(d)\le c(d)$ for every
dart $d$ of $G$ by the triangle inequality. Restricted to $G$, it is an $st$-flow of value
$\psi(f_2)$, since it returns $\psi(f_2)$ units from $t$ to $s$ through $e$. A shortest path $Q$
from $f_1$ to $f_2$ in $G^*-\{e^*,\rev(e)^*\}$ can be taken simple, and with $\rev(e)^*$ it forms a
simple dual cycle. Its darts leave one side of a cut of $G+e$, and since $\rev(e)$, from $s$ to $t$,
is among them, the darts of $Q$ form an $st$-cut of $G$ of capacity $\psi(f_2)$. So the flow is
maximum. Every vertex computes the flow on its darts from $\psi$ at its corners.

\emph{Determinism.} The two runs of \Exact{} on graphs derived from $\hat G$ use embeddings
that the vertices derive locally. All other steps are local or broadcasts.
\end{proof}

The \ThmFace{} improves the $(1-o(1))$-approximation for undirected networks in
$D\cdot n^{o(1)}$ rounds of~\cite[Thm.~1.3]{ADPW25}, whose proof uses the same method with an
approximate dual SSSP and finds the common face by part-wise aggregation. For any $s$ and $t$ in
undirected networks, the \ThmCut{} below gives the exact value of a minimum $st$-cut in
$\Ot(D)$ rounds, but not the flow on every edge.

\section{Minimum cuts in undirected networks}\label{sec:thmD}

\begin{named}{Minimum Cut Theorem}\label{thm:D}
Let $G$ be an undirected planar network with integer capacities in $\{0,\dots,n^{O(1)}\}$, hop
diameter $D$, and terminals $s,t$ whose identifiers all vertices know. Given the \HypL{}, the algorithm
\MinCut{} computes the value of a minimum $st$-cut, which is the value of a maximum $st$-flow, in
$\Ot(D)$ rounds. Every vertex learns this value, its side of a minimum $st$-cut, and which of its edges
cross that cut. It is deterministic relative to the labels.
\end{named}

\MinCut{} does not compute the flow on every edge; for that, the \ThmFlow{} is the best bound we know.

The exact value of a minimum $st$-cut takes $\Ot(D^2)$ rounds by~\cite[Thm.~1.2]{ADPW25}, for
directed and hence for undirected networks, or $D\cdot n^{o(1)}$ rounds by the \ThmFlow{}, both
with high probability. In $\Ot(D)$ rounds, a $(1-o(1))$-approximation for undirected networks is
computed deterministically in~\cite[Thm.~1.1]{ADW26b}, with approximate SSSP ``for the lack of
almost-$D$-rounds exact distributed SSSP''. \MinCut{} combines that algorithm with
\Exact{}. The combination needs two repairs. The auxiliary graphs of~\cite{ADW26b}, on
which we run \Exact{}, need a repair at $s$ and $t$; the \LemAnn{} and the \LemRep{}
state what the repaired graphs provide. The recursion of~\cite[Alg.~3]{ADW26b} can miss the minimum
cut, also with exact shortest paths; with two more shortest paths per level it is exact, by the
\LemWit{}, the \LemInc{} and the \LemRec{}. The proofs of these lemmas are in
Appendices~\ref{app:R}, \ref{app:O} and~\ref{app:S}, and Algorithm~\ref{alg:mincut} states the whole algorithm.

\paragraph{The algorithm of~\cite{ADW26b}.} It implements the algorithm of~\cite{R83} in the dual graph.
A shortest path $P$ is computed from a face of $s$ to a face of $t$, and the dual is cut open along
$P$, which gives an annulus in which every node $p_j$ of $P$ has two copies $p_j^0$ and $p_j^1$. For
the middle index $i$, a shortest path $P_i$ from $p_i^0$ to $p_i^1$ is computed, and the annulus is
cut open along $P_i$ into two annuli, one for the indices below $i$ and one for those above; this is
repeated on $O(\log n)$ levels. The nodes of an annulus are \emph{node-parts}, pieces of faces of
$G$~\cite[Def.~2.9]{ADW26b}, the \emph{face-part} $p(f)$ of a node-part $f$ is the part of the
boundary of its face that it keeps~\cite[Claim~2.13]{ADW26b}, the annuli of one level are connected
and disjoint~\cite[Claim~2.12]{ADW26b}, and every edge of $G^*$ is incised at most
once~\cite[Lemma~2.6]{ADW26b}, so it has at most two copies on one level. The algorithm reports the
least length of a computed path as the value of a minimum $st$-cut; we describe the recursion that
computes these paths, and repair it, below. All shortest paths are computed in lines~2, 7, 8
and~11 of~\cite[Alg.~3]{ADW26b}, on all annuli of one level at once (on level~$0$, which computes
$P$, the only annulus is $G^*$), through~\cite[Thm.~4.1]{ADW26b}: given a source node-part $f_i$ in
every annulus $A_i$ of level $\ell$, it computes a shortest-path tree $T_i$ of $A_i$ rooted at $f_i$
such that
\begin{enumerate}[(R1)]
\item $T_i$ obeys the approximate triangle inequality~\cite[Def.~A.1]{ADW26b};
\item for every node-part $g$ of $A_i$, every vertex of $p(g)$ knows the length of the path from
  $f_i$ to $g$ in $T_i$;
\item every vertex knows, for each of its edges $e$, in which trees $T_i$ a copy of $e^*$ lies, and
  the identifiers of its ends there.
\end{enumerate}
In~\cite{ADW26b} the trees are $(1+\eps)$-approximate, and (R1) means that every subpath of a tree path
is $(1+\eps)$-approximately shortest. The \LemTrees{} below gives exact trees, for which every
subpath of a tree path is a shortest path.

\paragraph{The graph $\hat G_\ell$.} For a level $\ell$ whose incision paths are stored at the
vertices as in~\cite[Def.~3.2]{ADW26b}, the graph $\hat G_\ell$ of~\cite[Thm.~3.1]{ADW26b} has, for
every vertex $v$ of $G$, a star centre, a copy $v_i$ for every corner $i$ of $v$, and copies $v_i^L$
and $v_i^R$ at corners next to incised edges. Its $E_S$ edges join the copies of one vertex in a tree
of depth two, its $E_R$ edges are copies of edges of $G$, and each $E_C$ edge joins two copies of one
vertex and represents an edge of the dual. For the graph as defined in~\cite{ADW26b}, the following
holds.
\begin{itemize}
\item Every vertex $v$ of $G$ has at most $3\deg(v)+1$ copies, and $\hat G_\ell$ is built in $O(1)$
  rounds~\cite[Lemma~3.15]{ADW26b}. A CONGEST round on $\hat G_\ell$ takes four rounds on $G$, since
  $E_S$ and $E_C$ edges join copies of one vertex~\cite[Lemma~3.16]{ADW26b}.
\item The graph is planar~\cite[Lemma~3.5]{ADW26b}. Its hop diameter is at most $4D$ between star
  centres~\cite[Lemma~3.4]{ADW26b}, and every copy of $v$ is within two hops of the star centre of
  $v$, so the hop diameter is at most $4D+4$.
\item[(1)] Every node-part $f$ of level $\ell$ is represented by a path $\hat p(f)$ of $E_R$ edges
  that contains a copy of every vertex of $p(f)$, and these paths are vertex-disjoint components of
  $\hat G_\ell[E_R]$~\cite[Thm.~3.1(1), Lemma~3.6]{ADW26b}.
\item[(2)] For every edge between node-parts $f$ and $g$ of a level-$\ell$ annulus there is exactly
  one $E_C$ edge between $\hat p(f)$ and $\hat p(g)$, and there is none if $f$ and $g$ are not
  adjacent on level $\ell$~\cite[Thm.~3.1(2), Lemma~3.13]{ADW26b}. The other $E_C$ edges represent
  dual edges outside every annulus.
\item Every vertex $v$ knows, for each node-part $f$ with $v\in p(f)$, an identifier of $f$, a
  leader vertex of $p(f)$, and the identifier of the annulus that contains
  $f$~\cite[Lemma~3.17]{ADW26b}, or that $f$ lies in no annulus~\cite[Claim~5.8]{ADW26b}.
\end{itemize}

\paragraph{Endpoint-case edges.} Let $P=(p_0,\dots,p_m)$, and let $e_1,\dots,e_m$ be its primal
edges, where $e_k$ has $p_{k-1}$ on one side and $p_k$ on the other. To cut the dual open along $P$,
the vertex $s$ chooses an \emph{endpoint-case edge} $e_0$, an edge of $s$ on the face $p_0$ whose dual
is not on $P$, and $t$ chooses $e_{\mathrm{last}}$ on $p_m$ in the same way~\cite[Sec.~5]{ADW26b}. The
face $p_0$ is split into the node-parts $p_0^0$ and $p_0^1$ at the corner of $s$ between $e_0$ and the
next edge, and $\hat G_\ell$ is adjusted at $s$ and $t$ for this in~\cite[Remark~3.3]{ADW26b}. As we read
that remark, the resulting graph can violate (1) or (2) at $s$ or $t$ in three ways, which a
simulation found (Section~\ref{sec:numerics}).
\begin{enumerate}[(D1)]
\item If an incision path of a later level incises $e_0$, the remark handles $e_0$ ``just like any
  other incised edge''. Then the copy $s_i$ at the corner $i$ that the cut along $P$ splits receives
  an $E_R$ copy of $e_0$ as well as the copy of the next edge of $s$, so that in a leaf annulus the
  pieces of $p_0^0$ and $p_0^1$ lie on one path $\hat p$ and have distance~$0$. The same happens at
  $t$. In the smallest instance found, $G$ has the vertices $0,1,2,3$, the edges $03$, $01$, $12$,
  $23$ of capacity~$1$ and $13$ of capacity~$5$, the rotations $(03,01)$, $(12,13,01)$, $(12,23)$
  and $(23,03,13)$ at $0$, $1$, $2$ and~$3$, and $s=1$, $t=2$. The faces are $031$, $3012$ and $213$,
  the path $P$ crosses $03$ and $12$, the endpoint-case edges are $13$ and $23$, and the incision
  path of level~$1$ incises $23$. The minimum cut has capacity~$2$, and the algorithm of
  the \ThmCut{} with this graph in place of the repaired one reports~$0$.
\item The $E_C$ edge of an endpoint-case edge $e_0$ that is not incised is always placed at the
  other end $u$ of $e_0$. If $u=t$ and $e_0=e_{\mathrm{last}}$, this edge
  does not join the paths of the two node-parts that the dual of $e_0$ joins. For the network that
  consists of the single edge $st$, both $E_C$ edges are loops, and the two node-parts of $p_0$ are
  joined only through $E_S$ edges.
\item The copy $s_i^L$ of the remark is placed on the face of the corner $i$ that follows
  $e_0=e_i$ at $s$, which is $p_0$ only if $e_0$ is chosen with this corner on $p_0$.
  The rule of~\cite[Sec.~5]{ADW26b} allows every edge of $s$ on $p_0$ whose dual is not on $P$. With
  that rule taken literally, $s_i^L$ was placed on the other face of $e_0$ in $3{,}724$ of $6{,}000$ runs.
\end{enumerate}
We use the following choice and repair; away from $e_0$ and $e_{\mathrm{last}}$, the graph is the
graph of~\cite{ADW26b}, built from the set $I_\ell$ below. As in~\cite{ADW26b}, the corner $i$ of $s$ lies between the edges $e_i$ and
$e_{i+1}$ of $s$.
\begin{itemize}
\item \emph{Choice.} The vertex $s$ fixes a corner $i$ of $s$ on $p_0$ and an edge
  $e_0\in\{e_i,e_{i+1}\}$ whose dual is not on $P$. If $e_0=e_i$, the copy of $s$ at the corner is the
  copy $s_i^L$ of~\cite[Remark~3.3]{ADW26b} (\emph{mode~L}); if $e_0=e_{i+1}$, it is a copy $s_i^R$
  with left and right exchanged (\emph{mode~R}). Such a choice exists (part~(5) of the \LemRep{}), any of
  them can be taken, and $s$ sends it to its neighbours once; it is the same on all levels. Every
  edge that the rule of~\cite[Sec.~5]{ADW26b} allows is $e_i$ or $e_{i+1}$ for a corner $i$ of $s$ on
  $p_0$, so it can be read as such a choice. The same holds for $e_{\mathrm{last}}$ at $t$, with $p_m$
  in place of $p_0$.
\item \emph{Repair (a).} If $e_0$ is incised at a later level, the $E_R$ copy of $e_0$
  attached to $s_i$ in~\cite{ADW26b}, at the corner $i$ that the cut along $P$ splits, is attached
  to a fresh copy of $s$ instead, which is joined to $s_i$ by an $E_S$ edge. The corner stays split.
  The same holds for $e_{\mathrm{last}}$ at $t$.
\item \emph{Repair (b).} Every $E_C$ edge connects two $E_R$ copies of an edge $e$ of $G$ at an end
  $w$ of $e$; it joins the two copies of $w$ to which these $E_R$ copies are attached.
\end{itemize}
The choice and the repair are local at $s$, at $t$ and at the other ends of $e_0$ and
$e_{\mathrm{last}}$, and they add at most two copies and two $E_S$ edges. From now on, $\hat G_\ell$
denotes the repaired graph. Appendix~\ref{app:R} defines it in terms of labelled darts and fixes
the end of each edge that carries its $E_C$ edges.

\paragraph{Pieces.} The \LemAnn{} describes the annuli of~\cite{ADW26b} by labelled darts. Let
$I_\ell$ be the set of edges of $G$ incised before level $\ell$: the edges of $P$ and of every
incised part $P_i'$ of a level below $\ell$. The vertices store this set, in the place of the whole
paths of~\cite[Def.~3.2]{ADW26b}, and $I_0:=\emptyset$, $I_1=E(P)$ and $I_\ell\subseteq I_{\ell+1}$. A dart
$\delta$ of an edge outside $I_\ell$ carries one \emph{token} $(\delta,0)$, and a dart of an edge in
$I_\ell$ carries two, $(\delta,1)$ and $(\delta,2)$. The tokens of an edge form \emph{pairs}:
$\{(\delta,0),(\rev\delta,0)\}$ if the edge is not in $I_\ell$, and $\{(\delta,1),(\rev\delta,2)\}$ and
$\{(\delta,2),(\rev\delta,1)\}$ if it is. A pair is a copy of the dual edge on level $\ell$. In the
boundary walk of every face, list the tokens in walk order, with $(\delta,1)$ before $(\delta,2)$, and
cut the list between $(\delta,1)$ and $(\delta,2)$ for every dart $\delta$ of an edge in $I_\ell$ and,
for $\ell\ge1$, at the two corners $c^*_s$ and $c^*_t$ that the choice above fixes at $s$ and $t$. A
\emph{piece} of level $\ell$ is a maximal run of tokens with no cut inside; a face without a cut is
one cyclic piece. The \emph{token graph} $G^*_\ell$ has the pieces as nodes and the pairs as edges, a
pair joining the pieces of its two tokens. The \emph{parent} of a token $x$ of $\delta$ is $x$ itself
if the edge of $\delta$ is in $I_{\ell-1}$, and $(\delta,0)$ otherwise. On level~$1$, the token graph
is the dual cut open along an arc $\gamma$ that leaves $s$ through the corner $c^*_s$, crosses
$e_1,\dots,e_m$ once each and enters $t$ through $c^*_t$. Every face $p_t$ of $P$ splits into two
pieces $p_t^0$ and $p_t^1$, on the two sides of $\gamma$, and every other face is one piece. An
annulus of level $\ell$ is \emph{live} if the recursion of~\cite[Alg.~3]{ADW26b} is called on it at
depth $\ell$.

\begin{named}{Annulus Lemma}\label{lem:N}
Let $\ell\ge1$.
\begin{enumerate}[(a)]
\item Every live annulus of level $\ell$ is a connected component of $G^*_\ell$. So its node-parts
  are pieces and its edges are pairs, and with a token it contains the other token of its pair. The
  other components of $G^*_\ell$ are called \emph{residual}.
\item Every piece of level $\ell$ refines one piece of level $\ell-1$: the parents of its tokens lie
  in that piece. The pieces of level~$0$ are the faces of $G$.
\item The live annulus of level~$1$ is $A_1:=G^*_1$. Let every pair have the capacity of its edge as
  its length, and let $\lambda$ be the value of a minimum $st$-cut. Every walk in $A_1$ from $p_t^0$ to
  $p_t^1$ has length at least $\lambda$, and for some $t$ there is such a path of length $\lambda$.
\end{enumerate}
All three parts hold for every choice of the corners $c^*_s$ and $c^*_t$.
\end{named}

\begin{named}{Repaired Graph Lemma}\label{lem:R}
Let $\ell\ge1$, and let the endpoint data be any choice allowed above. The repaired graph
$\hat G_\ell$ has the following properties.
\begin{enumerate}[(1)]
\item For every piece $f$ of level $\ell$, the $E_R$ edges of its tokens form a path $\hat p(f)$, a
  cycle if $f$ is cyclic, which contains a copy of every vertex of $p(f)$. The paths of distinct
  pieces are vertex-disjoint, and every copy except the star centres lies on one of them. With
  part~(a) of the \LemAnn{}, this is~\cite[Thm.~3.1(1)]{ADW26b} for every node-part of a live annulus.
\item The $E_C$ edges correspond one to one to the pairs, and the $E_C$ edge of a pair $\{x,y\}$
  joins $\hat p(f)$ and $\hat p(g)$, where $f$ and $g$ are the pieces of $x$ and $y$. So every edge
  of a live annulus of level $\ell$ has exactly one $E_C$ edge, which joins the paths of its two
  ends, and every other $E_C$ edge joins the paths of two pieces of one residual component. An $E_C$
  edge is a loop only if its host has degree~$1$, and then its pair is a dual loop.
\item The graph $\hat G_\ell$ is planar. Every vertex $v$ of $G$ computes a planar rotation at each
  of its copies (Appendix~\ref{app:tableR}) from its own rotation, from which of its edges are in
  $I_\ell$, from the identifiers of its neighbours and from the endpoint data of $s$ and $t$, within
  $O(1)$ rounds.
\item The hop diameter of $\hat G_\ell$ is at most $4D+4$. A vertex $v$ has at most $3\deg(v)+1$
  copies, and $s$ and $t$ have at most $3\deg+2$. Once $I_\ell$ is stored, $\hat G_\ell$ is built in
  $O(1)$ rounds, and a CONGEST round on $\hat G_\ell$ takes four rounds on $G$.
\item The choice of the endpoint data always exists: $s$ has a corner on $p_0$, and every such
  corner has a flank whose dual is not on $P$. The same holds for $t$ and $p_m$.
\item The statement of~\cite[Lemma~3.17]{ADW26b} holds for $\hat G_\ell$: every copy, the fresh
  ones included, learns the identifiers of its node-part and of its annulus, and the leader of
  $\hat p(f)$ is a copy of a vertex of $p(f)$. In the local phase of its proof, copies are matched
  across levels by their tokens instead of by name.
\end{enumerate}
This covers an endpoint edge in $I_\ell$, also one incised anew on level $\ell$ (D1);
$e_0=e_{\mathrm{last}}$, incised or not (D2); mode~R (D3); terminals of degree~$1$ or~$2$; parallel
edges; several corners of $s$ on $p_0$ and an endpoint edge with both darts on $p_0$; and $m=0$. On
level~$0$ no edge is incised and no corner is cut, and $\hat G_0$ is the graph
of~\cite[Sec.~3.1]{ADW26b}.
\end{named}

The statements about $\hat G_\ell$ that the algorithm uses besides these,
\cite[Lemma~3.14, Claim~3.30, Cor.~3.23, Thm.~3.26]{ADW26b}, are derived in~\cite{ADW26b} from them,
from the hop diameter and from planarity. The failures (D1) to (D3) may come from a misreading
of~\cite{ADW26b} on our side, and we have not checked whether they affect the approximate algorithm
of~\cite{ADW26b}.

\paragraph{Positive capacities.} Loops cross no $st$-cut and are deleted, and parallel edges are
merged into one edge whose capacity is their sum. This changes no cut value, and an edge crosses a
cut exactly when its merged edge does. So $G$ is simple from now on, with $|E|\le3n-6$ for $n\ge3$
and $|E|=1$ for $n=2$. The algorithm of~\cite{ADW26b} takes positive weights. Replace every capacity
$c(e)$ by $c'(e):=n^2c(e)+1$. An $st$-cut $C$ has $c'(C)=n^2c(C)+|C|$ with $|C|<n^2$, so a minimum
$c'$-cut is a minimum $c$-cut and its $c$-value is $\lfloor c'(C)/n^2\rfloor$. From now on all
capacities are the positive integers $c'$, and lengths in $G^*$ are capacities of the primal edges.

\paragraph{Exact paths lose nothing.} With exact trees, the three places where the algorithm of~\cite{ADW26b}
loses a factor $1+\eps$ lose nothing.
\begin{itemize}
\item \emph{Non-crossing paths.} In~\cite{ADW26b}, the distance between the two copies of a node of
  $P$ can grow by a factor $1+\eps$ from an annulus to its child~\cite[Lemma~2.7 and
  Claim~A.2]{ADW26b}, and the overlap rule of the recursion rests on~\cite[Claim~2.3]{ADW26b}. With
  exact paths, (INV) of the \LemInc{} below shows that in every live annulus the distance between
  its two copies of $p_t$ is the distance $\lambda_t$ in $A_1$, for every index $t$ of its range. Its
  proof (Claim~\ref{cl:Oinv}) replaces the part of a shortest walk between its first and its last node
  on the incised part $Q$ by the part of $Q$ in between, lifted to the called side, which holds exactly
  one part of each node of $Q$; the parts of the walk off $Q$ stay on their side. In place
  of~\cite[Claim~2.3]{ADW26b}, (C23) of the \LemInc{} (Claim~\ref{cl:Oc23}) shows that every common
  node of $P_i$ and a bound lies on a shortest walk between the two copies of $p_\mu$ for every
  index $\mu$ that the overlap rule serves, so that rule computes these distances exactly. Ties do no harm, since only the existence of an equally short
  walk is used. In place
  of~\cite[Claim~2.4]{ADW26b} we use the \LemWit{}. Whole paths $P_i$ can cross along parts that
  are not incised. This does no harm, since only the incised parts bound annuli.
\item \emph{The flow value.} In~\cite[Sec.~5.2]{ADW26b} the value of the cut is divided by $1+\eps$ to
  obtain a flow value. With exact paths the value of the minimum cut is the maximum flow value.
\item \emph{A simple cycle.} In~\cite[Sec.~5.2]{ADW26b} the computed path $P_t$ of least length is
  turned into a simple cycle of $G^*$. Since $P$ is a shortest path, this does not increase the
  length; we prove this under \emph{The cut} below.
\end{itemize}
So it suffices to prove~\cite[Thm.~4.1]{ADW26b} with exact trees in $\Ot(D)$ rounds, and to repair
the recursion. We do the first on the repaired graph $\hat G_\ell$, with weights in the spirit of
the \LemDD{}.

\begin{named}{Exact Trees Lemma}\label{lem:exact41}
The statement of~\cite[Thm.~4.1]{ADW26b} holds with exact shortest-path trees in $\Ot(D)$ rounds:
every path of $T_i$ from $f_i$ is a shortest path of $A_i$, and so is every subpath of it. The
algorithm is deterministic relative to the labels.
\end{named}

\begin{proof}
\emph{Multigraph.} The graph $\hat G_\ell$ can have parallel $E_C$ edges, at vertices of degree~$2$,
$E_C$ loops, at vertices of degree~$1$, and at a terminal of degree~$1$ an $E_C$ edge parallel to an
$E_S$ edge (part~(2) of the \LemRep{} and Appendix~\ref{app:R}), as $\hat G$ can have the first two
(Section~\ref{sec:flow}). Their host drops the loops and, of parallel edges of any type, keeps the
lightest; this is local and changes no distance below, and no hop distance. A dropped $E_C$ edge
belongs to a dual loop or to a parallel dual edge and is in no shortest-path tree.

\emph{Weights.} Let $K:=19n+1$ and $M:=K\bigl(2\sum_ec'(e)+1\bigr)+1$, where $\sum_ec'(e)$ is
aggregated in $O(D)$ rounds. By part~(4) of the \LemRep{}, $|V(\hat G_\ell)|\le6|E|+n+2$, which is less
than $K$ since $G$ is simple. Both arcs of every $E_R$ edge get length $0$. An $E_C$ edge whose ends
lie on paths $\hat p(f)$ and $\hat p(g)$ of node-parts $f,g$ of one live annulus represents an edge of
that annulus, the copy of the dual of a primal edge $e$; both of its arcs get length $Kc'(e)+1$. All
other arcs, those of $E_S$ and of the other $E_C$ edges, get length $M$. The host of an $E_C$ edge
sets its lengths locally, since it knows the annulus of each copy (part~(6) of the \LemRep{}) and whether
it is live. The recursion decides this as described under \emph{Naming and sides} below. All lengths are integers of polynomial size.

\emph{Distances.} Let $f,g$ be node-parts of one live annulus $A$ of level $\ell$, and let
\[
  d_K(f,g):=\min_Q\bigl(Kc'(Q)+|Q|\bigr)
\]
over all paths $Q$ from $f$ to $g$ in $A$, where $|Q|$ is the number of edges of $Q$. We claim that
$\dist_{\hat G_\ell}(x,y)=d_K(f,g)<M$ for all $x\in\hat p(f)$ and $y\in\hat p(g)$, and that every walk
of length less than $M$ that starts on some $\hat p(f)$ stays on the paths $\hat p(\cdot)$ of the
annulus of $f$. For the upper bound, replace every edge of a minimizing $Q$ by its $E_C$ edge, which
exists by part~(2) of the \LemRep{}, and join consecutive ones along the paths $\hat p(\cdot)$, whose arcs
have length $0$. Every $E_C$ length is at least $1$, so a minimizing $Q$ is a simple path. It has
fewer edges than $A$ has node-parts, and the paths $\hat p(\cdot)$ are disjoint and nonempty, so
$|Q|<|V(\hat G_\ell)|<K$. Every edge of $G^*$ has at most two copies on level $\ell$, so
$c'(Q)\le2\sum_ec'(e)$ and $Kc'(Q)+|Q|<M$. For the lower bound, take a walk $W$ of length less than
$M$ from a vertex of $\hat p(f)$. It uses only $E_R$ arcs and $E_C$ arcs of live annuli. An $E_R$ arc
stays on one path $\hat p(h)$, since by part~(1) of the \LemRep{} these paths are the components of
$\hat G_\ell[E_R]$ apart from the star centres. By part~(2) of the \LemRep{}, an $E_C$ arc of a live
annulus that leaves a vertex of $\hat p(h)$ goes to a path $\hat p(h')$ with $hh'$ an edge of that
annulus, and since the paths are vertex-disjoint, this is the annulus of $h$. So $W$ stays in the
annulus $A$ of $f$, and replacing its $E_C$ arcs by the edges of $A$ they represent gives a walk $Q$
in $A$ of length $Kc'(Q)+|Q|=|W|$. Finally, since a minimizing $Q$ has $|Q|<K$, it is a shortest path
for $c'$, and $\lfloor d_K(f,g)/K\rfloor=\dist_A(f,g)$.

\emph{SSSP.} Let $f_i$ be the source in $A_i$, for the annuli that have one in this call. Join a
virtual source $\sigma$ to the copy of the leader vertex of $p(f_i)$ on $\hat p(f_i)$ with offset $0$,
for every such $i$, and to every other vertex of $\hat G_\ell$ with offset $M$. By the claim, every
vertex $y$ on a path $\hat p(g)$, with $g$ a node-part of $A_i$, has $\dist(\sigma,y)=d_K(f_i,g)<M$,
since every walk from another source or through an offset $M$ has length at least $M$. \Exact{} covers virtual sources with integer offsets. We run it on $\hat G_\ell$, which is
planar by part~(3) of the \LemRep{}, has $O(n)$ vertices and hop diameter at most $4D+4$, in $\Ot(D)$
rounds on $\hat G_\ell$ and hence on $G$. Its embedding is the rotation of part~(3) of the \LemRep{},
which every vertex computes locally for its copies.

\emph{Trees.} All copies on one path $\hat p(g)$ have the same distance. Every node-part $g\ne f_i$
of $A_i$ picks as its parent the $E_C$ arc $(x,y)$ with $y\in\hat p(g)$ and
$\dist(\sigma,x)+w(x,y)=\dist(\sigma,y)$ of smallest identifier, by one part-wise aggregation over
the paths $\hat p(\cdot)$, in $\Ot(D)$ rounds~\cite[Claim~3.30, Cor.~3.23]{ADW26b}. Such an arc exists,
since a shortest walk from $\sigma$ to $y$ has length below $M$ and enters $\hat p(g)$ through an $E_C$
arc of $A_i$. Every $E_C$ length is positive, so distances strictly decrease along parents, and the
parents form a tree $T_i$ rooted at $f_i$. Its path from $f_i$ to $g$ is a path $Q$ of $A_i$ with
$Kc'(Q)+|Q|=d_K(f_i,g)$, hence a shortest path for $c'$, and so is every subpath of it. This is (R1)
with ratio~$1$. For (R2), every vertex of $p(g)$ reads $\lfloor\dist(\sigma,y)/K\rfloor$ at its copy $y$
on $\hat p(g)$. For (R3), the parent arc of $g$ belongs to an $E_C$ edge that one end of a primal edge
$e$ hosts; it tells the other end of $e$ in one round, with the identifiers of the node-parts
from part~(6) of the \LemRep{}.
\end{proof}

\paragraph{The recursion.} We use~\cite[Alg.~3]{ADW26b} in the following form. A piece of a live
annulus that refines $p_t^\sigma$, through part~(b) of the \LemAnn{} applied level by level, is a
\emph{piece of $p_t^\sigma$}; by part~(c) of the \LemInc{} below, a live annulus $A$ has at most one, which
we call $g_t^\sigma$. Every live annulus $A$ has an index range $[j_A,k_A]$~\cite[Sec.~5.1]{ADW26b},
and $A_1$ has the range $[0,m]$. If $k_A\le j_A+1$, then $A$ is a leaf and computes shortest paths
$P_{j_A}$ and $P_{k_A}$, where $P_t$ runs from $g_t^0$ to $g_t^1$. Otherwise let
$i:=\lfloor(j_A+k_A)/2\rfloor$. The annulus computes $P_i$, read from $g_i^0$; if the tree was rooted at $g_i^1$, the path is reversed
first. Let $i_1$ and $i_2$ be
the least and the greatest index of a piece of a face of $P$ on $P_i$, let $x$ be the index of the
last piece of some $p_t^0$ on $P_i$, and $y$ the index of the first piece of some $p_t^1$ after it.
The part $Q:=P_i'$ of $P_i$ between them runs from a piece $g_x$ of $p_x^0$ to a piece $g_y$ of
$p_y^1$, has no piece of a face of $P$ inside, and is the part that is
incised~\cite[Remark~2.5, Lemma~5.7]{ADW26b}. Incising $Q$ splits $A$ into subgraphs on two sides, the
$j$-side and the $k$-side. The annulus is bounded by at most two \emph{bounds}, the copies in $A$ of
incised parts of earlier levels, $P_a$ on its $j$-side and $P_b$ on its $k$-side; $A_1$ has neither.
If $P_i$ shares a node with $P_a$,
the $j$-side gets no call, and the \emph{overlap rule}~\cite[Alg.~3, lines~9 to~14]{ADW26b} grows a
shortest-path tree in $A$ from a common node and reads from it a path from $g_t^0$ to $g_t^1$ for
every $t\in[j_A,i]$. Otherwise the $j$-side is called with the range $[j_A,i_1]$, if this range is
not empty. The $k$-side is symmetric, with $P_b$, the range $[i_2,k_A]$ and the overlap rule for
$t\in[i,k_A]$. A subgraph that contains a \emph{bad edge}, an edge of $P_a$ or $P_b$ that is not on
$P_i$ and has an end on $P_i\cap P_a$ or on $P_i\cap P_b$, is discarded~\cite[Claim~5.8]{ADW26b}. The
called subgraphs are the live annuli of the next level. The algorithm outputs the least length of a
computed path $P_t$, so not of $P$, and a cut of that capacity is read off as in~\cite[Sec.~5.2]{ADW26b}.

Five points of this description are our reading of~\cite{ADW26b}; two further changes, the names and
sides below and the indices $i_1$ and $i_2$ in the proof of the \ThmCut{}, are stated where they are used.
\begin{itemize}
\item \emph{The bound is the incised copy.} The overlap test of~\cite[Alg.~3, lines~9 to~12; Claim~2.3;
  Def.~2.8]{ADW26b} is written with the whole path $P_j$ of an earlier level. We read it with the copy
  in $A$ of the incised part of that path, the bound $P_a$. Only under this reading does the bad-edge
  rule of~\cite[Claim~5.8]{ADW26b} make sense: with whole paths, the copies of edges of $P$ on the
  parts outside the incised part would count as bad edges. The test is then local: by the invariant of
  Appendix~\ref{app:inv}, the nodes of $P_a$ and $P_b$ are exactly the nodes of $A$ that carry an old
  token of an edge off $P$. The two readings differ, since $P_i$ can meet the parts of the earlier path
  outside its incised part.
\item \emph{No call on a side with overlap.} This follows~\cite[Def.~2.8]{ADW26b}. The implementation
  of~\cite{ADW26b} decides by bad edges~\cite[Claim~5.8]{ADW26b}, whose converse fails
  (Section~\ref{sec:numerics}), so we test overlap directly, as $P_i\cap P_a\ne\emptyset$.
\item \emph{No call on a side with an empty range.} In~\cite[Alg.~3, line~18]{ADW26b}, the side with
  the range $[j_A,i_1]$ is called also when $i_1<j_A$; its leaf step would then look for a piece of
  $p_{j_A}^0$ that the side need not have.
\item \emph{Incised parts are stored.} The set $I_\ell$ holds all incised parts of earlier levels, while
  in~\cite[Def.~3.2]{ADW26b} the vertices store whole paths and use those that bound annuli of level
  $\ell$. The token form of the \LemAnn{} needs our set (Section~\ref{sec:numerics}).
\item \emph{The range of the overlap rule.} It serves $[j_A,i]$, where~\cite[Alg.~3]{ADW26b} has
  $j\le m\le i$ with $j$ the index of the bound. Below $\max(x_a,y_a)$ the annulus has no pieces of
  these faces, so this form is needed.
\end{itemize}
The two call rules only remove calls. The whole-path reading of the overlap test gave the right
value on all test instances (Section~\ref{sec:numerics}), but the \LemInc{} does not cover it, and
we do not use it.

\paragraph{Naming and sides.} The names above must be computable. On level~$1$, let $p_t^0$ be the
piece of $p_t$ whose first token in walk order follows the cut at $c^*_s$, for $t=0$, or the cut inside
the dart of $e_t$ in $p_t$. The pair $\{(\delta,1),(\rev\delta,2)\}$ of $e_k$, with $\delta$ in
$p_{k-1}$, joins two such pieces, so the names agree along $P$. The enumeration of~\cite[Sec.~5]{ADW26b} with one aggregation over the paths
$\hat p(\cdot)$ on level~$1$ tells every corner of a face of $P$ its index and its name, and later pieces
inherit them. Let $\delta$ be the dart of a new edge of $Q$ at its end nearer to $g_x$. Then the copy
$\{(\delta,1),(\rev\delta,2)\}$ lies on the $j$-side and $\{(\delta,2),(\rev\delta,1)\}$ on the
$k$-side (Claim~\ref{cl:Oorient}).
The hosts store this with the edge, which also tells a node of a bound which of the two bounds it is on.
A level then runs in this order. On the annuli of the previous level, one aggregation computes the flags
for overlap, for an empty range and for a part $Q$ without a new edge. Then build $\hat G_\ell$, obtain the
identifiers of the components from part~(6) of the \LemRep{}, aggregate the stored sides and these flags
over every component, set the weights of the \LemTrees{}, and run SSSP. A called side is one subgraph that holds one copy of every edge of $Q$ (part~(b) of the \LemInc{}),
so a component is the called child of a side exactly when it holds a copy of a new edge of $Q$ of that
side. If $Q$ has no new edge, nothing is cut, and the one component lies on the side given by
part~(a) of the \LemInc{}.

\paragraph{The defect.} After computing $P_i$, the recursion drops every index $t$ with
$i_1<t<i_2$: the children get the ranges $[j_A,i_1]$ and $[i_2,k_A]$~\cite[Alg.~3, lines~15
to~19; Lemma~5.6]{ADW26b}. The reason given is~\cite[Claim~2.4]{ADW26b}: $P_{i_1}$ or $P_{i_2}$ is no
longer than $P_t$. The proof of~\cite[Lemma~5.6]{ADW26b} assumes that the subpath of $P$ from
$p_{j_A}$ to $p_{k_A}$ has two copies in $A$, that is, that $A$ holds only the copies of the faces of
its range. Then $i_1$ and $i_2$ lie in $[j_A,k_A]$, and the end of every range is computed. But an
annulus can hold copies of faces of $P$ outside its range. Then the witness $P_{i_1}$ or $P_{i_2}$
lies outside the range of $A$, and another annulus may have dropped it already. Two drops can rest
on each other, and then no computed path has the minimum length.

\begin{example}\label{ex:recursion}
Section~\ref{sec:numerics} gives a simple planar network with $11$ vertices and $22$ edges whose
minimum $st$-cut has capacity~$15$, while the recursion reports~$37$; lengths here are the capacities.
For the faces of $s$ and $t$ and the corners chosen in the run, $P$ has the eight faces
$p_0,\dots,p_7$, and the minimum cut is a dual triangle through $p_4$, $p_5$ and a face $f_7$ off $P$
(faces off $P$ are numbered as in the script). It uses the edge of $P$ between $p_4$ and $p_5$, so it
is a shortest walk for both indices $4$ and~$5$. On level~$1$, with the range $[0,7]$, the path
$P_3=p_3^0f_7p_5^1p_4^1p_3^1$ has length $99$, so $i_1=3$ and $i_2=5$. Index~$4$ is dropped with
$P_5$ as its witness, and the children have the ranges $[0,3]$ and $[5,7]$. On level~$2$, the annulus
with the range $[5,7]$ also holds copies of $p_3^0$ and $p_4^0$, and the distance between its pieces of
$p_5^0$ and $p_5^1$ is $15$. Its path $P_6=p_6^0p_5^0p_4^0f_4f_6p_7^1p_6^1$ has length $37$, so
$i_1=4<5$: index~$5$ is dropped with $P_4$ as its witness, which level~$1$ has dropped already. The
recursion reports $37$, which is $2.47$ times the minimum. For the lengths $w$ below, every path that the run computes is unique, so no tie-breaking rule changes
the outcome. The sub-ranges of~\cite[Sec.~5.1]{ADW26b} change only the overlap rule, whose one use in
the run serves the indices $1$ to $3$, with lengths at least $99$, in an annulus with no piece of
$p_4^0$ or $p_5^0$.
\end{example}

As we read~\cite{ADW26b}, its recursion is the same with approximate trees, and exact trees are a
special case of approximate ones. The step of~\cite[Sec.~5.2]{ADW26b} that says that the leaf annuli
cover all of $P$ is where the defect enters.

\paragraph{The repair.} In every live annulus $A$ that is not a leaf, once $P_i$, $i_1$ and $i_2$ are
known, also compute shortest paths $P_{i^-}$ and $P_{i^+}$ in $A$, where
\[
  i^-:=\max(i_1,j_A)\qquad\text{and}\qquad i^+:=\min(i_2,k_A),
\]
and skip one that equals $i$. Nothing else changes: the same indices are dropped, and the children,
the incisions and the graphs $\hat G_\ell$ stay the same. The two paths are computed like $P_i$, each
by one call of the \LemTrees{} with one source per annulus and all annuli of a level at once;
so a level makes two more calls. Every node of $A$ knows $i_1$ and $i_2$ after~\cite[Lemma~5.6]{ADW26b}
and $j_A$ and $k_A$ by~\cite[Sec.~5.1]{ADW26b}, so it computes $i^-$ and $i^+$ locally, and the source
is elected as for $P_i$. The two paths are marked as tree paths~\cite[Lemma~5.2]{ADW26b} and enter
the final minimum and the cut extraction of~\cite[Sec.~5.2]{ADW26b}, which gives a cut whose capacity
is the length of the path for them too. Among computed paths of least length, the one with the least
index $t$ is taken, then the first computed, then the one of the annulus with the least identifier.
\begin{algorithm}[t]
\caption{\MinCut{}: the recursion of~\cite{ADW26b} with exact paths and its repair}\label{alg:mincut}
\begin{algorithmic}[1]
\Require undirected planar network $G$, integer capacities $c$, terminals $s,t$
\Ensure every vertex learns the value of a minimum $st$-cut, its side and which of its edges cross
\State Delete loops, merge parallel edges, $c'\gets n^2c+1$, and use the lengths $w=Kc'+1$ with $K=19n+1$
\State Compute a shortest path $P=(p_0,\dots,p_m)$ in $G^*$ from a face of $s$ to a face of $t$; choose the corners and the endpoint-case edges
\State Cut $G^*$ open along $P$ into the annulus $A_1$, with the range $[0,m]$ and no bound
\For{each of the $O(\log n)$ levels $\ell\ge1$, while there is a live annulus}
  \State Build the repaired graph $\hat G_\ell$; every shortest path of this level comes from one call of the \LemTrees{} on it, for all live annuli at once
  \For{every live annulus $A$ with range $[j_A,k_A]$ and bounds $P_a,P_b$}
    \If{$k_A\le j_A+1$} compute $P_{j_A}$ and $P_{k_A}$; $A$ is a leaf
    \Else
      \State $i\gets\lfloor(j_A+k_A)/2\rfloor$; compute $P_i$ from $g_i^0$ to $g_i^1$
      \State Let $i_1,i_2$ be the least and the greatest index of a piece of a face of $P$ on $P_i$
      \State Compute $P_{i^-}$ and $P_{i^+}$ for $i^-=\max(i_1,j_A)$ and $i^+=\min(i_2,k_A)$ \Comment{the repair; skip one equal to $i$}
      \State Incise the part $Q$ of $P_i$ between the last $0$-piece and the first $1$-piece after it
      \If{$P_i$ meets $P_a$} read $P_t$ for all $t\in[j_A,i]$ from a tree grown at a common node
      \ElsIf{$[j_A,i_1]\ne\emptyset$} call the $j$-side with the range $[j_A,i_1]$
      \EndIf
      \If{$P_i$ meets $P_b$} read $P_t$ for all $t\in[i,k_A]$ from a tree grown at a common node
      \ElsIf{$[i_2,k_A]\ne\emptyset$} call the $k$-side with the range $[i_2,k_A]$
      \EndIf
    \EndIf
  \EndFor
\EndFor
\State Report the least $c'$-length of a computed path or walk $P_t$; its value for $c$ is $\lfloor\cdot/n^2\rfloor$; read off the cut
\end{algorithmic}
\end{algorithm}

\paragraph{Lengths.} In the \LemWit{}, the \LemInc{} and the \LemRec{}, every dual edge, and every
copy of it in an annulus, has the length $w:=Kc'+1$, with $K=19n+1$ as in the \LemTrees{}.
These lengths are positive, and $K$ exceeds the number of edges of every simple path in an annulus,
since every node-part has its own nonempty path $\hat p(\cdot)$ in $\hat G_\ell$. By
the \LemTrees{}, every computed path, $P$ included, is a $w$-shortest path in its annulus.
For an index $t$, let $\lambda_t$ be the $w$-distance in $A_1$ from $p_t^0$ to $p_t^1$, let
$\Lambda:=\min_t\lambda_t$, and let $\lambda'$ be the value of a minimum $c'$-cut. By
part~(c) of the \LemAnn{}, every walk in $A_1$ from $p_t^0$ to $p_t^1$ has $c'$-length at least $\lambda'$
and hence $w$-length at least $K\lambda'+1$, and for some $t$ there is a simple path of $c'$-length
$\lambda'$ with fewer than $K$ edges. So $\lfloor\Lambda/K\rfloor=\lambda'$. By the \LemAnn{}, every
walk in a live annulus maps to a walk in $A_1$ of the same length, and a piece of $p_t^\sigma$ maps
to $p_t^\sigma$.

\begin{named}{Witness Interval Lemma}\label{lem:S1}
Let $\bar P$ be a walk in $A_1$ from $p_i^0$ to $p_i^1$ of length $\lambda_i$. Let $x$ be the index of
the last piece of some $p_t^0$ on $\bar P$ and $y$ the index of the first piece of some $p_t^1$ after
it, and let $i_1$ and $i_2$ be the least and the greatest index of a piece of a face of $P$ on
$\bar P$. Then $i_1=\min(i,x,y)$ and $i_2=\max(i,x,y)$, and for every $t$ with $i_1<t<i_2$,
\[
  \lambda_q\le\lambda_t\ \text{for all}\ q\in[i_1,t]\qquad\text{or}\qquad
  \lambda_q\le\lambda_t\ \text{for all}\ q\in[t,i_2].
\]
\end{named}

This is an interval form of~\cite[Claim~2.4]{ADW26b}, and it is proved in $A_1$ alone. By (INV) of
the \LemInc{}, the path $P_i$ of a live annulus has length $\lambda_i$, so its image in $A_1$ is a
walk as in the \LemWit{}, with the same $x$, $y$, $i_1$ and $i_2$. The interval of the witnesses
reaches $i_1$ or $i_2$ itself, which can lie outside $[j_A,k_A]$; this is what the recursion
of~\cite{ADW26b} misses.

Appendix~\ref{app:O} draws $\gamma$ and the incised parts as curves in the sphere. There, $Q$ is drawn
as an arc $C_Q$ from a point $q_0$ of $\gamma$ in the face $p_x$ to a point $q_1$ of $\gamma$ in the
face $p_y$, and $J_Q:=C_Q\cup\gamma[q_0,q_1]$ is a simple closed curve. Its \emph{low side} is the side
that contains the part of $\gamma$ before $q_0$ and $q_1$, and its other side is the \emph{high side}.
The low side is the $j$-side of~\cite{ADW26b} and the high side the $k$-side; this is the reading
of~\cite[Def.~2.2, Remark~2.5]{ADW26b} that the \LemAnn{} also makes. The invariant (I) is defined
in Appendix~\ref{app:inv}. An edge, pair or token is \emph{old} on level $\ell$ if its edge is in
$I_\ell$, and \emph{new} otherwise.

\begin{named}{Incision Lemma}\label{lem:O}
Suppose that all live annuli of level $\ell$ satisfy the invariant (I), and let $A$ be one of them
that is not a leaf; the annulus $A_1$ satisfies (I), and so does every called child of an annulus
that satisfies it. Let
$P_i$, $x$, $y$, $Q$, $g_x$, $g_y$, $P_a$, $P_b$ and $J_Q$ be as above.
\begin{enumerate}[(a)]
\item \emph{Sides.} Every subgraph obtained by incising $Q$ lies on one side of $J_Q$. The pieces of
  $A$ of $p_t^0$ for $t<x$ and of $p_t^1$ for $t<y$ lie on the low side, and those for $t>x$ and for
  $t>y$ on the high side. If the edge of $Q$ at $g_x$ is new, $g_x$ splits into its part below the
  cut, on the low side, and its part above it, on the high side. If that edge is old, it is the
  first edge of $P_a$ or of $P_b$, and $g_x$ does not split; it lies on the high side if the edge is
  on $P_a$ and on the low side if it is on $P_b$. Then $P_i$ meets that path at $g_x$, so the side of
  that path gets no call, and the called side, if any, holds all of $g_x$. The same holds at $g_y$,
  with the last edge of $Q$. For $t$ strictly between $x$ and $y$ such that $A$ has pieces of
  $p_t^0$ and of $p_t^1$, every walk in $A$ between them meets a node of $Q$.
\item \emph{The called side.} A side on which $P_i$ does not meet the bounding path of that side is
  one subgraph. It has no bad edge, and it holds exactly one copy of each edge of $Q$. A bad edge in
  a subgraph of the $j$-side is an edge of $P_a$, and on the $k$-side an edge of $P_b$.
\item \emph{Pieces.} Every live annulus has at most one piece of each $p_t^\sigma$ and at most one
  piece of each face off $P$.
\item[(INV)] For every live annulus $A'$ and every $t$ such that $A'$ has pieces of $p_t^0$ and of
  $p_t^1$, in particular for every $t\in[j_{A'},k_{A'}]$, their distance in $A'$ is $\lambda_t$.
\item[(C23)] Let $x_a$ and $y_a$ be the indices of the ends of $P_a$, a piece of $p_{x_a}^0$ and a
  piece of $p_{y_a}^1$. If $P_i$ meets $P_a$ in a node $z$, then
  $\dist_A(g_\mu^0,z)+\dist_A(z,g_\mu^1)=\lambda_\mu$ for every $\mu\in[\max(x_a,y_a),i]$, an interval that
  contains $[j_A,i]$. Symmetrically, if $P_i$ meets $P_b$ in a node $z$, the same holds for every
  $\mu\in[i,\min(x_b,y_b)]\supseteq[i,k_A]$, where $x_b$ and $y_b$ are the indices of the ends of
  $P_b$.
\end{enumerate}
\end{named}

So the overlap rule computes $\lambda_\mu$ for every $\mu\in[j_A,i]$: the tree from a common node $z$
gives a walk from $g_\mu^0$ to $g_\mu^1$ of length at most $\dist_A(g_\mu^0,z)+\dist_A(z,g_\mu^1)=\lambda_\mu$,
and every such walk has length at least $\lambda_\mu$ by (INV). The equality in (C23) fails beyond $i$
in general (Section~\ref{sec:numerics}).

\begin{named}{Repaired Recursion Lemma}\label{lem:S2}
With the repair, every computed path or walk from a piece of $p_t^0$ to a piece of $p_t^1$ has
$w$-length at least $\lambda_t\ge\Lambda$, and some computed path has $w$-length $\Lambda$. So the
least $c'$-length of a computed path or walk is $\lfloor\Lambda/K\rfloor=\lambda'$, the value of a minimum
$c'$-cut.
\end{named}

\paragraph{The cut.} Let $W$ be a computed path or walk of least length, from a piece of $p_t^0$ to a
piece of $p_t^1$, and let $\lambda'$ be its $c'$-length. A walk of the overlap rule is a shortest walk
by (C23), and $w>0$, so $W$ is simple. Let $x$ be the index of the last piece of some $p_u^0$ on $W$ and
$y$ the index of the first piece of some $p_u^1$ after it, and write $W=U\cdot Q'\cdot V$ with $Q'$ the
part between them. Its image in $G^*$ has $U$ as a walk between $p_t$ and $p_x$ and $V$ as a walk between
$p_y$ and $p_t$. The path $P$ is a shortest path for $w$ with fewer than $K$ edges, hence for $c'$, so
$c'(U)\ge c'(P[p_t,p_x])$ and $c'(V)\ge c'(P[p_t,p_y])$, and the union of these two subpaths contains
$P[p_x,p_y]$. So the closed walk $C$ made of the image of $Q'$ and $P[p_x,p_y]$ has $c'$-length at
most $\lambda'$. The internal nodes of $Q'$ are off $P$ and, by part~(c) of the \LemInc{}, are pieces
of distinct faces. No edge of $Q'$ is a copy of an edge of $P$, since such a copy joins two pieces on
one side of $\gamma$ (Appendix~\ref{app:curves}), while an edge of $Q'$ has an end off $P$ or joins a
$0$-piece to a $1$-piece. The edges of $Q'$ are copies of distinct edges of $G$, since a live annulus has
at most one pair of each edge off $P$, by (Old) in Appendix~\ref{app:inv}. So $C$ is a simple dual cycle, or a
loop if $x=y$ and $Q'$ has one edge. Draw the pairs and nodes of $Q'$ as curves as in the proof of
part~(c) of the \LemAnn{}, and close the curve along $\gamma$ between its two ends, on one side of
$\gamma$. The closed curve meets $G$ only in edges of $C$ and crosses $\gamma$ exactly once, so $C$
separates $s$ from $t$. Hence $c'(C)=\lambda'$, and $C$ is a minimum $c'$-cut. Since $c'>0$, removing
its edges leaves exactly two components, which the bisection of~\cite[Sec.~5.2]{ADW26b} finds. The
vertices mark $C$ for it: $W$ is a tree path or two tree paths through a common node, so~\cite[Lemma~5.7]{ADW26b}
applies, the indices $x$ and $y$ are broadcast, and the edge $e_k$ knows $k$. The two ends of an edge
then compare their sides in one round.

\begin{proof}[Proof of the \ThmCut{}]
Delete loops, merge parallel edges and replace $c$ by $c'$ as above; a bound on $n$ is known, and one
aggregation gives $n$. \MinCut{} runs~\cite[Alg.~3]{ADW26b} on the
capacities $c'$, with the endpoint data chosen as above, on the repaired graphs $\hat G_\ell$, with
every call of~\cite[Thm.~4.1]{ADW26b} answered by the \LemTrees{}, and with the repair of the
recursion. By the \LemRec{}, the least $c'$-length of a computed path or walk is $\lambda'$. By the last
paragraph, its cut is a minimum $c'$-cut, hence a minimum $c$-cut, whose value every vertex obtains as
$\lfloor\lambda'/n^2\rfloor$. Ranges halve from level to level, so there are $O(\log n)$ levels, and
each makes $O(1)$ calls of the \LemTrees{}: those of~\cite[Alg.~3]{ADW26b}, and two more for the paths
$P_{i^-}$ and $P_{i^+}$ of the repair. Each call takes $\Ot(D)$ rounds. All other steps of~\cite{ADW26b}
are unchanged and take $\Ot(D)$ rounds, since they use only the properties of $\hat G_\ell$ in
the \LemRep{}: the simulation of minor aggregation~\cite[Thm.~3.26]{ADW26b}, the tree primitives
and the steps of~\cite[Lemmas~5.2 and~5.4 to~5.7]{ADW26b}, and the bisection. Two of them need a
remark. The indices $i_1$ and $i_2$ are taken over all pieces of faces of $P$ in $A$, by one aggregation,
in place of the assumption in the proof of~\cite[Lemma~5.6]{ADW26b} that $A$ holds only the faces of
its range. And~\cite[Lemma~5.5]{ADW26b} needs $O(1)$ values per face of $P$, which holds since the
ranges of one level share only end indices. The algorithm of~\cite{ADW26b} is deterministic given
the planar embedding~\cite[Thm.~1.1]{ADW26b}, and the repair of the recursion adds only local steps and
calls of the \LemTrees{}, so \MinCut{} adds no randomness to the labels and the embedding.
\end{proof}

The \ThmSSSP{} is not a minor-aggregation algorithm, so it cannot run on each annulus directly,
whose hop diameter can be large; this is why an approximate SSSP algorithm
through minor aggregation is used in~\cite{ADW26b}. One exact SSSP on all of $\hat G_\ell$, with the weights above, needs only
CONGEST on a planar network of hop diameter $O(D)$.
\section{Numerical checks}\label{sec:numerics}

The scripts are in the directory \texttt{sim/} of the repository named in Section~\ref{sec:lean}
and need Python~3 with \texttt{numpy}, \texttt{scipy} and \texttt{networkx}. They implement the
sequential meaning of the statements on finite families of instances. They do not test round
counts, they do not cover all inputs, and no proof in this paper relies on them. No check in
Table~\ref{tab:checks} failed. The notes and scripts in the repository use older names:
Theorems A, B, C, C$'$ and D for the \ThmSSSP{}, the \ThmNeg{}, the \ThmFlow{}, the \ThmFace{} and
the \ThmCut{}, and Lemmas V, D, N, R, S1, O and S2 for the \LemVS{}, the \LemDD{}, the \LemAnn{}, the
\LemRep{}, the \LemWit{}, the \LemInc{} and the \LemRec{}.

\begin{table}[t]\centering\small
\begin{tabular}{>{\raggedright\arraybackslash}p{0.31\textwidth}>{\raggedright\arraybackslash}p{0.23\textwidth}>{\raggedright\arraybackslash}p{0.36\textwidth}}
\toprule
Statement & Script & Runs, all without a failed check\\
\midrule
\LemVS{}, along chains, with values in $G$, in the bag and mixed; all-bags form & \texttt{check.py}; \texttt{attack/lemmaV.py} & $80$ instances; $300$ with adversarial overestimates\\
\LemDD{} & \texttt{check.py}, \texttt{attack/} & $80$ instances ($33$ distances, $47$ negative cycles); $600$ more\\
\ThmFlow{} & \texttt{check.py}, \texttt{attack/} & $80$ instances; $600$ more\\
\ThmFace{} & \texttt{check.py}, \texttt{attack/} & $69$ instances; about $440$ more\\
\MinCut{}, the reduction & \texttt{thmD/thmD.py} & $64{,}000$ instances in eight families\\
\MinCut{}, repaired graphs & \texttt{thmD/thmD.py} & $79{,}000$ runs; $1{,}848$ graphs planar with hop diameter at most $4D+4$\\
\LemAnn{}, \LemRep{} & \texttt{thmD/repair/} & $98{,}000$ instances with $260{,}625$ levels, and $12$ named networks\\
Adversarial search on these two lemmas & \texttt{thmD/attackR/} & $6.9$ million level-$1$ corner pairs, $1.3$ million graph builds, $3.5$ million instances of the recursion\\
Repaired recursion & \texttt{thmD/} & $1{,}440{,}000$ instances\\
\LemWit{}, \LemInc{}, invariant & \texttt{thmD/lemmaO/}, \texttt{refereeO/} & $342{,}000$ instances with $232{,}868$ incisions; $162{,}000$ with $118{,}856$ incisions\\
Referee simulator for the \ThmCut{} & \texttt{thmD/refereeD/} & $240{,}000$ runs\\
Adversarial pass on the recursion & \texttt{thmD/attackSO/} & $25{,}109{,}452$ runs\\
\bottomrule
\end{tabular}
\caption{Numerical checks. The script names are relative to \texttt{sim/}.}
\label{tab:checks}
\end{table}

The checks of the reduction, the repaired graphs and the adversarial search on them have negative
controls with a deliberately broken input or construction, which failed as they should. With
$\hat G_\ell$ as we read~\cite[Remark~3.3]{ADW26b}, $10{,}673$ of $30{,}000$, $1{,}617$ of $4{,}000$ and
$3{,}245$ of $15{,}000$ runs of the reduction failed, every one through (D1) or (D2). With the weight
$c$ in place of $Kc+1$, with $M=1$, and with $+1$ on $30\%$ of the $E_C$ arcs, the checks flagged
$1{,}561$, $2{,}913$ and $2{,}841$ of $3{,}000$ runs. The adversarial search also found the instance on
which the converse of~\cite[Claim~5.8]{ADW26b} fails (seed $308940$), and the defect of the recursion.

The runs of \texttt{thmD/thmD.py} in Table~\ref{tab:checks} predate the repair of the recursion, and
none of them met its defect. That script builds $\hat G_\ell$ from a smaller set than $I_\ell$: the edges
of $P$ and of the paths that bound the live annuli. The scripts in \texttt{thmD/repair/} use $I_\ell$. On
$256{,}056$ of their levels the two sets agree and both builders gave the same edges, and on the other
$4{,}569$ levels $I_\ell$ has $10{,}929$ more edges. With the smaller set in the place of $I_\ell$, $2{,}000$
instances gave $482$ tokens without a parent token and $272$ pieces that refine no single piece of the
previous level.

\paragraph{The adversarial pass on the recursion.} The directory \texttt{thmD/attackSO/} holds every run
of the recursion, up to a cap of runs per instance, on all connected loopless maps with at most $8$
edges and several capacity vectors; the cap was reached on one instance with $7$ edges and on $534$
with $8$. It also holds adversarial choices of corners, modes and ties, hill-climbing, the saved
instances on which the recursion without the repair fails, and direct checks of the \LemWit{} on
$3{,}185{,}392$ annuli. Its file \texttt{logs/summary.txt} has the counts. On the saved instances, the
recursion without the repair gives a wrong value in about a fifth of the runs.

\paragraph{The recursion.} The network of Example~\ref{ex:recursion} has the vertices
$0,\dots,10$ and the edges $0,\dots,21$ with the ends
$(0,1)$, $(0,2)$, $(2,7)$, $(1,3)$, $(7,8)$, $(8,4)$, $(4,5)$, $(8,9)$, $(7,10)$, $(4,9)$, $(5,8)$,
$(6,5)$, $(7,0)$, $(9,5)$, $(5,10)$, $(1,2)$, $(8,2)$, $(1,7)$, $(8,10)$, $(0,3)$, $(2,3)$, $(9,6)$
and the capacities $1$, $1$, $1$, $1$, $1$, $1$, $3$, $1$, $13$, $1$, $50$, $1$, $13$, $2$, $50$, $20$,
$1$, $30$, $47$, $20$, $30$, $2$. The dart $2e$ leaves the first end of edge $e$ and the dart $2e+1$
the second, and the rotations at the vertices $0,\dots,10$ are
$(0,38,2,25)$, $(1,34,30,6)$, $(3,40,31,4,33)$, $(7,41,39)$, $(11,12,18)$, $(13,28,20,27,23)$,
$(22,43)$, $(24,8,16,5,35)$, $(9,32,36,10,14,21)$, $(15,19,42,26)$ and $(17,29,37)$. The terminals are
$s=5$ and $t=3$. The minimum cut has capacity $15$ and consists of the edges $4$, $8$ and $16$; the
rotation system is planar. With the faces numbered as in \texttt{thmD/thmD.py}, the run of the example has
$P$ through the faces $12,10,9,8,5,3,0,1$. That script reports $37$ on $2$ of $80$ runs of this
network, which differ in the faces of $s$ and $t$ and in the corners. The file
\texttt{thmD/attackR/logs/trace\_min\_simple0.txt} holds a trace of the run. On the $25$ instances that
the adversarial search saved, $12$ runs each, the script gives a wrong value in $161$ of $300$ runs
without the repair and in none of $300$ with it, and on $1{,}200{,}000$ instances of its own families it
gives $6$ wrong values without the repair.

\paragraph{The incision lemma.} The equality of (C23) of the \LemInc{} held in all $415{,}821$ checks on
$[\max(x_a,y_a),i]$ and failed beyond $i$: in $1{,}669$ of $12{,}044$ checks on $(i,i_2]$ and in
$84{,}996$ of $184{,}078$ beyond $i_2$ (and symmetrically on the $k$-side). Under the whole-path
reading of the overlap test, $1{,}561$ sides in $162{,}000$ instances overlap that do not overlap
under our reading; a copy of the script with this test gave the right value on all of its $84{,}000$
instances and in the adversarial pass, $738{,}496$ further runs.

\section{Formal verification}\label{sec:lean}

The formalization is in the directory \texttt{lean/SsspFlow/} of the repository
\begin{center}
\url{https://github.com/ssenge/ExactShortestPathsAndFlowsOnPlanarGraphs},
\end{center}
at the tag \texttt{arxiv-v1}. It uses Lean~4 with the toolchain \texttt{leanprover/lean4:v4.36.0-rc1}
and Mathlib at the revision
\begin{center}
\texttt{d875a8de5222f62cf1c2fafc7984588adadc990a},
\end{center}
which is pinned in \texttt{lean/lakefile.toml} and in \texttt{lean/lake-manifest.json}. No proof
uses \texttt{sorry}. The command \texttt{lake env lean SsspFlow/Axioms.lean} prints the axioms of all
theorems named below. The output is recorded in \texttt{lean/axioms.txt}, and it lists only
\texttt{propext}, \texttt{Classical.choice} and \texttt{Quot.sound}. In a fresh clone,
\texttt{lake exe cache get} before \texttt{lake build} avoids compiling Mathlib from source. The
repository holds further Lean files, on the steps of its notes on~\cite{LTY26}, which this paper does
not use. Distances are infima over walks throughout.

\paragraph{\texttt{LemmaVChain.lean}.} The theorem \texttt{RHierarchy.lemmaV\_chain} of the namespace
\texttt{SsspFlow} is the \LemVS{}: a digraph with arc lengths in $[0,\infty]$, offsets in $[0,\infty]$, and a well-founded bag
hierarchy (\texttt{RHierarchy}, in \texttt{LemmaVRegion.lean}) whose bags carry arc sets (\texttt{arc}), where the root carries every
vertex and arc (\texttt{root\_bag}, \texttt{root\_arc}), arcs of a bag have their ends in it
(\texttt{arc\_bag}), non-hub vertices lie in exactly one child (\texttt{cover}, \texttt{uniq}), and an
arc of a bag between two non-hubs is an arc of some child (\texttt{edge}). Chains are the inductive
\texttt{Chain}. The values are hypotheses along chains, with the bounds of the lemma, and
$\eps\ne\infty$. No finiteness is assumed; $\infty$ stands for unreachable. Not formalized: the
CONGEST algorithm and its rounds, the construction of the hierarchy and of the values.

\paragraph{\texttt{LemmaD.lean}.} The structure \texttt{Setup} describes a graph with corner nodes,
star centres, a ring relation whose reflexive-transitive closure joins exactly the corners of each
face (\texttt{ring\_face}, \texttt{ring\_conn}), one arc per dual dart from a corner of its tail
face to a corner of its head face (\texttt{src\_face}, \texttt{dst\_face}), and spokes in both
directions; parallel arcs and loops are allowed. With $0\le\Lambda$, $-\Lambda\le\mu$ and
$|V(\hat G)|\cdot\Lambda<M$, \texttt{Setup.negCycle\_iff} is part~(i) of the \LemDD{} for closed walks,
and with no negative closed dual walk \texttt{Setup.walks} gives the lift and the projection. With $|\mu|\le\Lambda$, every face having
a corner, and no negative closed dual walk, \texttt{Setup.lemmaD} is part~(ii) of the \LemDD{}, with
distances in the extended reals. The definition \texttt{rotSetup} builds $\hat G$ from a
combinatorial map (corners are darts, faces are orbits of the face permutation), \texttt{rot\_pair}
shows that each edge gets one $E_C$ edge whose two arcs join the same two corners, and
\texttt{rot\_lemmaD} is the \LemDD{} for this graph. Planarity is not used. Not formalized:
the planarity, hop diameter and simulation properties of $\hat G$ from~\cite{ADPW25}, and the
algorithms run on it.

\paragraph{\texttt{Boost.lean}.} The structure \texttt{Run} is the run of
the \LemBoost{}: $\alpha\ge1$, $\eta\ge0$, every vertex reachable, integer input lengths, the
update with base $\ell_i+\eta$, and (O1) to (O3) as hypotheses on the estimates. The statements
\texttt{Run.len\_nonneg}, \texttt{Run.contract} and \texttt{Run.exact} are
parts~(a), (b) and~(e) of the \LemBoost{}, with (c) and (d) as \texttt{Run.decay} and
\texttt{Run.sum\_bounds}. The lemma \texttt{c1\_negative} is the one-arc instance in which the base
$\ell_i$ gives $-\eta$. Not formalized: the oracle (Algorithm~3 of~\cite{RHMGZ23} and the labels),
the choice of $I$, the rounding (W1), and the round count.

\paragraph{Proved on paper only.} The CONGEST algorithms and their round counts; the planarity,
diameter and simulation properties of $\hat G$; the \LemTrees{}; and the proofs of the \ThmSSSP{},
the \ThmNeg{}, the \ThmFlow{}, the \ThmFace{} and the \ThmCut{} from these parts. The \ThmCut{} is not
formalized in Lean: no part of its proof, including the \LemAnn{}, the \LemRep{}, the \LemWit{}, the
\LemInc{} and the \LemRec{}, has a Lean counterpart.

\section{Open questions}\label{sec:open}

\begin{itemize}
\item As we read~\cite{ADW26b}, its recursion with approximate shortest-path trees drops indices
  as in Example~\ref{ex:recursion}, so the approximation ratio of~\cite[Thm.~1.1]{ADW26b} may fail
  in the same way. Does it, and do the endpoint failures (D1) to (D3) affect it? The two extra paths
  of the repair can be added there as well, but our proof of the \LemRec{} uses exact distances,
  through (INV) and (C23) of the \LemInc{}.
\item The overlap test of~\cite[Alg.~3]{ADW26b} is written with the whole path of an earlier level;
  we read it with the incised part of that path (Section~\ref{sec:thmD}). Is the recursion with the
  whole-path test also exact? It gave the right value on all our test instances, but
  the \LemInc{} does not cover the common nodes outside the incised part.
\item The \ThmCut{} gives the value of a maximum $st$-flow in an undirected planar network and
  a minimum cut, but not the flow on every edge. Can the flow on every edge be computed in $\Ot(D)$
  rounds?
\item Directed global minimum cut in planar networks takes $\Ot(D^2)$ rounds~\cite[Thm.~1.5]{ADPW25}.
  It needs a shortest cycle of simple darts in the dual. Since $\hat G$ has cycles of length $0$,
  the girth of $\hat G$ does not give it. Can it be done in $\Ot(D)$ rounds?
\item The weighted girth of planar digraphs is computed in $\Ot(D^2)$ rounds with high
  probability~\cite[Lemma~4]{P20}. Can it be
  computed in $\Ot(D)$ rounds?
\item Can the factor $n^{o(1)}$ in the \ThmNeg{} and the \ThmFlow{} be removed? This would need
  a reduction for negative weights with polylogarithmic overhead.
\item Are the \ThmNeg{} and the \ThmFlow{} deterministic? This needs deterministic versions of
  the reduction of~\cite{A24} and of the remaining steps of~\cite{ADPW25}.
\end{itemize}

\section*{Use of AI}

The proofs, the code and the Lean files were developed with Claude Opus~5.5 (Anthropic), which was
also used to search the literature and to draft the text. They were reviewed by separate AI
referees, by adversarial AI agents that searched for counterexamples, and by the numerical scripts
of Section~\ref{sec:numerics}. No human expert has checked them yet.

\appendix

\section{The repaired graphs: proofs of the Annulus Lemma and the Repaired Graph Lemma}\label{app:R}

Throughout, $G$ is connected and planar with a fixed rotation system, $n\ge2$ and $s\ne t$. The
proof of the \ThmCut{} makes $G$ simple, but the proofs below allow parallel edges; loops are
excluded. Every edge has two darts, $\rev\delta$ is the reverse of the dart $\delta$, and
$\mathrm{pos}(\delta)$ is the index of $\delta$ in the rotation at its tail.

\paragraph{Corners and faces.} As in Section~\ref{sec:model}, the corner $c=(v,j)$ lies between
$r_v(j-1)$ and $r_v(j)$. Its \emph{in-dart} $\cin(c):=\rev r_v(j-1)$ enters $v$, and its
\emph{out-dart} $\cout(c):=r_v(j)$ leaves $v$; the edges of $r_v(j-1)$ and $r_v(j)$ are the two
\emph{flanks} of $c$. A dart $\delta$ has the \emph{tail corner}
$\tau\delta:=(\mathrm{tail}\,\delta,\mathrm{pos}(\delta))$, the corner $c$ with $\cout(c)=\delta$, and the
\emph{head corner} $\eta\delta:=(\mathrm{head}\,\delta,\mathrm{pos}(\rev\delta)+1)$, the corner $c$ with
$\cin(c)=\delta$. A face is a cyclic walk of darts in which $\delta$ is followed by
$\cout(\eta\delta)$, and the corner $\eta\delta$ lies between $\delta$ and its successor. So every dart
and every corner lies in exactly one face, and the walk of a face alternates darts and corners. The
corner $i$ of a vertex in~\cite{ADW26b} is the pair $(e_i,e_{i+1})$ of edges, so its in-dart runs
along $e_i$ and its out-dart along $e_{i+1}$.

\paragraph{The incised sets.} The path $P=(p_0,\dots,p_m)$ and the sets $I_\ell$ are as in
Section~\ref{sec:thmD}; $m=0$ is allowed. Here $P_i'$ is the maximal subpath of $P_i$ that is
internally disjoint from the copies of $P$~\cite[Alg.~3, line~16]{ADW26b}. Edges that $P_i'$ shares
with a path that bounds its annulus are not incised again~\cite[Remark~2.5]{ADW26b}, but they are in
$I_\ell$ already. Every vertex knows its edges in $I_\ell$ before the subgraphs of level $\ell$ are
built and tested for liveness.

\subsection{The definition}

\paragraph{Endpoint data.} The vertex $s$ fixes a corner $c^*_s$ of $s$ in $p_0$ and a flank $e_0$ of
$c^*_s$ that is not in $E(P)$. The \emph{mode} is $\mu_s=L$ if $e_0$ is the edge of $\cin(c^*_s)$, and
$\mu_s=R$ if it is the edge of $\cout(c^*_s)$; if $\deg(s)=1$, both flanks are $e_0$ and either mode
may be chosen. The vertex $s$ sends $(c^*_s,e_0,\mu_s)$ to its neighbours in one round, and the data
stay fixed for all levels. The same rule at $t$, with $p_m$ in place of $p_0$, gives $c^*_t$,
$e_{\mathrm{last}}$ and $\mu_t$. This is the choice of Section~\ref{sec:thmD}: mode~L gives the copy
$s_i^L$ of~\cite[Remark~3.3]{ADW26b}, and mode~R its mirror image $s_i^R$.

\paragraph{Tokens, pairs, cuts and pieces.} These are defined in Section~\ref{sec:thmD}. No piece is
empty: $c^*_s\ne c^*_t$, and a corner cut and a dart cut never sit at the same place of a walk, so a
token lies between any two cuts.

\paragraph{Copies.} Every vertex $v$ has a star centre $\sigma(v)$, a copy $D(c)$ for each corner $c$
of $v$, a copy $L(c)$ for each corner $c$ of $v$ such that the edge of $\cin(c)$ is in $I_\ell$, and a
copy $R(c)$ for each corner $c$ of $v$ such that the edge of $\cout(c)$ is in $I_\ell$. In addition,
there is one fresh copy $X(c^*)$ for each of $c^*_s$ and $c^*_t$.

\paragraph{$E_R$ edges.} Each token $x$ of a dart $\delta$ is one $E_R$ edge $\alpha(x)\beta(x)$, where
$\alpha(x)$ is a copy of the tail and $\beta(x)$ a copy of the head of $\delta$:
\[
  (\delta,0):\ D(\tau\delta)\,D(\eta\delta);\qquad
  (\delta,1):\ D(\tau\delta)\,L(\eta\delta);\qquad
  (\delta,2):\ R(\tau\delta)\,D(\eta\delta).
\]
\emph{Repair (a).} Let $c^*$ be a cut corner with mode $\mu$. If $\mu=L$, let $x^*$ be the token of
$\cin(c^*)$ with $\beta(x^*)=D(c^*)$, that is, $(\cin(c^*),0)$ or $(\cin(c^*),2)$, and set
$\beta(x^*):=X(c^*)$. If $\mu=R$, let $x^*$ be the token of $\cout(c^*)$ with $\alpha(x^*)=D(c^*)$, that
is, $(\cout(c^*),0)$ or $(\cout(c^*),1)$, and set $\alpha(x^*):=X(c^*)$. The token $x^*$ is the
\emph{moved token} of $c^*$; it is a token of the endpoint edge.

\paragraph{$E_S$ edges.} The edges $\sigma(v)D(c)$ for each corner $c$ of $v$, and $D(c)L(c)$,
$D(c)R(c)$ and $D(c^*)X(c^*)$ where these copies exist.

\paragraph{$E_C$ edges.} Every pair $\pi$ has a \emph{host} $h(\pi)$, an end of its edge. For an edge
in $I_\ell$, the pair $\{(\delta,1),(\rev\delta,2)\}$ is hosted at the tail of $\delta$. If
$e_0\notin I_\ell$, its pair is hosted at the end other than $s$, which is $t$ if
$e_0=e_{\mathrm{last}}$. If $e_{\mathrm{last}}\notin I_\ell$ and $e_{\mathrm{last}}\ne e_0$, its pair is
hosted at the end other than $t$. Every other pair is hosted at the end with the smaller identifier.
For a token $x$ and an end $h$ of its edge, $\mathrm{att}_h(x)$ is $\alpha(x)$ if $h$ is the tail of the
dart of $x$, and $\beta(x)$ otherwise. \emph{Repair (b).} The $E_C$ edge of $\pi=\{x,y\}$ is
$\mathrm{att}_h(x)\,\mathrm{att}_h(y)$ with $h=h(\pi)$; both of its ends are copies of $h$.

\paragraph{Remarks.} With no moved token, the $E_R$ edges are those of~\cite[Sec.~3.1]{ADW26b}, and
the $E_C$ edges are its edges $(u_{j-1},u_j)$ and $(v_{i-1},v_i)$; so the repair changes $\hat G_\ell$
only at the copies of $s$ and $t$. If $e_0\notin I_\ell$, the copy $X(c^*_s)$ is the copy $s_i^L$ or
$s_i^R$ of the choice; if $e_0\in I_\ell$, it is the fresh copy of repair~(a). The literal reading
of~\cite[Remark~3.3]{ADW26b} differs in three ways: if $e_0\in I_\ell$, no token moves and the corner
$c^*_s$ is not cut~(D1); the $E_C$ edge for $e_0$ is always $(u_{j-1},u_j)$, also when one of these
copies has lost its token to the cut at $u=t$~(D2); and there is only mode~L~(D3). For an edge
outside $I_\ell$, either end may host its pair: its $E_C$ edge is a rim edge at both ends (proof
of~(3)), and the proofs below do not use which end it is. For an edge in $I_\ell$ the host is forced:
hosted at the head of $\delta$, the pair $\{(\delta,1),(\rev\delta,2)\}$ would get the $E_C$ edge
$L(\eta\delta)R(\tau\rev\delta)$, which is not a rim edge, and $\hat G_\ell$ can be non-planar. The graph
$\hat G_0$ is this definition with $I_0=\emptyset$ and without endpoint data: no corner is cut, there
is no copy $X$, and every face is one cyclic piece. It is the graph $\hat G_0$
of~\cite[Sec.~3.1]{ADW26b}, and the proofs below cover it.

\subsection{The token graph and the Annulus Lemma}

The rotation of $G^*_\ell$ at a piece lists its tokens in walk order; for a linear piece, the
\emph{gap} between its last and its first token closes the list into a cycle. This is the embedding
that incisions inherit from $G^*$. A piece is \emph{linear} if its face has a cut, and \emph{cyclic}
otherwise. The \emph{ends} of a linear piece are its first and its last token. A token $(\delta,1)$ is
followed by the cut inside $\delta$, so it is the last token of its piece, and $(\delta,2)$ is a first
token. Hence every token of an edge in $I_\ell$ is an end of its piece.

We read a node-part of~\cite{ADW26b} as a run of the walk of its face, that is, as a sequence of
tokens. A node-part is described in~\cite[Def.~2.9]{ADW26b} by an interval of the dual edges at a
node; a dual loop at $f$, which is a bridge of $G$ with $f$ on both sides, occurs twice in that order,
once for each dart, and tokens keep the two occurrences apart. This reading is ours.

\paragraph{The incision along $P$.} In words, $A_1=G^*_1$ is the sphere cut along the arc $\gamma$
that leaves $s$ through the corner $c^*_s$ into $p_0$, crosses $e_1,\dots,e_m$ once each, and enters
$t$ through $c^*_t$. It has one segment in each face $p_k$, and the two ends of that segment are the
two cuts of $p_k$. So $p_k$ splits into two pieces, and every other face is one cyclic piece. If
$m=0$, then $\gamma$ is one segment of $p_0$ from $c^*_s$ to $c^*_t$. The arc $\gamma$ is simple, since
$P$ visits each face once, and it meets $G$ only at $s$, at $t$ and once in each $e_k$.
Appendix~\ref{app:curves} draws $\gamma$ explicitly and names the two pieces of $p_k$ as $p_k^0$ and
$p_k^1$ by the side of $\gamma$ on which they lie. This agrees with~\cite[Def.~2.2,
Remark~3.3]{ADW26b} when both modes are L and the other face of $e_0$ (of $e_{\mathrm{last}}$) differs
from $p_0$ (from $p_m$). It also covers the cases left open there: mode~R; several corners of $s$ on
$p_0$, where the corner fixes the point at which $\gamma$ meets $s$; an endpoint edge with both darts
on $p_0$, whose dual is a loop at $p_0$; and $\deg(s)=1$, where both flanks of $c^*_s$ are $e_0$.

\paragraph{Facts used from~\cite{ADW26b}.} Let $A$ be a live annulus of level $\ell-1\ge1$ that is
not a leaf, and let $P_a$ and $P_b$ be the incised parts that bound it; near $s$ or $t$ one of them is
missing, and on level~$1$ both are. In~\cite[Alg.~3]{ADW26b}, $P_i$ is computed in $A$, its part
$P_i'$ is incised, and some of the resulting subgraphs are discarded.
\begin{enumerate}[({N}1)]
\item \cite[Alg.~3, lines~8 and~15 to~17; Remark~2.5]{ADW26b}. The part $P_i'$ is a simple path in
  $A$. Its endpoints are node-parts of faces of $P$, and its internal nodes are not. Its edges that
  are not in $I_{\ell-1}$ are incised now; the others lie on $P_a$ or $P_b$ and are not incised again.
\item \cite[Def.~2.2, Remark~2.5, proof of Claim~2.10]{ADW26b}. Nodes off $P_i'$ do not change. An
  internal node of $P_i'$ splits into its parts on the two sides of $P_i'$, and each newly incised
  edge gets one copy on each side. An endpoint of $P_i'$ whose edge on $P_i'$ is newly incised splits
  into two runs that share only that edge; any other endpoint does not change. An internal node of
  $P_i'$ that an earlier incision split lies on $P_a$ or on $P_b$.
\item \cite[Claim~5.8; Alg.~3, lines~9 to~19]{ADW26b}. A resulting subgraph is discarded if it
  contains a bad edge. A side on which $P_i$ meets $P_a$ or $P_b$, and a side whose index range is
  empty, get no call either; these two rules, and $P_a$ and $P_b$ as the copies of incised parts, are
  our reading of~\cite{ADW26b} (Section~\ref{sec:thmD}). The remaining subgraphs are the live annuli of level $\ell$, and none of
  the others is incised again. The first sentence is stated in~\cite[Claim~5.8]{ADW26b} as an
  equivalence, but the converse fails: a side can overlap $P_a$ without a bad edge
  (Section~\ref{sec:numerics}). Only the direction stated here is used.
\item \cite[Claim~2.12]{ADW26b}. Every annulus is connected.
\item \cite[Lemma~2.6]{ADW26b}. An edge is incised at most once, so it has at most four tokens.
\end{enumerate}

\begin{proof}[Proof of parts~(a) and~(b) of the \LemAnn{}]
\emph{Refinement.} Merge $(\delta,1)$ and $(\delta,2)$ into $(\delta,0)$ for every dart $\delta$ of an
edge in $I_\ell\setminus I_{\ell-1}$. This maps the walk of each face on level $\ell$ onto its walk on
level $\ell-1$, by taking parents. Every cut of level $\ell-1$ is a cut of level $\ell$, since
$I_{\ell-1}\subseteq I_\ell$ and the corner cuts are the same on every level $\ell\ge1$ (level~$0$ has
none). So a run with no cut of level $\ell$ inside maps into a run with no cut of level $\ell-1$
inside. This proves~(b). We prove~(a) by induction on $\ell$.

\emph{Level~$1$.} We have $A_1=G^*_1$, so it remains to show that $G^*_1$ is connected; the cases
listed above lie outside~\cite[Claim~2.12]{ADW26b}.
\begin{itemize}
\item Fix a vertex $v$ and an index $j$, and let $\delta=r_v(j)$. The out-token of the corner $(v,j)$
  is the token of $\delta$ labelled $0$ or $1$, and the in-token of $(v,j+1)$ is the token of
  $\rev\delta$ labelled $0$ or $2$. These two tokens form a pair. So consecutive corners of $v$ are
  joined by pairs. At a corner that is not cut, the in-token and the out-token lie in one piece.
  Hence all pieces that hold a token at a corner of $v$ are connected: a cut corner only separates
  its own in-token and out-token, and the chain around $v$ reaches both.
\item If $e=uv$ is not in $I_1$, its token $(\delta,0)$ is the out-token of $\tau\delta$ at $u$ and
  the in-token of $\eta\delta$ at $v$. This joins the pieces around $u$ to those around $v$.
\item $E(P)$ is the edge set of a simple path of $G^*$, so it contains no cycle of $G^*$, hence no bond
  of $G$, and $G-E(P)$ is connected.
\item Every token is the in-token or the out-token of some corner.
\end{itemize}
So $G^*_1$ is connected.

\emph{Step from $\ell-1$ to $\ell\ge2$.} Every live annulus of level $\ell$ is a child of a live
annulus $A$ of level $\ell-1$ that is not a leaf, made by incising $Q:=P_i'$ in $A$. By induction, $A$
is a component of $G^*_{\ell-1}$. A new edge of $Q$, one not in $I_{\ell-1}$, has its pair
$\{(\delta,0),(\rev\delta,0)\}$ in $A$, so its two new cuts lie in pieces of $A$ that are nodes of
$Q$. Let $g=(x_1,\dots,x_r)$ be a piece of $A$. The two sides of $Q$ at an internal node $g$ of $Q$ are
the two arcs of the rotation of $g$ between the two tokens of $Q$ at $g$.

(i) \emph{$g$ is not on $Q$.} No new cut lies in $g$, so $g$ is a piece of $G^*_\ell$, and by (N2) it
does not change.

(ii) \emph{$g$ is a cyclic internal node of $Q$.} Its face has no cut, hence no incised dart, so both
tokens of $Q$ at $g$ are new. The two new cuts split $g$ into two runs, which are the two sides of $Q$
at $g$, and both are pieces of $G^*_\ell$. No endpoint of $Q$ is cyclic: by (N1) it is a node-part of
a face of $P$, and those faces are cut on level~$1$.

(iii) \emph{$g$ is an endpoint of $Q$.} Let $z$ be its token on $Q$. If $z$ is new, the cut inside its
dart splits $g$ into the run that ends with the first half of $z$ and the run that starts with the
second half. These are the two runs of (N2) that share only the edge of $z$, and both are pieces. If
$z$ is old, no new cut lies in $g$, and by (N2) $g$ does not change.

(iv) \emph{$g$ is a linear internal node of $Q$.} Let $u$ and $w$ be its two tokens on $Q$, with $u$
before $w$ in $g$. The face of $g$ has a cut, and by (N1) it is not a face of $P$; so the cut is
inside a dart that an earlier incision cut, and by (N2) $g$ lies on $P_a$, say. By (N1), $g$ is not a
node-part of a face of $P$, while the endpoints of $P_a$ are; so $g$ is an internal node of $P_a$.
Every edge of $P_a$ is in $I_{\ell-1}$, so both tokens of $P_a$ at $g$ are ends of $g$. They are
distinct, so they are $x_1$ and $x_r$, and both are old. The two sides of $Q$ at $g$ are
$S_{\mathrm{in}}=(u,\dots,w)$ and $S_{\mathrm{out}}=(w,\dots,x_r,x_1,\dots,u)$, which runs through the
gap. If $u=x_1$ and $w=x_r$, both are old, no new cut lies in $g$, and $g$ does not change. Otherwise
$S_{\mathrm{out}}$ contains $x_1\ne u$ or $x_r\ne w$. That token is neither $u$ nor $w$, since $u$ comes
before $w$, so it belongs to an edge of $P_a$ at $g$ that is not on $P_i$, because $P_i$ meets $g$ only
in $u$ and $w$. Since $g\in P_i\cap P_a$, this edge is bad, and by (N3) the subgraph that contains
$S_{\mathrm{out}}$ is discarded. The run $S_{\mathrm{in}}$ starts right after the new cut in the dart of
$u$ if $u$ is new; if $u$ is old, it starts at $x_1=u$, since an old token is an end and $u$ comes
before $w$. It ends at $w$ in the same way. No cut lies inside $S_{\mathrm{in}}$, so it is a piece of
$G^*_\ell$.

\emph{Pairs.} Pairs of old edges, and of edges not on $Q$, do not change. Let $e$ be a new edge of
$Q$, with its dart $\delta$ in the node $g_1$ and $\rev\delta$ in $g_2$. In the rotation of $A$, the
sector of $g_1$ just before $\delta$ is at the corner $\tau\delta$, and the sector of $g_2$ just after
$\rev\delta$ is at the corner $\eta\rev\delta$. Both are corners of the tail of $\delta$ next to $e$, so
both sectors lie on the side of $e^*$ at the tail of $\delta$, and hence on the same side of $Q$. On
that side, $g_1$ keeps the run that ends with $(\delta,1)$ and $g_2$ the run that starts with
$(\rev\delta,2)$. So the copy of $e^*$ on the side of the tail of $\delta$ is
$\{(\delta,1),(\rev\delta,2)\}$, and the other copy is $\{(\delta,2),(\rev\delta,1)\}$. These are the
pairs of level $\ell$.

\emph{Conclusion.} Let $A'$ be a live annulus of level $\ell$, a child of $A$. By (i) to (iv), its
node-parts are pieces of $G^*_\ell$, and by the last paragraph its edges are pairs. Each token of a
node-part of $A'$ lies in a pair that is an edge of $A'$ at that node-part, and the other token of
the pair lies in the node-part at the other end. So the pieces of $A'$ are closed under pairs, and
$A'$ is a union of components of $G^*_\ell$. By (N4) it is connected, so it is one component.
\end{proof}

\begin{proof}[Proof of part~(c) of the \LemAnn{}]
($\ge$) Let $W$ be a walk in $A_1$ from $p_k^0$ to $p_k^1$. Draw each pair of $W$ as a curve across
its edge: across an edge outside $I_1$ at any point, and for the pair $\{(\delta,1),(\rev\delta,2)\}$ of
an edge of $P$ between the tail of $\delta$ and the point where $\gamma$ crosses the edge. Inside a
piece, join consecutive crossings by a curve on the side of $\gamma$ of that piece. Close $W$ by a
curve in $p_k$ that crosses $\gamma$ once. The closed curve crosses $\gamma$ exactly once and meets $G$
only in edges of $W$. A closed curve that crosses an arc from $s$ to $t$ an odd number of times
separates $s$ from $t$. So every path from $s$ to $t$ in $G$ uses an edge of $W$, and the length of
$W$ is at least $\lambda$.

($\le$) A minimum $st$-cut is a simple dual cycle $C$ that separates $s$ from
$t$~\cite[Fact~1.2]{ADW26b}. By the exchange argument of~\cite[Claim~2.1]{ADW26b}, $C$ can be chosen to
cross $\gamma$ only once. If $C$ crosses $\gamma$ more than once, it does so at least three times, and
since the algebraic crossing number is $\pm1$, two crossings consecutive along $C$ have the part of $C$
between them on one side of $\gamma$. That part is replaced by the part of $P$ between the same faces,
which is no longer, since $P$ is a shortest path. Cut at that crossing, in a face
$p_k$, the cycle $C$ lifts to a walk in $A_1$ from one piece of $p_k$ to the other, of length
$\lambda$. A shortest such walk is a path, and by ($\ge$) its length is $\lambda$.

Both directions use $\gamma$ only as a simple arc from $s$ to $t$ that runs through $p_0,\dots,p_m$
and meets $G$ only at $s$, at $t$ and once in each $e_k$. Every corner of $s$ in $p_0$ and of $t$ in
$p_m$ gives such an arc, so~(c) holds for every choice of the endpoint data. The same holds for~(a)
and~(b), whose proofs do not use the choice.
\end{proof}

\subsection{The rotation table of the copies}\label{app:tableR}

All lists follow the rotation sense of $G$.
\begin{itemize}
\item $\sigma(v)$: the copies $D(c)$ of the corners of $v$, in rotation order.
\item $D(c)$, for a corner $c$ of $v$ with $a=\cin(c)$ and $b=\cout(c)$: the list $\sigma(v)$, $C_A$,
  $A_1$, $A_2$, $B_2$, $B_1$, $C_B$, with absent entries skipped. Here $C_A$ is the $E_C$ edge of the
  pair whose token of $a$ ends at $D(c)$, listed only if that pair is hosted at $v$, and $C_B$ is the
  $E_C$ edge of the pair whose token of $b$ starts at $D(c)$, listed only if that pair is hosted at
  $v$. These are the $E_C$ edges at $D(c)$ across the flank of $a$ and across the flank of $b$; if
  $\deg(v)=1$, they are the two ends of one loop. If the edge of $a$ is not in $I_\ell$, then $A_1$ is
  the token of $a$ and $A_2$ is absent; otherwise $(A_1,A_2)$ is $((a,2),L(c))$ under option~(i) and
  $(L(c),(a,2))$ under option~(ii). If the edge of $b$ is not in $I_\ell$, then $B_1$ is the token of
  $b$ and $B_2$ is absent; otherwise $(B_1,B_2)$ is $(R(c),(b,1))$ under option~(i) and
  $((b,1),R(c))$ under option~(ii).
\item At a cut corner $c^*$ of the terminal $z$, the entry of the moved token at $D(c^*)$ is the $E_S$
  edge to $X(c^*)$. The end of the $E_C$ edge that moved to $X(c^*)$ is not listed at $D(c^*)$; this
  is $C_A$ in mode~L and $C_B$ in mode~R. If $\deg(z)=1$, then $C_A$ and $C_B$ are the two ends of one
  rim loop; only the end on the side of $x^*$ moves, and the other end stays at $D(c^*)$, as $C_B$ in
  mode~L and as $C_A$ in mode~R.
\item $L(c)$ and $R(c)$ have two edges each, so their order is immaterial.
\item $X(c^*)$: the list $D(c^*)$, $C$, $x^*$ in mode~L, and $D(c^*)$, $x^*$, $C$ in mode~R, where $C$
  is the $E_C$ edge of the pair of $x^*$, present if that pair is hosted at the vertex of $c^*$.
\item Option~(ii) is used for exactly one kind of dart: the dart $\cout(c^*)$ of a cut corner $c^*$
  in mode~R, when its edge is in $I_\ell$. Every other dart of an edge in $I_\ell$ uses option~(i).
  The option is chosen per dart and is the same at both of its ends.
\end{itemize}

\subsection{Proof of the Repaired Graph Lemma}

\paragraph{(5) The choice.} The face $p_0$ is a face at $s$~\cite[Alg.~1, line~1, and Alg.~3,
line~1]{ADW26b}, so $s$ has a corner in $p_0$. Let $c$ be such a corner, and let $A$ and $B$ be its
flanks, the edges of $\cin(c)$ and $\cout(c)$. Both have $p_0$ on the side of $c$. An edge $e_k$ of $P$
has $p_{k-1}$ and $p_k$ on its two sides, and $P$ is simple, so $e_1$ is the only edge of $P$ with
$p_0$ on a side, and there is none if $m=0$. Suppose that both flanks are in $P$. Then $A=B=e_1$.
Since $G$ has no loops, $e_1$ has one dart at $s$, so the darts $r_s(j-1)$ and $r_s(j)$ at the
corner $c=(s,j)$ coincide and $\deg(s)=1$. Then both sides of $e_1$ lie in the face of $c$, which is
$p_0$, so $e_1^*$ is a dual loop, which a simple dual path with $m\ge1$ does not use. Hence one flank
is not in $P$, and the choice can take it, with mode~L if it is $A$ and mode~R if it is $B$. If $m=0$,
both flanks are allowed.

In both modes the cut is at $c^*_s\in p_0$; the mode only decides which flank is $e_0$ and which token
moves. So the pieces and the annuli depend on $c^*_s$ but not on the mode. Under~\cite[Sec.~5]{ADW26b},
$s$ may pick any edge $e$ of $s$ outside $P$ with $p_0$ on one side. One of the two corners of $s$ next
to $e$ lies in $p_0$; the choice reads $e$ as mode~L at that corner if $e$ is its in-flank, and as
mode~R if $e$ is its out-flank. So every choice allowed in~\cite{ADW26b} has a legal reading. The
formula of~\cite[Remark~3.3]{ADW26b} is the mode-L reading in both cases; in the second case it cuts
the corner on the other side of $e$, which lies in the other face of $e$, and this is~(D3). The
argument for $t$ is the same, with $p_m$ and $e_m$.

\paragraph{(1) Pieces become disjoint paths.}

\begin{claim}[$E_R$ degrees]\label{cl:Rdeg}
The star centres have no $E_R$ edge. The copy $D(c)$ carries the \emph{in-token} of $c$, the token of
$\cin(c)$ labelled $0$ or $2$, and the \emph{out-token} of $c$, the token of $\cout(c)$ labelled $0$
or $1$. These are two distinct $E_R$ edges, except at a cut corner, where $D(c^*)$ keeps one of them
and the moved token sits at $X(c^*)$. The copies $L(c)$, $R(c)$ and $X(c^*)$ carry exactly one $E_R$
edge each.
\end{claim}

\begin{proof}
By the definition of $E_R$, $\beta(x)=D(c)$ only for the token of $\cin(c)$ labelled $0$ or $2$, and
$\alpha(x)=D(c)$ only for the token of $\cout(c)$ labelled $0$ or $1$; each dart has exactly one such
token. Also, $\beta(x)=L(c)$ only for $(\cin(c),1)$, $\alpha(x)=R(c)$ only for $(\cout(c),2)$, and
$X(c^*)$ is an end only of $x^*$. Repair~(a) moves $x^*$ from $D(c^*)$ to $X(c^*)$. The darts $\cin(c)$
and $\cout(c)$ are distinct, and $\alpha(x)\ne\beta(x)$ for every token, since $G$ has no loops.
\end{proof}

\begin{claim}[links]\label{cl:Rlinks}
Let $y$ and $z$ be consecutive tokens of a face walk with no cut between them. Then they share the
copy $D(c)$ of the corner $c$ between them. Conversely, the two tokens at a copy with two $E_R$ edges
are consecutive in one piece.
\end{claim}

\begin{proof}
The cut between $(\delta,1)$ and $(\delta,2)$ is always present, so $y$ is the last token of a dart
$\delta$ and $z$ the first token of $\delta'=\cout(\eta\delta)$, and $c=\eta\delta$ is not a cut corner.
Then $y$ is the in-token and $z$ the out-token of $c$, and repair~(a) did not move them. By
Claim~\ref{cl:Rdeg}, a copy with two $E_R$ edges is $D(c)$ for a corner $c$ that is not cut, and its
tokens are the in-token and the out-token of $c$. These are consecutive in the walk of the face of
$c$, with no cut between them.
\end{proof}

Let $f=(x_1,\dots,x_k)$ be a linear piece. The edges $x_1,\dots,x_k$ form the walk $\alpha(x_1)$,
$\beta(x_1)=\alpha(x_2)$, \dots, $\beta(x_k)$, by Claim~\ref{cl:Rlinks}. Its interior vertices are the
$D(c)$ of the $k-1$ corners between consecutive tokens, and these corners are distinct, since a face
walk passes each corner once. The piece starts just after a cut, so $\alpha(x_1)$ carries one $E_R$
edge: after a dart cut, $\alpha(x_1)$ is $R(\tau\delta)$, and after the corner cut $c^*$ it is $D(c^*)$
in mode~L and $X(c^*)$ in mode~R. Likewise $\beta(x_k)$ carries one $E_R$ edge. So both ends differ
from the interior vertices, which carry two. They also differ from each other: a copy that equals
both $\alpha(x_1)$ and $\beta(x_k)$ carries $x_1$ and $x_k$, so $k=1$ and $\alpha(x_1)=\beta(x_1)$,
which is impossible. Hence $\hat p(f)$ is a path with $k$ edges and $k+1$ distinct vertices. A cyclic
piece is a whole face without cuts, and its tokens close up into a cycle through the $D(c)$ of all its
corners. By Claim~\ref{cl:Rlinks}, the paths of different pieces share no copy, and by
Claim~\ref{cl:Rdeg}, every copy other than a star centre has an $E_R$ edge and lies on exactly one
$\hat p(f)$. Every vertex $v$ of $p(f)$ is an end of a dart of $f$, so the $E_R$ edge of that token ends
in a copy of $v$. With part~(a) of the \LemAnn{}, this gives~\cite[Thm.~3.1(1)]{ADW26b} for every
node-part of a live annulus.

This step replaces the proof of~\cite[Lemma~3.11]{ADW26b} near $s$ and $t$. That proof treats the
endpoint case ``as if $v_i$ had two incident edge copies in $\hat G_\ell$ and one of them (the one
corresponding to the endpoint-case edge) got deleted'', which fails
when $e_0\in I_\ell$. In mode~L, the piece $\{(\cin(c^*),2)\}$ then lies between the cut inside
$\cin(c^*)$ and the cut at $c^*$. Under repair~(a) its path is the single edge
$R(\tau\cin(c^*))X(c^*)$. Without the repair it ends at $D(c^*)$, which also carries the out-token of
$c^*$, and the two pieces merge into one path, as in~(D1).

\paragraph{(2) $E_C$ edges.} Every pair gets exactly one $E_C$ edge, and distinct pairs get distinct
edges, so the $E_C$ edges correspond one to one to the pairs. The copy $\mathrm{att}_h(x)$ is an end of
the $E_R$ edge $x$, so by~(1) it lies on $\hat p(f)$ for the piece $f$ of $x$. Hence the $E_C$ edge of
$\{x,y\}$ joins $\hat p(f)$ and $\hat p(g)$. By part~(a) of the \LemAnn{}, every live annulus $A$ of level
$\ell$ is a component of $G^*_\ell$. So if $x$ lies in a node-part of $A$, then $y$ lies in $A$ as well,
and the pair is a dual edge of $A$; conversely, each dual edge of $A$ is such a pair. If one token of
a pair lies in a residual piece, so does the other, since residual pieces form whole components.
The residual components are the leaves and discarded subgraphs of earlier levels, the subgraphs
discarded on this level~\cite[Claim~5.8, Cor.~2.11]{ADW26b}, and the sides that get no call (N3); a
discarded node-part of~\cite{ADW26b} can consist of two pieces ($S_{\mathrm{out}}$ in case~(iv) of the
proof of the \LemAnn{}).

For the loops: $\mathrm{att}_h(x)$ and $\mathrm{att}_h(y)$ are copies at the two corners of $h$ that
flank the edge, or $X(c^*)$ for a moved token. They coincide only if $\deg(h)=1$ and neither token
moved at $h$; then $x$ and $y$ are consecutive at the single corner of $h$, so they lie in one piece,
and the pair is a dual loop. If the pair of $x^*$ is hosted at the terminal $z$ of $c^*$ and
$\deg(z)=1$, its $E_C$ edge is $D(c^*)X(c^*)$, parallel to the $E_S$ edge $D(c^*)X(c^*)$. This happens
for $e_0\in I_\ell$ with $\deg(s)=1$, for $e_{\mathrm{last}}\in I_\ell$ with $\deg(t)=1$, and for an
unincised $e_0=e_{\mathrm{last}}$ with $\deg(t)=1$.

This step replaces the endpoint paragraph of the proof of~\cite[Lemma~3.13]{ADW26b}, which assumes
that the $E_C$ edge $(u_{j-1},u_j)$ at the other end $u$ is unaffected. That fails when $u=t$ and
$e_0=e_{\mathrm{last}}$~(D2). For the single edge $st$, $(u_{j-1},u_j)$ is the loop $D(t,0)D(t,0)$,
while under repair~(b) the $E_C$ edge is $X(c^*_t)D(t,0)$, which joins the pieces $\{(s\to t,0)\}$ and
$\{(t\to s,0)\}$.

\paragraph{(3) Planarity.} We extend the proof of~\cite[Lemma~3.5]{ADW26b}, which turns $\hat G_0$
into $\hat G_\ell$ through four local steps and leaves the embedding implicit, and we give the
embedding explicitly.

\emph{Step A: the graph $H^+$.} Let $H^+$ be the graph of the definition with no moved token and no
copy $X$, and with, at every end $w$ of every edge $e$, the \emph{rim edge} that joins the copies $D(c)$
and $D(c')$ of the two corners $c,c'$ of $w$ next to $e$. These are the $E_C$ edges of~\cite{ADW26b}
at both ends of every edge, so $H^+$ is the graph after steps~1 to~3 of the proof
of~\cite[Lemma~3.5]{ADW26b}, with step~1 adding all rim edges, which its argument covers. We draw $H^+$
as follows.
\begin{itemize}
\item Fix a planar drawing of $G$ that realizes its rotation system. Take small disjoint disks $B_v$
  around the vertices, each meeting only the edges at $v$, in radial segments, and thin disjoint
  strips around the edges outside the disks. For a face $h$, let $h^\circ$ be $h$ minus the disks and
  strips. It is an open disk, and its boundary follows the walk of $h$ once: along one side of the
  strip of each dart of $h$, and along the arc of $\partial B_v$ inside each corner of $h$.
\item Put $\sigma(v)$ at $v$ and $D(c)$ on the arc of $\partial B_v$ inside $c$. Each edge
  $\sigma(v)D(c)$ is a radial segment.
\item The rim edge at $v$ across an edge $e$ runs inside $B_v$ between the copies of the two corners
  next to $e$. It crosses the radial segment of $e$ and nothing else, and it reaches each of its ends
  from the side of $e$.
\item An unincised token $(\delta,0)$ runs along $\partial h^\circ$ from $D(\tau\delta)$ to
  $D(\eta\delta)$, where $h$ is the face of $\delta$.
\item An incised dart $\delta$ gets a lens: a near arc along $\partial h^\circ$ and a far arc parallel
  to it inside $h^\circ$, both from the corner arc of $\tau\delta$ to that of $\eta\delta$. Under
  option~(i), the near arc is $(\delta,2)$, from $R(\tau\delta)$ to $D(\eta\delta)$, and the far arc is
  $(\delta,1)$, from $D(\tau\delta)$ to $L(\eta\delta)$; the copy $R(\tau\delta)$ lies on $\partial
  h^\circ$ between $D(\tau\delta)$ and the strip of $\delta$, and $L(\eta\delta)$ lies just inside
  $h^\circ$ next to $D(\eta\delta)$, on the side of $\delta$. Under option~(ii), the near arc is
  $(\delta,1)$, from $D(\tau\delta)$ to $L(\eta\delta)$, and the far arc is $(\delta,2)$, from
  $R(\tau\delta)$ to $D(\eta\delta)$; then $L(\eta\delta)$ lies on $\partial h^\circ$ between the strip
  of $\delta$ and $D(\eta\delta)$, and $R(\tau\delta)$ just inside $h^\circ$ next to $D(\tau\delta)$, on
  the side of $\delta$. Under both options $L(c)$ lies at the corner $c$ on the side of $\cin(c)$, and
  $R(c)$ on the side of $\cout(c)$; the rotation table depends on this. The edges $D(c)L(c)$ and $D(c)R(c)$ are
  short segments.
\end{itemize}
Nothing crosses. Radial segments and rim edges lie inside the disks and meet $\partial B_v$ only at
the copies $D(c)$. Near arcs lie on the curves $\partial h^\circ$, which are disjoint for distinct
faces, and the near arcs of one face use disjoint stretches of $\partial h^\circ$. Far arcs of one face
run at one small distance from $\partial h^\circ$, along disjoint stretches. At a corner $c$, the
entries on the side of $\cin(c)$ leave $D(c)$ towards one flank and those on the side of $\cout(c)$
towards the other. So $H^+$ is planar, and reading the drawing around each copy gives the rotation table, with
both rim edges at every $D(c)$.

\emph{Step B: the repair.} This replaces step~4 of the proof of~\cite[Lemma~3.5]{ADW26b}. For each cut
corner $c^*$ at a terminal $z$:
\begin{enumerate}[1.]
\item Subdivide the $E_R$ edge $x^*$ next to $D(c^*)$ by the new vertex $X(c^*)$. The part between
  $D(c^*)$ and $X(c^*)$ becomes the $E_S$ edge $D(c^*)X(c^*)$. In the drawing, $X(c^*)$ is a point of
  the corner arc of $c^*$, so it sits in the corner that it cuts. Subdivision keeps planarity.
\item If the pair $\pi^*$ of $x^*$ is hosted at $z$, its $E_C$ edge is the rim edge at $D(c^*)$ on the
  side of $x^*$: $C_A$ in mode~L, where $x^*$ is a token of $\cin(c^*)$, and $C_B$ in mode~R. In
  the rotation table this rim edge is next to $x^*$ at $D(c^*)$. In mode~L, $x^*$ is $A_1$: the token of an
  unincised $\cin(c^*)$, or $(\cin(c^*),2)$ under option~(i). In mode~R, $x^*$ is $B_1$: the token of an
  unincised $\cout(c^*)$, or $(\cout(c^*),1)$ under option~(ii); this is why option~(ii) is used at
  corners in mode~R. So the rim edge can be rerouted along $x^*$ to end at $X(c^*)$ without crossing
  anything. This is repair~(b) at $c^*$, and the rotation at $X(c^*)$ is then the one in the rotation table.
\item Delete every rim edge that is not an $E_C$ edge of the definition. Deleting edges keeps
  planarity.
\end{enumerate}
Each $E_C$ edge of the definition is a rim edge of $H^+$. For an unincised edge this holds with either
end as host: at the tail of $\delta$ the $E_C$ edge is $D(\tau\delta)D(\eta\rev\delta)$, and at the head
it is $D(\eta\delta)D(\tau\rev\delta)$, both rim edges. For an incised edge it needs the host rule: at
the tail of $\delta$, the pair $\{(\delta,1),(\rev\delta,2)\}$ gets the rim edge
$D(\tau\delta)D(\eta\rev\delta)$, while at the head it would get $L(\eta\delta)R(\tau\rev\delta)$, which is
not a rim edge of $H^+$. At most one end of an $E_C$ edge is rerouted, because two moved tokens at one
host would need two cut corners at one vertex. The result is $\hat G_\ell$, so $\hat G_\ell$ is planar.

The two options never conflict. A dart $\delta$ needs option~(ii) only as $\cout(c^*)$ of a corner in
mode~R, which is its tail corner. It would need option~(i) at its head only if it were also
$\cin(c^{*\prime})$ of a corner in mode~L with $\delta$ incised. Then $c^*=\tau\delta$ and
$c^{*\prime}=\eta\delta$ both lie in the face of $\delta$, so $p_0=p_m$ and $m=0$. With $m=0$ the
annulus of level~$1$ is a leaf~\cite[Def.~2.8; Alg.~2, line~5]{ADW26b}, so no edge is ever incised.

\emph{Local computation.} The rotation table at $v$ uses the rotation at $v$; which edges at $v$ are in $I_\ell$;
the host of each edge at $v$, which needs the identifiers of the neighbours and whether the edge is
$e_0$ or $e_{\mathrm{last}}$; whether $v$ is $s$ or $t$, with its corner $c^*$ and mode $\mu$; and, for
the option at the head of a dart, whether the dart is $\cout(c^*)$ of a corner in mode~R at its tail.
The terminals send their endpoint data to their neighbours in one round. So each vertex computes the
rotations of all its copies within $O(1)$ rounds, and the \ThmSSSP{} receives an embedding of
$\hat G_\ell$ that is computed locally from that of $G$.

\paragraph{(4) Diameter, size and simulation.} A copy of $v$ is within distance $2$ of $\sigma(v)$:
$D(c)$ at distance $1$, and $L(c)$, $R(c)$ and $X(c^*)$ at distance $2$ through $D(c)$. Let $e=uv$ be an
edge of $G$. For a token $x$ of $e$, the path $\sigma(u),\mathrm{att}_u(x),\mathrm{att}_v(x),\sigma(v)$,
with the $E_S$ steps through $D$ where needed, has $3$ hops plus one for each end of $x$ at a copy
$L$, $R$ or $X$. Unincised tokens have copies $D$ at both ends unless moved, and incised tokens have
exactly one end at $L$ or $R$ before repair~(a). Repair~(a) moves at most one token of $e$ at $s$ and
one at $t$, and it moves only ends at copies $D$. If $e$ is unincised, either one token is moved at
both ends and the other has no end at $L$, $R$ or $X$, or each token has at most one such end; either
way some token gives at most $4$ hops. If $e$ is incised, at most two of its four tokens are moved, so
an unmoved token gives $4$ hops. So the distance between $\sigma(u)$ and $\sigma(v)$ is at most $4$ for
every edge $uv$ of $G$, and the hop diameter of $\hat G_\ell$ is at most $4D+4$; the bound
in~\cite[Lemma~3.4]{ADW26b} is $4D$, without the two hops at each end. The $E_C$ edges only shorten
distances.

A vertex $v$ has one star centre, $\deg(v)$ copies $D$, at most $\deg(v)$ copies $L$, one per corner
with an incised in-dart, at most $\deg(v)$ copies $R$, and a copy $X$ only if $v\in\{s,t\}$. So
$|V(\hat G_\ell)|\le6|E|+n+2$.

For the construction~\cite[Lemma~3.15]{ADW26b}, besides the stored set $I_\ell$, the vertex $v$ needs
the index of each of its edges in the rotation of the other end (one round) and the endpoint data of
$s$ and $t$ (one round). The identifier of a copy is the identifier of $v$ followed by a number in
$[0,3\deg(v)+1]$. For the simulation~\cite[Lemma~3.16]{ADW26b}: $E_S$ edges join copies of one vertex,
and by the host rule so do $E_C$ edges; each edge of $G$ has at most four tokens (N5), so a round on
$\hat G_\ell$ takes four rounds on $G$. Repair~(a) only changes which copy of $s$ or $t$ holds a token.

\paragraph{(6) The statement of~\cite[Lemma~3.17]{ADW26b}.} \emph{Global phase.} The global phase
of~\cite[Lemma~3.17]{ADW26b} runs connectivity on $\hat G_\ell[E_R\cup E_C]$ and on $\hat G_\ell[E_R]$,
with $\Ot(1)$ part-wise aggregations on $\hat G_\ell$~\cite[Cor.~3.23]{ADW26b}, which needs only~(3)
and~(4). By~(1), the components of $\hat G_\ell[E_R]$, apart from the isolated star centres, are
exactly the paths $\hat p(f)$ of the pieces, and $X(c^*)$ lies on $\hat p(f)$ for the piece $f$ of
$x^*$. By~(2) and part~(a) of the \LemAnn{}, the component of $\hat G_\ell[E_R\cup E_C]$ that contains
$\hat p(f)$, for a node-part $f$ of a live annulus $A$, is the union of the paths $\hat p(g)$ for the
node-parts $g$ of $A$ and of the $E_C$ edges of $A$. So $X(c^*)$ receives the right identifiers of its
node-part and of its annulus, with no new case. The leader of $\hat p(f)$ is its copy of largest
identifier. If that copy is $X(c^*)$, the leader vertex is $s$ or $t$, which lies in $p(f)$ because
$x^*$ is a token of an edge at that terminal. The \ThmCut{} uses exactly these outputs: the
annulus of each copy, for the weights and the test for liveness, and the leader copy of $\hat p(f_i)$,
for the virtual source.

\emph{Local phase, item~1} (the node-part and annulus of each incident edge). The vertex $v$ reads
them at the copy $\mathrm{att}_v(x)$, for each token $x$ of an edge at $v$, $X(c^*)$ included. There is
no new case.

\emph{Local phase, item~2} (from level $\ell-1$ to level $\ell$). The local phase
of~\cite{ADW26b} matches copies of $\hat G_{\ell-1}$ and $\hat G_\ell$ by name ($v_i$, $v_i^L$ and
$v_{i-1}^R$; cases (a) to (c) there). We replace this by a rule on tokens. \emph{Rule~M:} for a token
$x$ of level $\ell$ at $v$, with parent $\bar x$, the vertex $v$ declares the node-part of
$\mathrm{att}^\ell_v(x)$ in $\hat G_\ell$ to be contained in the node-part of
$\mathrm{att}^{\ell-1}_v(\bar x)$ in $\hat G_{\ell-1}$. Since $I_{\ell-1}\subseteq I_\ell$, the parent
$\bar x$ is a token of level $\ell-1$ for every $x$, so the rule is defined everywhere.

\begin{claim}\label{cl:Rmatch}
Let $x$ be a token of a piece $f$ of level $\ell$, and let $g$ be the piece of level $\ell-1$ that $f$
refines (part~(b) of the \LemAnn{}). Then $\mathrm{att}^{\ell-1}_v(\bar x)$ lies on $\hat p(g)$. If $f$ is a
node-part of a live annulus $A'$, then $g$ is a node-part of the parent annulus of $A'$.
\end{claim}

\begin{proof}
The parent $\bar x$ lies in $g$ by part~(b) of the \LemAnn{}, and $\mathrm{att}^{\ell-1}_v(\bar x)$ is an end
of the $E_R$ edge $\bar x$, which lies on $\hat p(g)$ by~(1) on level $\ell-1$. For $\ell=1$, $g$ is a
face and $\hat p(g)$ its cycle in $\hat G_0$. For the second part, the pieces of $A'$ arise from those
of its parent $A$ by the new cuts of the step in the proof of part~(a) of the \LemAnn{}, so $g$ is a piece
of $A$.
\end{proof}

So Rule~M is correct for every piece, residual or not; it is~\cite[Claim~2.15]{ADW26b} in the form of
tokens. It is local, since $v$ knows $I_{\ell-1}$, $I_\ell$ and the endpoint data, and hence the
attachments on both levels. Away from $s$ and $t$ it reproduces cases~(a) to~(c) of~\cite{ADW26b}.
At $s$ in mode~L, with $a=\cin(c^*)$, it gives:
\begin{center}\small
\begin{tabular}{@{}lp{0.6\linewidth}@{}}
\toprule
$e_0$ on levels $\ell-1$ and $\ell$ & copies of $s$ at $c^*$ in $\hat G_\ell$, and their match in $\hat G_{\ell-1}$\\
\midrule
unincised, unincised & $X(c^*)\mapsto X(c^*)$\\
unincised, incised anew & $X(c^*)$, holding $(a,2)$, and $L(c^*)$, holding $(a,1)$, both $\mapsto X(c^*)$, which held $(a,0)$\\
incised, incised & $X(c^*)\mapsto X(c^*)$ and $L(c^*)\mapsto L(c^*)$\\
$\ell=1$ ($\hat G_0$ has no $X$) & $X(c^*)\mapsto D(c^*)$, the copy of $x^*$ on the cycle of $p_0$ in $\hat G_0$\\
any & $D(c^*)\mapsto D(c^*)$\\
\bottomrule
\end{tabular}
\end{center}
Since $I_\ell$ grows with $\ell$, an incised $e_0$ stays incised. Mode~R is the mirror image, with $R$
and $b=\cout(c^*)$ in place of $L$ and $a$: if $e_0$ is incised anew, $X(c^*)$ holds $(b,1)$ and
$R(c^*)$ holds $(b,2)$, and both map to $X(c^*)$, which held $(b,0)$. This copy is the mirrored
$s_i^R$ of the choice.

The endpoint paragraph of case~2(c) in the proof of~\cite[Lemma~3.17]{ADW26b} says that the paths of
$v_i^L$ and $v_i$ in $\hat G_\ell$ both correspond to the path of $v_i^L$ in $\hat G_{\ell-1}$. Under
the repair the second copy is $X(c^*)$, not $v_i=D(c^*)$, and $D(c^*)$ corresponds to $D(c^*)$. In the
graph as literally defined, $v_i$ carries both $(a,2)$ and the out-token of $c^*$, whose node-parts
have different parents, so the stated correspondence is wrong there; this is~(D1) again.

\emph{Local phase, items~3 and~4} (the parent annulus, and the identifiers of all ancestor
node-parts) follow from items~1 and~2 by the argument of~\cite{ADW26b}, unchanged. This proves~(6).

\paragraph{Degenerate cases.} An endpoint edge in $I_\ell$ moves $(\cin(c^*),2)$ or $(\cout(c^*),1)$
to $X(c^*)$; Claim~\ref{cl:Rdeg} still holds, since $L(c^*)$ or $R(c^*)$ carries the other token, and
step~B uses option~(i) or~(ii). If $e_0=e_{\mathrm{last}}$, the $E_C$ edges use $\mathrm{att}_h$: an
unincised edge has one pair, hosted at $t$ and ending at $X(c^*_t)$, and an incised edge has each pair
hosted at its own end, where it may end at $X$. An endpoint edge incised anew is covered because the
definition is per level and $I_\ell$ grows with $\ell$ (Rule~M). If $\deg(s)=1$, the two flanks of
$c^*_s$ are the same edge $e_0$, and $\cin(c^*_s)$ and $\cout(c^*_s)$ are its two darts; nothing above
assumes distinct flanks, and in the \LemAnn{}, $e_0$ is a bridge with both darts in $p_0$. If
$\deg(s)=2$, there are two rim edges between the two copies $D$ of $s$, hence parallel $E_C$ edges.
Parallel edges of $G$ give $2$-cycles $\hat p$ for faces of length~$2$, and $n=2$ is allowed. For
several corners of $s$ in $p_0$, the endpoint data fix one, and $\gamma$ meets $s$ there. An endpoint
edge with both darts in $p_0$ has a dual loop at $p_0$, and tokens keep its two darts apart. If $m=0$,
then $p_0=p_m$ carries both corner cuts, $\gamma$ is one segment of $p_0$, and no edge is ever
incised.

\section{Proof of the Incision Lemma}\label{app:O}

The proof is an induction over the levels. Along with the \LemInc{} it proves an invariant (I)
of the live annuli (Appendix~\ref{app:inv}), and the \LemInc{} is stated for annuli that satisfy
(I). The model is a drawing of the token graph in the sphere $S^2$, in which every incision is a curve
and every side is a side of a simple closed curve.

\subsection{Setting}

We use the notation of Section~\ref{sec:thmD} and Appendix~\ref{app:R}. The graph $G$ is connected
and drawn in $S^2$, $P=(p_0,\dots,p_m)$ is the shortest path of the algorithm, and the dual of $e_k$
joins $p_{k-1}$ and $p_k$. Lengths are $w=Kc'+1$, and $\lambda_t$ is the distance in $A_1$ from $p_t^0$
to $p_t^1$.
\begin{itemize}
\item On level $\ell$, every old token is an end of its piece (Appendix~\ref{app:R}).
\item A \emph{$P$-piece} is a piece of a face of $P$. A piece of level $\ell$ is a \emph{piece of
  $p_t^\sigma$} if it refines the piece $p_t^\sigma$ of $A_1$, through part~(b) of the \LemAnn{} applied
  level by level. Other pieces are \emph{off $P$}.
\item The annulus $A_1=G^*_1$ is the only live annulus of level~$1$.
\item \emph{Bounding paths.} The annulus $A_1$ has none. Let $A$ be live and not a leaf, and $Q$ the
  part that it incises. The child on the low side ($j$-side, range $[j_A,i_1]$) has the $j$-bound of
  $A$ as its $j$-bound and the copy of $Q$ on the low side as its $k$-bound. The child on the high side
  ($k$-side, range $[i_2,k_A]$) has the copy of $Q$ on the high side as its $j$-bound and the $k$-bound
  of $A$ as its $k$-bound. We write $P_a$ and $P_b$ for the $j$-bound and the $k$-bound of $A$; either
  may be absent. A bound is the copy in $A$ of the incised part of the path of an ancestor, not of
  its whole path; this is a reading of~\cite{ADW26b} (Section~\ref{sec:thmD}). The ends of $P_a$ are a piece of $p_{x_a}^0$ and a piece of $p_{y_a}^1$, and those of
  $P_b$ a piece of $p_{x_b}^0$ and a piece of $p_{y_b}^1$ (Claim~\ref{cl:Opass}). If $P_a$ is absent,
  set $x_a=y_a=0$; if $P_b$ is absent, set $x_b=y_b=m$.
\item \emph{The step in $A$.} Let $A$ be live and not a leaf, with range $[j_A,k_A]$ and
  $i=\lfloor(j_A+k_A)/2\rfloor$. Then $P_i$ is a shortest path in $A$ from $g_i^0$ to $g_i^1$, which
  exist and are unique by (Pieces) below, computed exactly and read from $g_i^0$. The indices $x$ and
  $y$, the part $Q=P_i'$ from $g_x$ to $g_y$, and $i_1$ and $i_2$ are as in Section~\ref{sec:thmD}.
  Since $g_i^0$, $g_x$ and $g_y$ lie on $P_i$, we have $i_1\le\min(i,x,y)$ and
  $\max(i,x,y)\le i_2$.
\item \emph{The part $Q$.} It is a subpath of a shortest path, so it is a simple path and a shortest
  path between any two of its nodes. Its internal nodes are off $P$, by the choice of $x$ and $y$. No
  edge of $Q$ is a copy of an edge of $P$: such a copy joins two pieces on the same side of $\gamma$
  (Appendix~\ref{app:curves}), while an edge of $Q$ has an end off $P$ or joins a $0$-piece to a
  $1$-piece.
\item \emph{Incision in token form.} Another annulus $A''$ of level $\ell$ cuts only darts in its own
  pieces: a new edge of its incised part has its only pair in $A''$, and $A''$ is a component of
  $G^*_\ell$ by (Comp), so no piece of $A$ holds a dart of that edge. So the cuts of level $\ell+1$ in the pieces of $A$ are its
  cuts of level $\ell$ and one cut inside each dart of each new edge of $Q$. Let $A^Q$ be the graph
  whose nodes are the pieces of level $\ell+1$ that refine pieces of $A$, and whose edges are the pairs
  of level $\ell+1$ whose tokens lie in these pieces. The \emph{subgraphs obtained by incising $Q$}
  are the components of $A^Q$. Nodes of $A$ off $Q$ and old pairs do not change, and a new edge of $Q$
  gets the two pairs $\{(\delta,1),(\rev\delta,2)\}$ and $\{(\delta,2),(\rev\delta,1)\}$. This is the
  incision of~\cite[Remark~2.5]{ADW26b} in the reading of the \LemAnn{}.
\item \emph{Calls.} As in Section~\ref{sec:thmD}: if $P_i$ meets $P_a$, that is, shares a node with
  it, the $j$-side gets no call and the overlap rule computes, from a common node, a path for every
  $t\in[j_A,i]$; the $k$-side and $P_b$ are symmetric. By (Old) in Appendix~\ref{app:inv}, the nodes of $P_a$ and $P_b$ are
  exactly the nodes of $A$ that carry an old token of an edge off $P$, so every node knows locally
  whether it lies on a bound. Overlap is tested directly, and not through bad edges. A side without
  overlap is called if its range is not empty. The two call rules are readings of~\cite{ADW26b}.
  Claim~\ref{cl:Ocalled} shows that a called side is one subgraph, so the call is well defined.
\end{itemize}

\subsection{The curve model}\label{app:curves}

All curves are polygonal, for a fixed piecewise linear structure on $S^2$ in which the drawing of $G$
is polygonal. An \emph{arc} is a simple curve with two distinct ends, and a \emph{Jordan domain} is a
component of the complement of a simple closed curve. We use four standard facts.
\begin{itemize}
\item \emph{Jordan.} A simple closed curve $J$ splits $S^2\setminus J$ into two connected open sets,
  its \emph{sides}, each with boundary $J$. A connected set disjoint from $J$ lies in one of them.
\item \emph{$\theta$-curve.} Let $\Delta$ be a Jordan domain, and let $\kappa$ be an arc with
  distinct ends on $\partial\Delta$ and interior in $\Delta$. Then $\Delta\setminus\kappa$ has two
  components, and each is a Jordan domain bounded by $\kappa$ and one of the two arcs of $\partial\Delta$
  between the ends of $\kappa$.
\item \emph{Parity.} Two closed curves in general position cross an even number of times.
\item \emph{One crossing.} If a curve crosses a simple closed curve $J$ transversally at one point and
  meets $J$ nowhere else, its two ends lie on different sides of $J$.
\end{itemize}
Moreover, two distinct points of the boundary of a Jordan domain are joined by an arc whose interior
lies in the domain.

\paragraph{The thickened drawing.} Replace each vertex $v$ of $G$ by a small closed disk $B_v$ and each
edge $e$ by a thin closed strip $S_e$ that joins the disks of its ends. Each face $h$ of $G$ becomes a
component $h^\circ$ of the complement of the union of these sets. Since $G$ is connected, $h^\circ$ is
an open disk and $\partial h^\circ$ is a simple closed curve, also when the boundary walk of $h$
repeats vertices or edges. The curve $\partial h^\circ$ follows the walk of $h$: one \emph{side
segment} of $\partial S_e$ for each dart of $h$, and one \emph{corner arc} of $\partial B_v$ for each
corner. The boundaries of distinct faces are disjoint.
\begin{itemize}
\item A token $(\delta,0)$ is the side segment of $\delta$. For an incised dart $\delta$, the \emph{cut
  point} $c(\delta)$ is the midpoint of its side segment, $(\delta,1)$ is the half before $c(\delta)$ in
  walk order, and $(\delta,2)$ the half after it; for a token $x$ of $\delta$, $c(x):=c(\delta)$. The
  corner cuts at $c^*_s$ and $c^*_t$ are points of the corner arcs of these corners.
\item For an incised edge $e$ with darts $\delta$ and $\rev\delta$, the \emph{cut line} $\mathrm{cl}(e)$
  is a segment across $S_e$ from $c(\delta)$ to $c(\rev\delta)$. The pair
  $\{(\delta,1),(\rev\delta,2)\}$ consists of the two halves at the tail of $\delta$, and the pair
  $\{(\delta,2),(\rev\delta,1)\}$ of the two halves at its head. So the two pairs of an incised edge lie
  in the two components of $S_e\setminus\mathrm{cl}(e)$, the \emph{half-strips}.
\item The \emph{arc} $\mathrm{arc}(g)$ of a piece $g$ is the closed sub-arc of $\partial h^\circ$ made of
  the segments and halves of its tokens, the corner arcs between consecutive tokens and, for a piece
  next to $c^*_s$ or $c^*_t$, the part of the corner arc up to that corner cut; for a cyclic piece it
  is $\partial h^\circ$. Its ends are cut points, and the arcs of distinct pieces meet at most in their
  ends.
\end{itemize}

\paragraph{The arc $\gamma$.} It starts at the corner cut of $c^*_s$, runs through $p_0^\circ$ to the
cut point of the dart of $e_1$ in $p_0$, along $\mathrm{cl}(e_1)$, through $p_1^\circ$ to the cut
point of the dart of $e_2$ in $p_1$, and so on, and it ends through $p_m^\circ$ at the corner cut of
$c^*_t$. Its part $\gamma_t$ in $p_t$ is an arc from its \emph{entry} (the cut at $c^*_s$, or the cut
point of $e_t$ in $p_t$) to its \emph{exit} (the cut point of $e_{t+1}$ in $p_t$, or the cut at
$c^*_t$), with interior in $p_t^\circ$. The arc $\gamma$ is simple, because $P$ visits each face once.
We orient $\gamma$ from $s$ to $t$ and say that a point of $\gamma$ \emph{precedes} another.

The arc $\gamma$ has two local sides. By the $\theta$-curve fact, $\gamma_t$ splits the closed disk of
$p_t$ into two \emph{halves}, each bounded by $\gamma_t$ and the arc of one of the two pieces of $p_t$.
The drawing realizes the rotation system, so walk order keeps $h^\circ$ on one hand of $\partial h^\circ$
for every face $h$; we call this hand the left. Side~$0$ of $\gamma$ is its right side and side~$1$ its
left side. Let $H_t^\sigma$ be the half on side $\sigma$, and let $p_t^\sigma$ be the piece whose arc
bounds it. The boundary of $H_t^0$, with $H_t^0$ on the left, runs along the arc of $p_t^0$ and back along
$\gamma_t$, so this arc runs in walk order from the entry to the exit of $\gamma_t$. This is the naming of
Section~\ref{sec:thmD}. For a piece $g$ of $p_t^\sigma$ on a later level, the
\emph{low end} of $\mathrm{arc}(g)$ is its end nearer to the entry of $\gamma_t$ along
$\mathrm{arc}(p_t^\sigma)$, and the other end is its \emph{high end}. The pair of $e_k$ whose halves lie
on side $\sigma$ of $\mathrm{cl}(e_k)\subseteq\gamma$ joins $p_{k-1}^\sigma$ and $p_k^\sigma$. So in
$A_1$, the copies of $P$ form two paths, $p_0^\sigma,\dots,p_m^\sigma$ for $\sigma\in\{0,1\}$, and on
later levels a copy of an edge of $P$ joins two pieces on the same side of $\gamma$, since $e_k$ has
been in $I_\ell$ since level~$1$ and its tokens refine their pieces of $A_1$.

\paragraph{Incision curves, slots, collars and drawings.} Every incision of a part $Q$ in a live
annulus adds a curve $C_Q$ (Claim~\ref{cl:Ocurve}). Let $K_\ell$ be the union of $\gamma$ and of the
curves of all incisions on levels below $\ell$, in all annuli; so $K_1=\gamma$.
\begin{itemize}
\item A \emph{chord} of $K_\ell$ in a face $h$ is an arc of some $C_Q$ whose interior lies in
  $h^\circ$ and whose ends are cut points on $\partial h^\circ$ or points of $\gamma$. The set $K_\ell$
  meets the closed disk of $h$ in its chords and, if $h=p_t$, in $\gamma_t$. It meets a strip $S_e$ in
  $\mathrm{cl}(e)$ if $e\in I_\ell$, and not at all otherwise.
\item The \emph{slot} $\Delta(g)$ of a piece $g$ of a live annulus is the component of
  $h^\circ\setminus K_\ell$ that contains the points of $h^\circ$ near the interior of
  $\mathrm{arc}(g)$. Since $K_\ell$ meets $\partial h^\circ$ only in cut points, such points exist and
  lie in one component.
\item The \emph{collar} $\mathrm{col}(g)$ is a thin closed band in $\Delta(g)$ along $\mathrm{arc}(g)$
  minus small neighbourhoods of its ends, or along all of $\partial h^\circ$ if $g$ is cyclic.
\item The \emph{curve} of a pair with tokens in the pieces $g$ and $g'$ runs from $\mathrm{col}(g)$
  across the strip of its edge to $\mathrm{col}(g')$: anywhere across the strip if the edge is new, and
  inside the half-strip that holds the pair if the edge is old. It enters the collars only near the
  middle of its two token segments.
\item The \emph{drawing} $\Gamma(W)$ of a walk $W$ is the union of the collars of its nodes and the
  curves of its pairs. It is connected and avoids $K_\ell$, and inside it a curve runs from any point of
  the collar of the first node to any point of the collar of the last node. The drawing of a subgraph
  is defined in the same way.
\end{itemize}

\begin{claim}[disjoint drawings]\label{cl:Odisj}
Let $W$ and $W'$ be walks in a live annulus $A$ with no common node. Then
$\Gamma(W)\cap\Gamma(W')=\emptyset$.
\end{claim}

\begin{proof}
Collars of distinct pieces are disjoint: in distinct faces they lie in distinct disks, and in one
face they lie along arcs that meet at most in their ends, which collars avoid. The curve of a pair
meets only the collars of its own two pieces. The walks have no common pair, since a common pair would
give a common node. The curves of pairs of distinct edges lie in distinct strips, and the two pairs of
one old edge lie in the two half-strips.
\end{proof}

\subsection{The invariant}\label{app:inv}

A live annulus $A$ of level $\ell$ satisfies the invariant \emph{(I)} if the following hold. We write
$g_t^\sigma$ for the piece of $p_t^\sigma$ in $A$, where (Pieces) makes it unique.
\begin{itemize}
\item[(Comp)] $A$ is a component of $G^*_\ell$.
\item[(Geom)] (a) If a piece $g$ of $A$ off $P$ is cyclic, then $\Delta(g)=h^\circ$. If it is linear,
  $\Delta(g)$ is a Jordan domain bounded by $\mathrm{arc}(g)$ and one chord of $K_\ell$ that joins the
  two ends of $\mathrm{arc}(g)$.

  (b) If $g$ is a piece of $p_t^\sigma$, then $\Delta(g)$ is a Jordan domain whose boundary consists of
  $\mathrm{arc}(g)$, a sub-arc $\gamma_t[r_{\mathrm{lo}},r_{\mathrm{hi}}]$ with
  $r_{\mathrm{lo}}\ne r_{\mathrm{hi}}$, the \emph{$\gamma$-part} of $g$, and up to two chords: one from
  the low end of $\mathrm{arc}(g)$ to $r_{\mathrm{lo}}$, and one from the high end to $r_{\mathrm{hi}}$.
  A chord is absent where the end of $\mathrm{arc}(g)$ is the entry or the exit of $\gamma_t$, which is
  then $r_{\mathrm{lo}}$ or $r_{\mathrm{hi}}$.

  (c) A component $\Omega_A$ of $S^2\setminus K_\ell$ contains the slots of all pieces of $A$ and, for
  every pair of $A$, the part of its strip on its side of the cut line, or the whole open strip if its
  edge is new. Distinct live annuli of level $\ell$ have distinct such components.

  (d) If $P_a$ exists, it was created by an incision whose curve $C_a$ contains $\mathrm{cl}(e)$ for
  every edge $e$ of $P_a$ and runs from a point $q_0^a$ of $\gamma_{x_a}$, which it reaches from
  side~$0$, to a point $q_1^a$ of $\gamma_{y_a}$, which it reaches from side~$1$. The curve
  $J_a:=C_a\cup\gamma[q_0^a,q_1^a]$ is simple and closed, and $\Omega_A\subseteq\mathrm{Hi}(J_a)$.
  Symmetrically for $P_b$, with $C_b$, $q_0^b$, $q_1^b$ and $J_b$, and with
  $\Omega_A\subseteq\mathrm{Lo}(J_b)$. (The sides $\mathrm{Lo}$ and $\mathrm{Hi}$ are defined in
  Claim~\ref{cl:Osides}.)
\item[(Bound)] $P_a$ is a path in $A$ from $g_{x_a}^0$ to $g_{y_a}^1$, its internal nodes are off $P$,
  and the cut points of its first and its last token are the low ends of the arcs of these two pieces.
  The same holds for $P_b$, with high ends. The paths $P_a$ and $P_b$ have no common node.
\item[(Pieces)] $A$ has at most one piece of each face off $P$ and at most one piece of each
  $p_t^\sigma$. It has a piece of $p_t^0$ exactly for $t\in[x_a,x_b]$ and a piece of $p_t^1$ exactly for
  $t\in[y_a,y_b]$, and $\max(x_a,y_a)\le j_A\le k_A\le\min(x_b,y_b)$.
\item[(Old)] Every old pair of $A$ is a copy of an edge of $P$ or a pair of $P_a$ or of $P_b$, and $A$
  has at most one pair of each old edge off $P$.
\item[(Dist)] For every $t$ such that $A$ has $g_t^0$ and $g_t^1$, their distance in $A$ is
  $\lambda_t$.
\end{itemize}

\begin{claim}\label{cl:OA1}
The annulus $A_1$ satisfies (I).
\end{claim}

\begin{proof}
(Comp) holds since $A_1=G^*_1$ is connected by part~(a) of the \LemAnn{}. (Geom): $K_1=\gamma$; a face off
$P$ has one cyclic piece and no chord, so its slot is $h^\circ$, and the slot of $p_t^\sigma$ is the open
half $H_t^\sigma$, bounded by $\mathrm{arc}(p_t^\sigma)$ and $\gamma_t$, with no chord. The set
$S^2\setminus\gamma$ is connected, so it is $\Omega_{A_1}$, and there are no bounding paths. (Bound)
is empty. (Pieces): $A_1$ has one piece of each $p_t^\sigma$ and of each other face, and
$[x_a,x_b]=[y_a,y_b]=[0,m]=[j_{A_1},k_{A_1}]$. (Old): the old edges on level~$1$ are the edges of
$P$. (Dist) is the definition of $\lambda_t$.
\end{proof}

From now on, $A$ is a live annulus of level $\ell$ that satisfies (I) and is not a leaf, and $P_i$, $x$,
$y$, $Q$, $g_x$, $g_y$, $i_1$ and $i_2$ are as above. Let $\varepsilon_0$ be the token of the first edge
of $Q$ at $g_x$ and $\varepsilon_1$ the token of its last edge at $g_y$. For an internal node $g$ of
$Q$, let $u$ and $w$ be its two tokens on $Q$.

\subsection{One incision}

\begin{claim}[the curve of $Q$]\label{cl:Ocurve}
Draw
\begin{itemize}
\item for each new edge $e$ of $Q$, a cut line $\mathrm{cl}(e)$ across $S_e$ (for an old edge,
  $\mathrm{cl}(e)$ lies in $K_\ell$ already);
\item for each internal node $g$ of $Q$, a chord $\mathrm{ch}(g)$ from $c(u)$ to $c(w)$ with interior in
  $\Delta(g)$;
\item a chord $\mathrm{ch}(g_x)$ from $c(\varepsilon_0)$ to a point $q_0$ inside the $\gamma$-part of
  $g_x$, with interior in $\Delta(g_x)$, and a chord $\mathrm{ch}(g_y)$ from $c(\varepsilon_1)$ to a
  point $q_1$ inside the $\gamma$-part of $g_y$, with interior in $\Delta(g_y)$. The last part of
  $\mathrm{ch}(g_x)$ and of $\mathrm{ch}(g_y)$ is a straight segment perpendicular to $\gamma$. The
  points $q_0$ and $q_1$ are distinct and differ from the ends of all chords of $K_\ell$, and the
  points $q_0$ and $q_1$ chosen by all annuli of level $\ell$ are pairwise distinct.
\end{itemize}
Let $C_Q$ be the union of these chords and of the cut lines of all edges of $Q$, and let
$J:=J_Q:=C_Q\cup\gamma[q_0,q_1]$. Then:
\begin{enumerate}[(a)]
\item the drawing is possible, $C_Q$ is an arc from $q_0$ to $q_1$ with $C_Q\cap\gamma=\{q_0,q_1\}$,
  and $J$ is a simple closed curve;
\item $C_Q$ lies in the closure of $\Omega_A$;
\item $J$ meets a strip $S_e$ only in $\mathrm{cl}(e)$, and only if $e$ is an edge of $Q$ or an edge
  of $P$ whose cut line lies in $\gamma[q_0,q_1]$; and $J$ meets the closed disk of a face $h$ only in
  the chords of $C_Q$ in $h$, two if $x=y$ and $h=p_x$ and at most one otherwise, and in
  $\gamma[q_0,q_1]$.
\end{enumerate}
\end{claim}

\begin{proof}
(a) For an internal node $g$, $u\ne w$, and $c(u)$ and $c(w)$ are distinct points of
$\mathrm{arc}(g)$: the midpoint of the segment of a new token lies inside $\mathrm{arc}(g)$, and the
cut point of an old token, which is an end of $g$, is an end of $\mathrm{arc}(g)$. By (Geom),
$\Delta(g)$ is a Jordan domain whose boundary contains $\mathrm{arc}(g)$, so $\mathrm{ch}(g)$ exists.
The same holds at $g_x$ and $g_y$, whose $\gamma$-parts have interior points by (Geom b). The
internal nodes of $Q$ are distinct pieces off $P$, so by (Pieces) they lie in distinct faces off $P$.
The pieces $g_x$ and $g_y$ lie in faces of $P$, in distinct halves if $x=y$. So the chords lie in
pairwise disjoint slots. The edges of $Q$ are copies of distinct edges of $G$: a new edge has one pair,
an old edge off $P$ has one pair in $A$ by (Old), and $Q$ is simple. So the cut lines lie in distinct
strips. A chord meets $\partial h^\circ$ only in its ends, and a cut line meets the face disks only in
its ends. Along $Q$, consecutive parts of $C_Q$ share one cut point, and all these cut points are
distinct, since they belong to distinct darts. So $C_Q$ is an arc. It meets $\gamma$ only in $q_0$ and
$q_1$: the edges of $Q$ are not edges of $P$, so their strips miss $\gamma$; the faces of the internal
nodes are off $P$; and the interiors of $\mathrm{ch}(g_x)$ and $\mathrm{ch}(g_y)$ lie in slots, which
avoid $\gamma\subseteq K_\ell$.

(b) Slots of pieces of $A$ lie in $\Omega_A$, and so does the open strip of a new edge of $Q$ (Geom c).
The cut line of an old edge of $Q$ bounds the half-strip of its pair in $A$, which lies in $\Omega_A$.
The ends of the chords lie on the boundaries of their slots.

(c) follows from the construction.
\end{proof}

\begin{claim}[sides along $\gamma$]\label{cl:Osides}
Let $\mathrm{Lo}(J)$ be the side of $J$ that contains the points of $\gamma$ before $\min(q_0,q_1)$,
the \emph{low side}, and $\mathrm{Hi}(J)$ the other side, the \emph{high side}. Then:
\begin{enumerate}[(a)]
\item the points of $\gamma$ after $\max(q_0,q_1)$ lie in $\mathrm{Hi}(J)$;
\item let $r$ be a point of $\gamma$ other than $q_0$, $q_1$ and the ends of chords. Points near $r$ on
  side~$0$ of $\gamma$ lie in $\mathrm{Lo}(J)$ if $r$ precedes $q_0$, and in $\mathrm{Hi}(J)$ if $q_0$
  precedes $r$. Points near $r$ on side~$1$ lie in $\mathrm{Lo}(J)$ if $r$ precedes $q_1$, and in
  $\mathrm{Hi}(J)$ if $q_1$ precedes $r$.
\end{enumerate}
\end{claim}

\begin{proof}
Let $U$ be a closed disk around $\gamma$, so thin that $J\cap U$ consists of $\gamma[q_0,q_1]$, a part
of the final segment of $\mathrm{ch}(g_x)$ that reaches $q_0$ from side~$0$, and a part of the final
segment of $\mathrm{ch}(g_y)$ that reaches $q_1$ from side~$1$; both segments are perpendicular to
$\gamma$. The other parts of $J$ are compact and
disjoint from $\gamma$, so such a $U$ exists. Its boundary $\partial U$ is a simple closed curve made
of the $0$-bank, parallel to $\gamma$ on side~$0$, the $1$-bank, and two short end arcs around the ends
of $\gamma$. The curve $J$ crosses $\partial U$ exactly twice, transversally: on the $0$-bank by
$\mathrm{ch}(g_x)$, and on the $1$-bank by $\mathrm{ch}(g_y)$. So $\partial U\setminus J$ consists of two
arcs, $E_s$ through the end arc at $s$ and $E_t$ through the end arc at $t$, and by the one-crossing
fact they lie on different sides of $J$. Points of $\gamma$ before $\min(q_0,q_1)$ are joined to $E_s$
by short arcs inside $U$ that miss $J$, and points after $\max(q_0,q_1)$ are joined to $E_t$ in the same
way. This gives~(a) and shows that $E_s$ lies in $\mathrm{Lo}(J)$. For~(b), fix $r$ and shrink $U$, if
needed, so that its width is less than half the distance from $r$ to $q_0$ and to $q_1$; the sides of
$J$ do not depend on $U$. Inside $U$, the final segments of $\mathrm{ch}(g_x)$ and $\mathrm{ch}(g_y)$ lie
within the width of $U$ from $q_0$ and $q_1$. A point near $r$ on side~$0$ is joined to the $0$-bank by
a segment perpendicular to $\gamma$, which lies within the width of $U$ from $r$. So the segment misses
$\gamma$ and both final segments. The part
of the $0$-bank before its crossing with $\mathrm{ch}(g_x)$ lies in $E_s$, and the part after it in
$E_t$. Side~$1$ is the same, with $q_1$.
\end{proof}

The low side is the $j$-side and the high side the $k$-side. Claims~\ref{cl:Oends} and~\ref{cl:Ocalled}
justify the names: the low side holds the $P$-pieces with the small indices, and when it is called it
gets the range $[j_A,i_1]$.

\begin{claim}[every subgraph on one side]\label{cl:Oone}
Draw $A^Q$ with collars inside the slots of level $\ell$ and outside $C_Q$, and with curves of pairs as
above for the incised set $I_{\ell+1}$. This drawing is disjoint from $J$. Hence the drawing of each
component of $A^Q$ lies in $\mathrm{Lo}(J)$ or in $\mathrm{Hi}(J)$, and we say that the component, its
pieces and its pairs lie on that side.
\end{claim}

\begin{proof}
Let $g'$ be a piece of $A^Q$ that refines the piece $g$ of $A$. If $g$ is not a node of $Q$, no part of
$C_Q$ lies in $\Delta(g)$ (Claim~\ref{cl:Ocurve}(c)), and $\Delta(g)$ misses $\gamma$. If $g$ is a node
of $Q$, then $J$ meets $\Delta(g)$ only in the interior of $\mathrm{ch}(g)$, and $\mathrm{ch}(g)$ meets
$\partial h^\circ$ only in its ends there. These ends are ends of the arcs of the parts of $g$: a new cut
point is where $g$ splits, and an old cut point is an end of $\mathrm{arc}(g)$. So a collar along the
interior of $\mathrm{arc}(g')$ can be chosen in $\Delta(g)\setminus\mathrm{ch}(g)$. The curve of a pair
lies in its strip away from the cut line, and it enters collars only near the middle of token
segments, away from the chords; by Claim~\ref{cl:Ocurve}(c) it misses $J$. The drawing of a component
is connected, so the Jordan fact applies.
\end{proof}

\begin{claim}[$P$-pieces and the ends of $Q$]\label{cl:Oends}
\begin{enumerate}[(a)]
\item For $t\ne x$, $g_t^0$ lies in $\mathrm{Lo}(J)$ if $t<x$ and in $\mathrm{Hi}(J)$ if $t>x$. For
  $t\ne y$, $g_t^1$ lies in $\mathrm{Lo}(J)$ if $t<y$ and in $\mathrm{Hi}(J)$ if $t>y$.
\item If $\varepsilon_0$ is new, $g_x$ splits at $c(\varepsilon_0)$ into two pieces; the one whose arc
  runs from the low end of $\mathrm{arc}(g_x)$ to $c(\varepsilon_0)$ lies in $\mathrm{Lo}(J)$, and the
  other in $\mathrm{Hi}(J)$. If $\varepsilon_0$ is old, $g_x$ does not change; it lies in
  $\mathrm{Hi}(J)$ if $c(\varepsilon_0)$ is the low end of $\mathrm{arc}(g_x)$, and in $\mathrm{Lo}(J)$ if
  it is the high end. The same holds for $g_y$, $\varepsilon_1$ and $q_1$.
\item If $\varepsilon_0$ is old, its pair is the first edge of $P_a$ or of $P_b$, and $g_x$ is the first
  node of that path; $g_x$ lies in $\mathrm{Hi}(J)$ if the path is $P_a$, and in $\mathrm{Lo}(J)$ if it
  is $P_b$. If $\varepsilon_1$ is old, its pair is the last edge of $P_a$ or of $P_b$; $g_y$ lies in
  $\mathrm{Hi}(J)$ if the path is $P_a$, and in $\mathrm{Lo}(J)$ if it is $P_b$.
\end{enumerate}
\end{claim}

\begin{proof}
(a) For $t\ne x$, $g_t^0$ is not a node of $Q$, since the only $0$-piece of $Q$ is $g_x$. So it does not
change, and $\Delta(g_t^0)$ is disjoint from $J$: it misses $\gamma$, and the chords of $C_Q$ lie in
the closures of the slots of the nodes of $Q$, which are other slots (Claim~\ref{cl:Ocurve}(c)). By
(Geom b), its boundary contains a $\gamma$-part in $\gamma_t$ with interior points, so
$\Delta(g_t^0)$ contains points on side~$0$ near interior points $r$ of $\gamma_t$. All of $\gamma_t$
precedes $q_0\in\gamma_x$ if $t<x$, and follows it if $t>x$. Claim~\ref{cl:Osides}(b) gives the side.
The pieces of the $p_t^1$ are the same, with $y$ and $q_1$.

(b) By (Geom b), $\partial\Delta(g_x)$ consists of $\mathrm{arc}(g_x)$ from its low end to its high end,
then the chord to $r_{\mathrm{hi}}$ or nothing, then $\gamma_x$ back from $r_{\mathrm{hi}}$ to
$r_{\mathrm{lo}}$, then the chord to the low end or nothing. The chord $\mathrm{ch}(g_x)$ joins
$c(\varepsilon_0)$ on $\mathrm{arc}(g_x)$ to $q_0$ inside the $\gamma$-part. By the $\theta$-curve fact,
$\Delta(g_x)\setminus\mathrm{ch}(g_x)$ has two components: $\Delta_{\mathrm{lo}}$, bounded by
$\mathrm{ch}(g_x)$, the part of $\mathrm{arc}(g_x)$ from its low end to $c(\varepsilon_0)$, the old chord
from the low end to $r_{\mathrm{lo}}$ if there is one, and the part of the $\gamma$-part from
$r_{\mathrm{lo}}$ to $q_0$; and $\Delta_{\mathrm{hi}}$, bounded by $\mathrm{ch}(g_x)$, the part of
$\mathrm{arc}(g_x)$ from $c(\varepsilon_0)$ to its high end, the old chord from the high end to
$r_{\mathrm{hi}}$ if there is one, and the part of the $\gamma$-part from $q_0$ to $r_{\mathrm{hi}}$. Both are connected and miss $J$
(Claim~\ref{cl:Ocurve}(c)). The domain $\Delta_{\mathrm{lo}}$ contains points on side~$0$ near points of
$\gamma_x$ before $q_0$, so it lies in $\mathrm{Lo}(J)$, and $\Delta_{\mathrm{hi}}$ lies in
$\mathrm{Hi}(J)$ for the same reason. If $\varepsilon_0$ is new, the collar of each part of $g_x$ lies in
the component along its arc. If $\varepsilon_0$ is old, $c(\varepsilon_0)$ is an end of
$\mathrm{arc}(g_x)$, and all of $\mathrm{arc}(g_x)$ lies on the boundary of $\Delta_{\mathrm{hi}}$ if it
is the low end, and of $\Delta_{\mathrm{lo}}$ if it is the high end.

(c) Since $\varepsilon_0$ is old, by (Old) its pair is a copy of an edge of $P$ or a pair of $P_a$ or of
$P_b$, and it is not a copy of an edge of $P$, since it is an edge of $Q$. Say it is a pair of $P_a$.
Then $g_x$ is a node of $P_a$, and the only $0$-piece on $P_a$ is its first node (Bound). So the pair is
the first edge of $P_a$, and $\varepsilon_0$ is the first token of $P_a$, whose cut point is the low end
of $\mathrm{arc}(g_x)$ by (Bound). Now apply~(b). On $P_b$ the cut point is the high end. At $g_y$, the
only $1$-piece on a bounding path is its last node, and (Bound) gives the low end on $P_a$ and the high
end on $P_b$.
\end{proof}

\begin{claim}[local parity]\label{cl:Oparity}
\begin{enumerate}[(a)]
\item If $\mathrm{cl}(e)\subseteq J$, the two half-strips of $S_e$ lie on different sides of $J$. In
  particular, the two pairs of a new edge of $Q$ lie on different sides.
\item If a node $g$ of $Q$ is cut at a new cut point $c$, the two parts of $g$ that meet at $c$ lie on
  different sides of $J$.
\end{enumerate}
\end{claim}

\begin{proof}
(a) A short segment in $S_e$ that crosses $\mathrm{cl}(e)$ once meets $J$ nowhere else
(Claim~\ref{cl:Ocurve}(c)); apply the one-crossing fact. (b) Near $c$, $\mathrm{ch}(g)$ leaves
$\partial h^\circ$ into $\Delta(g)$ and separates the ends of the two collars near $c$. A short arc near
$c$ joins them, crosses $\mathrm{ch}(g)$ once and meets $J$ nowhere else; apply the one-crossing fact.
\end{proof}

\begin{claim}[the copies of a new edge]\label{cl:Oorient}
Let $e$ be a new edge of $Q$, and let $\delta$ be its dart in the node of $Q$ nearer to $g_x$. Then the
pair $\{(\delta,1),(\rev\delta,2)\}$ lies in $\mathrm{Lo}(J)$ and the pair $\{(\delta,2),(\rev\delta,1)\}$
in $\mathrm{Hi}(J)$.
\end{claim}

\begin{proof}
Traverse $C_Q$ from $q_0$ to $q_1$. It leaves $q_0$ into side~$0$, the right side of $\gamma$,
perpendicular to $\gamma$, so the points on side~$0$ just before $q_0$ lie on its right. They lie in
$\mathrm{Lo}(J)$ by Claim~\ref{cl:Osides}(b). Since $J$ is a simple closed curve, $\mathrm{Lo}(J)$ lies on
the right of $C_Q$ all along. The curve crosses $S_e$ along $\mathrm{cl}(e)$ from $c(\delta)$ to
$c(\rev\delta)$. In walk order the face of $\delta$ lies on the left of its side segment, so the strip lies
on the right, and $\mathrm{cl}(e)$ turns right into the strip. So the half-strip at the tail of $\delta$
lies on the right of $\mathrm{cl}(e)$. It holds the pair $\{(\delta,1),(\rev\delta,2)\}$, and it misses
$J$ apart from $\mathrm{cl}(e)$ (Claim~\ref{cl:Ocurve}(c)), so this pair lies in $\mathrm{Lo}(J)$. The other
pair lies in $\mathrm{Hi}(J)$ by Claim~\ref{cl:Oparity}(a).
\end{proof}

\begin{claim}[the bounding paths]\label{cl:Obound}
Suppose that $P_a$ exists. Then:
\begin{enumerate}[(a)]
\item $q_0^a$ precedes $q_0$, $q_1^a$ precedes $q_1$, and $\mathrm{Lo}(J_a)\subseteq\mathrm{Lo}(J)$;
\item the pair of $A$ of an edge of $P_a$ that is not on $Q$ lies in $\mathrm{Lo}(J)$, and so does every
  node of $P_a$ that is not on $Q$;
\item the pair of $A$ of an edge of $P_a$ that is on $Q$ lies in $\mathrm{Hi}(J)$.
\end{enumerate}
Symmetrically, if $P_b$ exists, then $q_0$ precedes $q_0^b$, $q_1$ precedes $q_1^b$,
$\mathrm{Hi}(J_b)\subseteq\mathrm{Hi}(J)$, the pairs and nodes of $P_b$ off $Q$ lie in $\mathrm{Hi}(J)$, and
the pair of $A$ of an edge of $P_b$ on $Q$ lies in $\mathrm{Lo}(J)$.
\end{claim}

\begin{proof}
(a) The slot $\Delta(g_x)$ lies in $\Omega_A\subseteq\mathrm{Hi}(J_a)$ (Geom c, d) and contains points
on side~$0$ near $q_0$. By Claim~\ref{cl:Osides}(b) for $J_a$, which applies since $C_a$ reaches its ends
as $C_Q$ does, these points lie in $\mathrm{Lo}(J_a)$ if $q_0$ precedes $q_0^a$. Since $q_0\ne q_0^a$,
$q_0^a$ precedes $q_0$; side~$1$ is the same. Now $J\cap\mathrm{Lo}(J_a)=\emptyset$: $C_Q$ lies in the
closure of $\Omega_A$, which lies in $\mathrm{Hi}(J_a)\cup J_a$; and $\gamma[q_0,q_1]$ lies after
$\min(q_0^a,q_1^a)$, where $\gamma$ lies either on $J_a$, between $q_0^a$ and $q_1^a$, or in
$\mathrm{Hi}(J_a)$, after both (Claim~\ref{cl:Osides}(a) for $J_a$). The set $\mathrm{Lo}(J_a)$ is
connected, so it lies on one side of $J$. It contains the points of $\gamma$ before
$\min(q_0^a,q_1^a)$, which precede $\min(q_0,q_1)$ and so lie in $\mathrm{Lo}(J)$.

(b), (c) Let $e$ be an edge of $P_a$. By (Geom d), $\mathrm{cl}(e)\subseteq J_a$, so by
Claim~\ref{cl:Oparity}(a) for $J_a$ the half-strip of $S_e$ opposite to the pair of $A$ lies in
$\mathrm{Lo}(J_a)$, hence in $\mathrm{Lo}(J)$. The edge $e$ is not an edge of $P$, since a pair of $P_a$
has an end off $P$ or joins a $0$-piece to a $1$-piece. If $e$ is not on $Q$, then $S_e$ misses $J$
(Claim~\ref{cl:Ocurve}(c)), so the pair of $A$ lies in $\mathrm{Lo}(J)$ too. If $e$ is on $Q$, then
$\mathrm{cl}(e)\subseteq J$, and by Claim~\ref{cl:Oparity}(a) the pair of $A$ lies on the other side,
$\mathrm{Hi}(J)$. A node of $P_a$ that is not on $Q$ does not change, and its collar meets the curve of
an edge of $P_a$ at it, which is not on $Q$; so it lies in $\mathrm{Lo}(J)$ as well. The case of $P_b$
is the mirror image, with $\mathrm{Hi}(J_b)\subseteq\mathrm{Hi}(J)$.
\end{proof}

\begin{claim}[nodes and edges of $Q$]\label{cl:Onodes}
Suppose that $Q$ has no node of $P_b$. Then every node of $Q$ has exactly one part in $\mathrm{Hi}(J)$,
and every edge of $Q$ has exactly one pair in $A^Q$ that lies in $\mathrm{Hi}(J)$. These \emph{high}
parts and pairs form a path $Q^{\mathrm{Hi}}$ from a piece of $p_x^0$ to a piece of $p_y^1$.
Symmetrically, if $Q$ has no node of $P_a$, the same holds with $\mathrm{Lo}(J)$, which gives a path
$Q^{\mathrm{Lo}}$.
\end{claim}

\begin{proof}
\emph{Edges.} A new edge has two pairs, on different sides (Claim~\ref{cl:Oparity}(a)). An old edge of
$Q$ is, by (Old), a pair of $P_a$ or of $P_b$, and not of $P_b$, since its ends would be nodes of $P_b$.
The annulus $A$ has one pair of it (Old), and that pair lies in $\mathrm{Hi}(J)$
(Claim~\ref{cl:Obound}(c)).

\emph{Internal nodes.} Let $g$ be an internal node of $Q$, in the face $h$. If $g$ is cyclic, its face
has no cut, so $u$ and $w$ are new. Then $g$ splits at $c(u)$ and $c(w)$ into two parts that meet at
both points, and by Claim~\ref{cl:Oparity}(b) they lie on different sides. Let $g$ be linear. Its end
tokens are old, since the cuts at its ends are dart cuts: corner cuts lie in faces of $P$. Since $g$ is
off $P$, by (Old) their pairs are pairs of $P_a$ or of $P_b$, and not of $P_b$, since $g$ is not a node
of $P_b$. So $g$ is a node of $P_a$, and an internal one (Bound). The two tokens of $P_a$ at $g$ are
distinct, old, and hence ends of $g$, so they are its first token $x_1$ and its last token $x_r$, with
$r\ge2$. Let $u$ come before $w$ in $g$. If $u=x_1$ and $w=x_r$, both edges of $Q$ at $g$ are edges of
$P_a$, no new cut lies in $g$, and $g$ lies in $\mathrm{Hi}(J)$, because the pair of $u$ does
(Claim~\ref{cl:Obound}(c)) and its curve meets $\mathrm{col}(g)$. Otherwise $g$ splits into
$S_{\mathrm{in}}$, from $u$ to $w$, and one or two pieces of $S_{\mathrm{out}}$: the run from $x_1$ to the
cut in $u$ if $u\ne x_1$, and the run from the cut in $w$ to $x_r$ if $w\ne x_r$. A token of
$S_{\mathrm{out}}$ among $x_1$ and $x_r$ belongs to an edge of $P_a$ that is not on $P_i$, since $P_i$
meets $g$ only in $u$ and $w$. By Claim~\ref{cl:Obound}(b), each piece of $S_{\mathrm{out}}$ lies in
$\mathrm{Lo}(J)$. If $u\ne x_1$, then $u$ is not an end of $g$, so it is new, and $S_{\mathrm{in}}$
meets a piece of $S_{\mathrm{out}}$ at the new cut point $c(u)$; by Claim~\ref{cl:Oparity}(b),
$S_{\mathrm{in}}$ lies in $\mathrm{Hi}(J)$. If $u=x_1$, then $w\ne x_r$, and the same holds at $c(w)$.

\emph{Ends.} If $\varepsilon_0$ is new, $g_x$ has one part on each side (Claim~\ref{cl:Oends}(b)). If
$\varepsilon_0$ is old, its pair is on $P_a$, not on $P_b$, and $g_x$ lies in $\mathrm{Hi}(J)$ as a
whole (Claim~\ref{cl:Oends}(c)). The same holds at $g_y$.

\emph{Path.} The curve of the high pair of an edge of $Q$ lies in $\mathrm{Hi}(J)$ and meets the collars
of the parts that hold its tokens, so these parts are the high parts of the two ends. The high parts
are distinct pieces, so $Q^{\mathrm{Hi}}$ is a path. Its first node is a part of $g_x$, a piece of
$p_x^0$, and its last node a part of $g_y$, a piece of $p_y^1$.
\end{proof}

\begin{claim}[separation on level $\ell$]\label{cl:Osep}
Let $t$ lie strictly between $x$ and $y$, and suppose that $A$ has $g_t^0$ and $g_t^1$. Then every walk
in $A$ from $g_t^0$ to $g_t^1$ meets a node of $Q$.
\end{claim}

\begin{proof}
Let $W$ be such a walk with no node on $Q$. Then its nodes and pairs do not change, $W$ is a walk in
$A^Q$, and its drawing is connected and misses $J$ (Claim~\ref{cl:Oone}). If $x<t<y$, then $g_t^0$ lies
in $\mathrm{Hi}(J)$ and $g_t^1$ in $\mathrm{Lo}(J)$ (Claim~\ref{cl:Oends}(a)); if $y<t<x$, it is the other
way round. Either way $W$ joins the two sides, a contradiction.
\end{proof}

\begin{claim}[the called side]\label{cl:Ocalled}
Suppose that $P_i$ has no node of $P_b$. Then:
\begin{enumerate}[(a)]
\item every component of $A^Q$ contains a part of a node of $Q$;
\item the high side is one subgraph $A'$, which contains $Q^{\mathrm{Hi}}$, so it holds exactly one
  pair of each edge of $Q$;
\item $A'$ has no bad edge;
\item a bad edge in a subgraph of the $j$-side is an edge of $P_a$.
\end{enumerate}
Symmetrically, if $P_i$ has no node of $P_a$, the low side is one subgraph, which contains
$Q^{\mathrm{Lo}}$, has no bad edge, and holds one pair of each edge of $Q$; and a bad edge on the
$k$-side is an edge of $P_b$.
\end{claim}

\begin{proof}
(a) Let $C$ be a component of $A^Q$ and $v$ a node of $C$ that is not a part of a node of $Q$. Then $v$
is a node of $A$ off $Q$. The annulus $A$ is connected (Comp); let $R$ be a shortest path in $A$ from
$v$ to a node of $Q$. Its nodes before the last are off $Q$, so $R$ without its last node is a walk in
$A^Q$, and its last pair, which has an end off $Q$ and is not an edge of $Q$, is a pair of $A^Q$ that
joins it to a part of the last node.

(b) By (a) and Claim~\ref{cl:Onodes}, every component on the high side contains the high part of some
node of $Q$. All high parts lie on the connected path $Q^{\mathrm{Hi}}$. So there is one component on
the high side, and it contains $Q^{\mathrm{Hi}}$. Its pairs of edges of $Q$ are those of
$Q^{\mathrm{Hi}}$, one for each edge, since the other pair of each new edge is low
(Claim~\ref{cl:Onodes}).

(c) Since $P_i$ has no node of $P_b$, no edge of $P_b$ is bad. A bad edge of $P_a$ is an edge of $P_a$
that is not on $P_i$, so not on $Q$, and by Claim~\ref{cl:Obound}(b) it lies in $\mathrm{Lo}(J)$.

(d) An edge of $P_b$ that is not on $P_i$ is not on $Q$, so it lies in $\mathrm{Hi}(J)$ by
Claim~\ref{cl:Obound}.
\end{proof}

Parts (a) and (d) hold without the hypothesis on $P_b$. When $P_i$ meets $P_a$, the low side can
consist of several subgraphs; it gets no call.

\begin{claim}[pieces of the children]\label{cl:Opieces}
Every called child $A'$ of $A$ has at most one piece of each $p_t^\sigma$ and of each face off $P$. If
it is the high side, it has a piece of $p_t^0$ exactly for $t\in[x,x_b]$ and a piece of $p_t^1$ exactly
for $t\in[y,y_b]$; if it is the low side, exactly for $t\in[x_a,x]$ and $t\in[y_a,y]$.
\end{claim}

\begin{proof}
Take the high side; the low side is symmetric. Since it is called, $P_i$ has no node of $P_b$. By
part~(b) of the \LemAnn{}, each piece of $A'$ refines a piece of $A$, and $A$ has at most one piece of each
kind (Pieces). A piece of $A$ off $Q$ does not change. A node of $Q$ has exactly one part in
$\mathrm{Hi}(J)$ (Claim~\ref{cl:Onodes}), and $A'$ lies in $\mathrm{Hi}(J)$. So $A'$ has at most one piece
of each kind. The $0$-pieces of $A$ are the $g_t^0$ with $t\in[x_a,x_b]$ (Pieces), and $x$ lies in this
interval, since $g_x$ is one of them. Those with $t>x$ lie in $\mathrm{Hi}(J)$
(Claim~\ref{cl:Oends}(a)) and hence in $A'$ (Claim~\ref{cl:Ocalled}(b)); $g_x$ has a high part
(Claim~\ref{cl:Onodes}), which lies on $Q^{\mathrm{Hi}}\subseteq A'$; and those with $t<x$ lie in
$\mathrm{Lo}(J)$. The $1$-pieces are the same, with $y$.
\end{proof}

\begin{claim}[distances in the called side]\label{cl:Oinv}
Suppose that $P_i$ has no node of $P_b$, and let $A'$ be the high side. If $A'$ has a piece of $p_t^0$
and a piece of $p_t^1$, their distance in $A'$ is $\lambda_t$. The low side is symmetric.
\end{claim}

\begin{proof}
($\ge$) By part~(b) of the \LemAnn{}, pieces of $A'$ refine pieces of $A_1$, and pairs are copies of edges of
the same length, so a walk in $A'$ between the two pieces maps to a walk in $A_1$ from $p_t^0$ to
$p_t^1$ of the same length, which is at least $\lambda_t$.

($\le$) The pieces of $p_t^0$ and $p_t^1$ in $A'$ refine $g_t^0$ and $g_t^1$, so by (Dist) $A$ has a walk
$W$ from $g_t^0$ to $g_t^1$ of length $\lambda_t$. For a node $v$ of $A$, let $\varphi(v)=v$ if $v$ is
off $Q$, and let $\varphi(v)$ be the high part of $v$ if $v$ is on $Q$ (Claim~\ref{cl:Onodes}); for an
edge of $Q$, let $\varphi$ of it be its high pair.
\begin{itemize}
\item If $W$ has no node on $Q$, it is a walk in $A^Q$ that misses $J$, and it contains $g_t^0$, which
  does not change and is the piece of $p_t^0$ in $A'$. So $W$ lies in $A'$.
\item Otherwise let $\pi_1$ be the first and $\pi_2$ the last position of $W$ at a node of $Q$, with
  nodes $z_1$ and $z_2$, and split $W=W_1W_2W_3$ at $\pi_1$ and $\pi_2$. Let $R$ be the part of $Q$
  between $z_1$ and $z_2$, a single node if $z_1=z_2$. Since $R$ is a shortest path in $A$,
  $W':=W_1RW_3$ has length at most $\lambda_t$.
\item \emph{Lifting $W_1$.} If $\pi_1>0$, the nodes of $W_1$ before $z_1$ are off $Q$, and its pairs
  have an end off $Q$, so $W_1$ without $z_1$ is a walk in $A^Q$ that misses $J$ and contains
  $g_t^0\in A'$; hence it lies in $A'$. Its last pair joins a node of $A'$ to a part of $z_1$, so that
  part lies in $A'\subseteq\mathrm{Hi}(J)$ and is $\varphi(z_1)$, the only high part of $z_1$. If
  $\pi_1=0$, then $g_t^0=z_1$ is a $0$-piece on $Q$, so $t=x$ and $z_1=g_x$, and $\varphi(g_x)$ is the
  piece of $p_x^0$ in $A'$ (Claim~\ref{cl:Opieces}). The part $W_3$ is the same at the other end; if
  $\pi_2$ is the last position, then $t=y$ and $\varphi(g_y)$ is the piece of $p_y^1$ in $A'$.
\item $\varphi(R)$ is the part of $Q^{\mathrm{Hi}}$ between $\varphi(z_1)$ and $\varphi(z_2)$, a path in
  $A'$ of the same length.
\end{itemize}
So $\varphi(W')$, with the ends of $W_1$ and $W_3$ attached to $\varphi(z_1)$ and $\varphi(z_2)$, is a walk
in $A'$ between its pieces of $p_t^0$ and $p_t^1$ of length at most $\lambda_t$.
\end{proof}

This covers $t=i_2=x$ or $t=i_2=y$, where the new walk starts or ends with $Q^{\mathrm{Hi}}$, and a walk
that meets $Q$ in one node, where $R$ is that node and both $W_1$ and $W_3$ attach to its unique high
part. The proof uses that the called side holds exactly one part of every node of $Q$, which fails on
a side with overlap.

\begin{claim}[the invariant passes on]\label{cl:Opass}
Every called child of $A$ satisfies (I).
\end{claim}

\begin{proof}
Take the high child $A'$, so that $P_i$ has no node of $P_b$; the low child is symmetric. Its bounding
paths are $P_a':=Q^{\mathrm{Hi}}$ and $P_b':=P_b$, with $x_a'=x$ and $y_a'=y$, and its range is
$[i_2,k_A]$.

(Comp) $A^Q$ is a union of components of $G^*_{\ell+1}$: a pair of level $\ell+1$ with a token in a
piece of $A^Q$ is a copy of an edge with a pair in $A$, and $A$ is closed under pairs (Comp for $A$).
By Claim~\ref{cl:Ocalled}(b), $A'$ is one component of $A^Q$, so it is a component of $G^*_{\ell+1}$.

(Geom) The set $K_{\ell+1}$ adds to $K_\ell$ the curves $C_Q$ of all incisions of level $\ell$. The curve
of another annulus $A''$ lies in the closure of $\Omega_{A''}$ (Claim~\ref{cl:Ocurve}(b)), which misses
$\Omega_A$ (Geom c). So the only new chords in the slots of the pieces of $A$ are those of $C_Q$, and
each lies in the slot of its own node. Let $g'$ be a piece of $A'$ that refines $g$.
\begin{itemize}
\item If $g$ is not on $Q$, then $\Delta(g')=\Delta(g)$, and (Geom a, b) carry over.
\item If $g$ is an internal node of $Q$, then $\Delta(g')$ is the component of
  $\Delta(g)\setminus\mathrm{ch}(g)$ along $\mathrm{arc}(g')$. If $g$ is cyclic, then by the
  $\theta$-curve fact it is bounded by $\mathrm{arc}(g')$ and $\mathrm{ch}(g)$. If $g$ is linear,
  $\Delta(g)$ is bounded by $\mathrm{arc}(g)$ and one chord, and by the $\theta$-curve fact the
  component along $S_{\mathrm{in}}$ is bounded by $\mathrm{arc}(S_{\mathrm{in}})$, from $c(u)$ to $c(w)$,
  and $\mathrm{ch}(g)$. This includes the case $u=x_1$, $w=x_r$.
\item If $g=g_x$, then $\Delta(g')$ is $\Delta_{\mathrm{hi}}$ of Claim~\ref{cl:Oends}. Its boundary is
  the part of $\mathrm{arc}(g_x)$ from $c(\varepsilon_0)$ to the high end, which is all of
  $\mathrm{arc}(g_x)$ if $\varepsilon_0$ is old, the old chord at the high end or none, the $\gamma$-part
  $\gamma_x[q_0,r_{\mathrm{hi}}]$ with $q_0\ne r_{\mathrm{hi}}$, and the new chord $\mathrm{ch}(g_x)$ from
  $c(\varepsilon_0)$ to $q_0$. So (Geom b) holds with $\mathrm{ch}(g_x)$ as the low chord. The piece
  $g_y$ is the same.
\end{itemize}
For (Geom c), the slots and half-strips of $A'$ lie in $\Omega_A$ and miss $K_{\ell+1}$; $A'$ is
connected, and a slot and the half-strip of a pair at it are joined across the open token segment,
which misses $K_{\ell+1}$. So they lie in one component $\Omega_{A'}$ of $S^2\setminus K_{\ell+1}$, and
$\Omega_{A'}\subseteq\Omega_A\cap\mathrm{Hi}(J)$, since it is connected, misses $K_\ell\cup J$ and meets
$\Omega_A\cap\mathrm{Hi}(J)$. Two children of one annulus lie on different sides of $J$, and children of
distinct annuli lie in the distinct sets $\Omega_A$. (Geom d): the $j$-curve of $A'$ is $C_Q$, which
contains the cut lines of all edges of $Q$ and runs from $q_0\in\gamma_x$, reached from side~$0$, to
$q_1\in\gamma_y$, reached from side~$1$, and $\Omega_{A'}\subseteq\mathrm{Hi}(J)$; the $k$-curve is $C_b$,
and $\Omega_{A'}\subseteq\Omega_A\subseteq\mathrm{Lo}(J_b)$.

(Bound) $Q^{\mathrm{Hi}}$ is a path in $A'$ from the piece of $p_x^0$ in $A'$ to that of $p_y^1$, with
internal nodes off $P$ (Claims~\ref{cl:Onodes} and~\ref{cl:Ocalled}). If $\varepsilon_0$ is new, the high
part of $g_x$ has its arc from $c(\varepsilon_0)$ to the high end, and the token of the first pair of
$Q^{\mathrm{Hi}}$ there is the half of the segment of $\varepsilon_0$ next to $c(\varepsilon_0)$, so its
cut point is the low end. If $\varepsilon_0$ is old, it is the first token of $P_a$, at the low end of
$g_x$ (Claim~\ref{cl:Oends}(c)), and $g_x$ does not change. The same holds at $g_y$. The path $P_b$ has
no node on $Q$, so its nodes and pairs do not change; they lie in $\mathrm{Hi}(J)$
(Claim~\ref{cl:Obound}) and hence in $A'$, and (Bound) for $P_b$ carries over. The paths $Q^{\mathrm{Hi}}$ and $P_b$
have no common node, because $Q$ has none on $P_b$.

(Pieces) Claim~\ref{cl:Opieces} gives the pieces, with $[x,x_b]$ and $[y,y_b]$. For the range,
$\max(x,y)\le i_2=j_{A'}$ and $k_{A'}=k_A\le\min(x_b,y_b)$.

(Old) The old pairs of $A'$ are copies of edges of $P$; pairs of $A$ of old edges, which are pairs of
$P_a$ or $P_b$, where those of $P_a$ off $Q$ lie in $\mathrm{Lo}(J)$ (Claim~\ref{cl:Obound}(b)) and are
not in $A'$, those of $P_a$ on $Q$ lie on $Q^{\mathrm{Hi}}$, and those of $P_b$ are on $P_b$; and the
high pairs of the new edges of $Q$, which lie on $Q^{\mathrm{Hi}}$. The child $A'$ has one pair of each
edge of $Q$ (Claim~\ref{cl:Ocalled}(b)) and at most one pair of every other old edge off $P$, since $A$
has.

(Dist) is Claim~\ref{cl:Oinv}.
\end{proof}

\begin{claim}[C23]\label{cl:Oc23}
Suppose that $P_a$ exists and that $P_i$ meets $P_a$ in a node $z$. Then for every
$\mu\in[\max(x_a,y_a),i]$,
\[
  \dist_A(g_\mu^0,z)+\dist_A(z,g_\mu^1)=\lambda_\mu .
\]
By (Pieces), $[\max(x_a,y_a),i]\supseteq[j_A,i]$, and every common node $z$ works. Symmetrically, if
$P_i$ meets $P_b$ in $z$, the equality holds for every $\mu\in[i,\min(x_b,y_b)]$.
\end{claim}

\begin{proof}
For $\mu\in[\max(x_a,y_a),i]$, $A$ has $g_\mu^0$ and $g_\mu^1$ (Pieces), and their distance in $A$ is
$\lambda_\mu$ (Dist). The left side is at least $\lambda_\mu$, since two shortest paths through $z$ form a
walk from $g_\mu^0$ to $g_\mu^1$. For $\mu=i$, $z$ lies on $P_i$, which has length $\lambda_i$, so equality
holds. Let $\mu<i$, and let $W$ be a shortest walk in $A$ from $g_\mu^0$ to $g_\mu^1$, of length $\lambda_\mu$.

\emph{Step~1: $W$ meets $P_i[g_i^0,z]$.} Suppose not. Build a closed curve $Z^0$ from four parts.
\begin{itemize}
\item $\zeta$: a curve in the drawing of $P_i[g_i^0,z]$ from $\mathrm{col}(g_i^0)$ to $\mathrm{col}(z)$.
\item The connector: from $\mathrm{col}(z)$ along the curve of a pair $\pi$ of $P_a$ at $z$, of the edge
  $e$, then inside the half-strip of $\pi$ to a point of $\mathrm{cl}(e)\subseteq C_a$.
\item $C_a$, followed from that point backwards to $q_0^a$.
\item The $0$-run: from $q_0^a$ forward along $\gamma$ on side~$0$, at small distance, to a point $\rho$
  inside the $\gamma$-part of $g_i^0$, then into $\Delta(g_i^0)$ and to the start of $\zeta$ in
  $\mathrm{col}(g_i^0)$. The $\gamma$-part of $g_i^0$ follows $q_0^a$, because
  $\Delta(g_i^0)\subseteq\mathrm{Hi}(J_a)$ (Claim~\ref{cl:Osides}(b) for $J_a$).
\end{itemize}
Build a second closed curve: a curve $\omega$ in $\Gamma(W)$ from $\mathrm{col}(g_\mu^0)$ to
$\mathrm{col}(g_\mu^1)$, closed by an arc $\beta$ in the face $p_\mu$. The arc $\beta$ runs inside the half
$H_\mu^0$ from $\mathrm{col}(g_\mu^0)$ to a point near $\gamma_\mu$ on side~$0$, crosses $\gamma_\mu$ once at a
point $\xi$, and runs inside $H_\mu^1$ to $\mathrm{col}(g_\mu^1)$. Choose $\xi$ inside $\gamma_\mu$ after
$q_0^a$ if $\mu=x_a$, and after $q_1^a$ if $\mu=y_a$. This is possible: if $\mu=x_a$, the chord of $C_a$ in
$H_\mu^0$ splits it, by the $\theta$-curve fact, into a part before and a part after $q_0^a$, and
$\Delta(g_\mu^0)$ lies in the part after, since $g_\mu^0$ lies in $\mathrm{Hi}(J_a)$; so $\beta$ avoids that
chord. The same holds in $H_\mu^1$ if $\mu=y_a$. The parts of $\beta$ outside a thin neighbourhood of
$\gamma$ avoid it.

We count the crossings of $\omega\cup\beta$ with $Z^0$.
\begin{itemize}
\item $\omega$ and $\zeta$: $W$ and $P_i[g_i^0,z]$ have no common node, so their drawings are disjoint
  (Claim~\ref{cl:Odisj}).
\item $\omega$ and the connector: $W$ does not visit $z$, so it does not use $\pi$, and $A$ has no
  other pair of $e$ (Old). So $\omega$ misses $\mathrm{col}(z)$ and $S_e$.
\item $\omega$ and $C_a$: $\omega$ lies in $\Omega_A$, which misses $K_\ell\supseteq C_a$.
\item $\omega$ and the $0$-run: collars avoid small neighbourhoods of cut points, and curves of pairs
  of edges of $P$ cross their strips away from $\gamma$, so $\omega$ misses a thin neighbourhood of
  $\gamma$. In $\Delta(g_i^0)$, the run meets only the collar of $g_i^0$, which $W$ does not visit.
\item $\beta$ and $\zeta$, and $\beta$ and the connector: inside $p_\mu$, $A$ has only the pieces $g_\mu^0$
  and $g_\mu^1$ (Pieces), and $P_i[g_i^0,z]$ visits neither, since $W$ would meet it; for the same reason
  $z$ is neither.
\item $\beta$ and $C_a$: the parts of $C_a$ in $p_\mu$ are its chords at the ends of $P_a$, if $\mu=x_a$ or
  $\mu=y_a$, and $\beta$ avoids them.
\item $\beta$ and the $0$-run: $\beta$ crosses $\gamma$ once, at $\xi$, and the run passes $\xi$ on
  side~$0$, because $q_0^a$ precedes $\xi$ ($\mu>x_a$, or the choice of $\xi$ if $\mu=x_a$) and $\xi$
  precedes $\rho$ ($\mu<i$). So $\beta$ crosses the run exactly once.
\end{itemize}
So the closed curves $\omega\cup\beta$ and $Z^0$ cross exactly once, against the parity fact.

\emph{Step~2: $W$ meets $P_i[z,g_i^1]$.} Build $Z^1$ in the same way from the drawing of
$P_i[z,g_i^1]$, the connector at $z$, $C_a$ from $\mathrm{cl}(e)$ to $q_1^a$, and the $1$-run on side~$1$
from $q_1^a$ forward to the $\gamma$-part of $g_i^1$. The same count, with $\xi$ after $q_1^a$ if
$\mu=y_a$, gives exactly one crossing.

\emph{Step~3: exchange.} Let $v_1$ be a node of $W$ on $P_i[g_i^0,z]$ and $v_2$ one on
$P_i[z,g_i^1]$. Replace the part of $W$ between its visits to $v_1$ and $v_2$, in whichever order $W$
makes them, by the part of $P_i$ between $v_1$ and $v_2$, read in the same direction. That part passes
through $z$, and as a subpath of the shortest path $P_i$ it is no longer than the part it replaces. So
there is a walk from $g_\mu^0$ to $g_\mu^1$ through $z$ of length at most $\lambda_\mu$, and the left side is
at most $\lambda_\mu$.

For $P_b$, mirror the construction: the runs go from the face $p_i$ forward to $q_0^b$ and $q_1^b$, and
need $\mu\le x_b$ and $\mu\le y_b$.
\end{proof}

\subsection{Proof of the Incision Lemma}

By Claim~\ref{cl:OA1}, $A_1$ satisfies (I), and by Claim~\ref{cl:Opass}, every called child of an
annulus that satisfies (I) does too; so by induction every live annulus satisfies (I). Let $A$ be one
that is not a leaf.
\begin{itemize}
\item[(a)] Every subgraph obtained by incising $Q$ lies on one side of $J$ (Claim~\ref{cl:Oone}). The
  sides of the pieces of the $p_t^\sigma$ are given by Claim~\ref{cl:Oends}(a), and those of $g_x$ and
  $g_y$ by Claim~\ref{cl:Oends}(b) and~(c). If the end edge at $g_x$ is old and lies on $P_a$, say, then
  $P_i$ meets $P_a$ at $g_x$, so the $j$-side gets no call, and the high side, if called, holds all of
  $g_x$; the other cases are the same. The separation is Claim~\ref{cl:Osep}.
\item[(b)] is Claim~\ref{cl:Ocalled}.
\item[(c)] is (Pieces) for every live annulus.
\item[(INV)] is (Dist) for every live annulus, and (Pieces) gives the pieces for every
  $t\in[j_{A'},k_{A'}]$.
\item[(C23)] is Claim~\ref{cl:Oc23}.
\end{itemize}
The \LemWit{} is not used.

\paragraph{Degenerate cases.} If $x=y$, then $g_x$ and $g_y$ are the two pieces of $A$ in $p_x$, in
different halves, and $q_0$ and $q_1$ lie in $\gamma_x$ in either order; Claim~\ref{cl:Osides} does not
depend on the order, and Claim~\ref{cl:Osep} is empty. If $Q$ has one edge, $C_Q$ is
$\mathrm{ch}(g_x)\cup\mathrm{cl}(e)\cup\mathrm{ch}(g_y)$; if that edge is old, it is the first edge of
$P_a$ (or $P_b$) at $g_x$ and its last edge at $g_y$, so the bounding path is this single edge and
equals $Q$. If $Q$ runs along $P_a$ or $P_b$, its old edges are edges of these paths (Old), its linear
internal nodes lie on them (Claim~\ref{cl:Onodes}), and $J$ shares cut lines, and possibly ends of
chords, with $J_a$ or $J_b$, which Claims~\ref{cl:Ocurve} and~\ref{cl:Obound} allow. If $Q=P_a$, all of
$Q$ lies on the high side, the low side is empty (Claim~\ref{cl:Ocalled}(a)), the $j$-side overlaps, and
the $k$-side child is $A$ itself with the range $[i_2,k_A]$ and the same bounding paths. The invariant
records all $P$-pieces of $A$, also those with an index outside $[j_A,k_A]$;
Claims~\ref{cl:Oends} and~\ref{cl:Opieces} hold for all indices, and Claim~\ref{cl:Oc23} reaches below
$j_A$. The low end of $\mathrm{arc}(p_0^\sigma)$ is the cut at $c^*_s$, the entry of $\gamma_0$, and the
high end of $\mathrm{arc}(p_m^\sigma)$ is the cut at $c^*_t$; chords of incision curves end at midpoints
of side segments of edges off $P$, never at these cuts. The mode and the moved token change
$\hat G_\ell$, not the token graph, so they do not enter. If $m\le1$, then $A_1$ is a leaf and nothing
is incised. The thickened drawing gives every dart its own side segment, so a bridge with one face on
both sides has two segments in that face, and the corner arc of a vertex of degree~$1$ runs around the
end of its strip. If $P_a$ or $P_b$ is absent, the claims about it are empty, and with $x_a=y_a=0$ or
$x_b=y_b=m$, (Pieces) still holds. On level~$1$ all internal nodes of $Q$ are cyclic, and both ends
split. An end piece of $Q$ that consists of one old token cannot occur in $A$. A cut lies inside an incised
dart or at $c^*_s$ or $c^*_t$, so a piece made of one old token has $c^*_s$ or $c^*_t$ as an end of its
arc and lies in $p_0$ or $p_m$. In $p_0$, the cut at $c^*_s$ is the entry of $\gamma_0$ and so the low
end of the arc; the cut point of the token is the high end, and by Claim~\ref{cl:Oends}(c), (Bound)
and (Old) its pair lies on $P_b$, with this piece as an end. Then $x_b=0$ or $y_b=0$, so $k_A\le0$ by
(Pieces), and $A$ is a leaf. In $p_m$ the mirror argument puts the pair on $P_a$, so $j_A\ge m$.

\paragraph{What is taken from~\cite{ADW26b}.} The structure of the recursion~\cite[Alg.~3]{ADW26b}:
in a live annulus that is not a leaf, $P_i$ is computed for the middle index, its part $P_i'$ is
incised, the overlap rule of lines~9 to~14 runs on a side where $P_i$ meets the bounding path, and the
two sides are called with the ranges $[j_A,i_1]$ and $[i_2,k_A]$. Our readings are listed in
Section~\ref{sec:thmD}. Three of them are used here: a bound is the copy of an incised part, not of a whole path; a side with
overlap gets no call, and overlap is tested directly; and a side with an empty range gets no call.
The two call rules only remove calls. Incision as a split of nodes~\cite[Def.~2.2, Remark~2.5]{ADW26b}, used in the token form of
the \LemAnn{}, with the two sides of~\cite{ADW26b} read as the low and the high side of $J_Q$; this
reading is also made by the \LemAnn{}. The proof does not use~\cite[Claim~2.3]{ADW26b}, nor (N2),
(N4) or (N5), nor the converse of~\cite[Claim~5.8]{ADW26b}: the parts of them that are needed are
proved above. In particular, the internal nodes of $P_i'$ that an earlier incision split and the old
edges of $P_i'$ lie on $P_a$ or $P_b$ (Claims~\ref{cl:Oends} and~\ref{cl:Onodes}), live annuli are
components of the token graph ((Comp) and Claim~\ref{cl:Ocalled}), and a discarded part lies on the
side of its bounding path (Claim~\ref{cl:Ocalled}(c), (d)).

\section{Proofs of the Witness Interval Lemma and the Repaired Recursion Lemma}\label{app:S}

\subsection{Setting}

Lengths are $w=Kc'+1$ on dual edges and on their copies. They are positive, and $K$ exceeds the
number of edges of every simple path in an annulus (Section~\ref{sec:thmD}). The path
$P=(p_0,\dots,p_m)$ is a $w$-shortest path in $G^*$ (the \LemTrees{}). For indices $u$ and $v$,
let $d(u,v)=d(v,u)$ be the $w$-length of the part of $P$ between $p_u$ and $p_v$. It is additive
along $P$: $d(u,v)=d(u,r)+d(r,v)$ if $r$ lies between $u$ and $v$, and $d(u,v)>0$ if $u\ne v$.

The annulus $A_1$ has one piece $p_t^0$ and one piece $p_t^1$ of each face $p_t$, and the copies of $P$
in it form two paths $p_0^\sigma,\dots,p_m^\sigma$, $\sigma\in\{0,1\}$ (Appendix~\ref{app:curves}). Let
$B^\sigma[u\to v]$ be the part of the path on side $\sigma$ from $p_u^\sigma$ to $p_v^\sigma$, of length
$d(u,v)$. A piece of a face $h$ of $A_1$ maps to $h$ and a pair of an edge $e$ to $e^*$, so a walk in
$A_1$ maps to a walk in $G^*$ of the same length. By part~(b) of the \LemAnn{}, a walk in a live annulus maps
to a walk in $A_1$ of the same length, and a piece of $p_t^\sigma$ maps to $p_t^\sigma$; we use these
maps without naming them.

For a walk $\bar P$ in $A_1$ from $p_i^0$ to $p_i^1$, let $x$ be the index of its last visit to a
$0$-piece and $y$ the index of its first visit to a $1$-piece after it; it has one, since it ends at
$p_i^1$. The part $\bar Q$ of $\bar P$ between these two visits runs from $p_x^0$ to $p_y^1$ and has no
piece of a face of $P$ inside. For the image of $P_i$, which is read from $g_i^0$, these are the $x$
and $y$ of Section~\ref{sec:thmD}, and $\bar Q$ is the image of $Q=P_i'$.

\subsection{Four facts in \texorpdfstring{$A_1$}{A1}}

\begin{claim}[P]\label{cl:SP}
A walk in $A_1$ from a piece of $p_u$ to a piece of $p_v$ has length at least $d(u,v)$.
\end{claim}

\begin{proof}
Its image is a walk in $G^*$ from $p_u$ to $p_v$, and the part of $P$ between them is a shortest path.
\end{proof}

\begin{claim}[B]\label{cl:SB}
If $Z$ is a walk in $A_1$ from $p_u^0$ to $p_v^1$, then $\lambda_q\le d(q,u)+w(Z)+d(v,q)$ for every
index $q$.
\end{claim}

\begin{proof}
$B^0[q\to u]\cdot Z\cdot B^1[v\to q]$ is a walk from $p_q^0$ to $p_q^1$.
\end{proof}

\begin{claim}[W]\label{cl:SW}
Let $W$ be a walk in $A_1$ from $p_t^0$ to $p_t^1$ that visits $p_u^0$ and $p_v^1$, in either order.
Then $\lambda_q\le w(W)$ for every $q$ between $\min(t,u,v)$ and $\max(t,u,v)$.
\end{claim}

\begin{proof}
Let $Z$ be the part of $W$ between a visit to $p_u^0$ and a visit to $p_v^1$, read from $p_u^0$ to
$p_v^1$. The two other parts of $W$ run from $p_t^0$ to one of $p_u^0$ and $p_v^1$ and from the other
to $p_t^1$, so by Claim~\ref{cl:SP}, $w(W)\ge d(t,u)+w(Z)+d(v,t)$. By Claim~\ref{cl:SB},
$\lambda_q\le d(q,u)+w(Z)+d(v,q)$. The function $q\mapsto d(q,u)+d(q,v)$ is constant between $u$ and
$v$ and grows away from them, so its value at $q$ is at most its value at $t$ for every $q$ in the
stated range.
\end{proof}

\begin{claim}[Sep]\label{cl:Ssep}
Let $\bar Q$ be a walk in $A_1$ from $p_x^0$ to $p_y^1$ with no piece of a face of $P$ inside, and let
$t$ lie strictly between $x$ and $y$. Then every walk $W$ in $A_1$ from $p_t^0$ to $p_t^1$ meets a node
of $\bar Q$.
\end{claim}

\begin{proof}
We use the curve model of Appendix~\ref{app:curves} on level~$1$, where $K_1=\gamma$, the slot of
$p_t^\sigma$ is the open half $H_t^\sigma$, and $A_1$ satisfies (I) (Claim~\ref{cl:OA1}). Suppose that
$W$ meets no node of $\bar Q$. By Claim~\ref{cl:Odisj}, $\Gamma(W)\cap\Gamma(\bar Q)=\emptyset$; this
covers the two copies of an edge of $P$, whose curves lie in the two half-strips of its strip, and the
two pieces of a face of $P$, whose collars lie in its two halves. Collars avoid small neighbourhoods
of the ends of their arcs, and the curve of a copy of an edge $e$ of $P$ stays in its half-strip away
from $\mathrm{cl}(e)$. So there is a thin closed neighbourhood $N$ of $\gamma$ that meets no collar and
no curve of a pair, and hence misses $\Gamma(W)$ and $\Gamma(\bar Q)$.

Let $\omega$ be a curve in $\Gamma(W)$ from a point of $\mathrm{col}(p_t^0)$ to a point of
$\mathrm{col}(p_t^1)$, and close it by an arc $\beta$ in the face $p_t$: from $\mathrm{col}(p_t^0)$
through $H_t^0$ into $N$, across $\gamma_t$ once at a point $\xi$, and through $H_t^1$ to
$\mathrm{col}(p_t^1)$. Let $\kappa$ be a curve in $\Gamma(\bar Q)$ from a point of
$\mathrm{col}(p_x^0)$ to a point of $\mathrm{col}(p_y^1)$, and close it by a curve $\rho$: from
$\mathrm{col}(p_y^1)$ through $H_y^1$ into $N$, inside $N$ on side~$1$ of $\gamma$ along $\gamma$ to the
face $p_x$, across $\gamma_x$ once inside $N$, and through $H_x^0$ to $\mathrm{col}(p_x^0)$. Both closed
curves may be non-simple, since $W$ and $\bar Q$ may repeat nodes. We count their crossings.
\begin{itemize}
\item $\omega$ and $\kappa$ are disjoint.
\item $\omega$ and $\rho$: $\omega$ misses $N$. Outside $N$, $\rho$ lies in $H_y^1$ and $H_x^0$, where
  $\Gamma(W)$ could meet only the collars of $p_y^1$ and $p_x^0$, which are nodes of $\bar Q$ and so not
  visited by $W$. A visit of $W$ to $p_x^1$ or $p_y^0$ puts collars in the other halves, which $\rho$
  enters only inside $N$.
\item $\beta$ and $\kappa$: $\beta$ lies in the closed disk of $p_t$, and $\Gamma(\bar Q)$ misses it,
  since $\bar Q$ visits no piece of $p_t$: $t$ differs from $x$ and $y$, and $\bar Q$ has no piece of a
  face of $P$ inside. Curves of pairs lie in strips.
\item $\beta$ and $\rho$: outside $N$ they lie in distinct faces, since $t\ne x,y$. Inside $N$, $\beta$
  is a short segment across $\gamma_t$. Since $t$ lies strictly between $x$ and $y$, the run of $\rho$
  on side~$1$ passes all of $\gamma_t$, and its crossing of $\gamma$ lies in $\gamma_x$. So $\beta$
  crosses $\rho$ exactly once, near $\xi$ on side~$1$.
\end{itemize}
So the two closed curves cross exactly once, against the parity fact, which also holds for closed
curves that are not simple.
\end{proof}

\subsection{Proof of the Witness Interval Lemma}

Write $\bar P=U\cdot\bar Q\cdot V$, with $U$ from $p_i^0$ to $p_x^0$ and $V$ from $p_y^1$ to $p_i^1$. As
$\bar P$ has length $\lambda_i$, it is no longer than any walk in $A_1$ from $p_i^0$ to $p_i^1$.

(a) \emph{$w(U)=d(i,x)$ and $w(V)=d(y,i)$.} The bounds $\ge$ hold by Claim~\ref{cl:SP}. The walks
$B^0[i\to x]\cdot\bar Q\cdot V$ and $U\cdot\bar Q\cdot B^1[y\to i]$ run from $p_i^0$ to $p_i^1$, which
gives the bounds $\le$. So $\lambda_i=w(\bar P)=w(\bar Q)+d(i,x)+d(i,y)$.

(b) \emph{The indices.} A visit of $U$ to a piece of $p_r$ splits $U$ into two walks, and
Claim~\ref{cl:SP} gives $d(i,r)+d(r,x)\le w(U)=d(i,x)$. Since $d$ is additive and positive between
distinct indices, $r$ lies between $i$ and $x$. This holds for $1$-pieces on $U$ as well. In the same
way, the pieces of faces of $P$ on $V$ have indices between $y$ and $i$, and $\bar Q$ has no such piece
inside. Since $p_i^0$, $p_x^0$ and $p_y^1$ lie on $\bar P$, we get $i_1=\min(i,x,y)$ and
$i_2=\max(i,x,y)$.

Let $m_1:=\min(x,y)$, $m_2:=\max(x,y)$, let $J$ be the set of indices between $m_1$ and $m_2$, and let
$f(q):=d(q,x)+d(q,y)$. Then $f$ is constant on $J$, and off $J$, $f(q)=f(m_1)+2d(q,r)$, where $r$ is the
end of $J$ nearer to $q$.

(c) \emph{Two bounds.} By Claim~\ref{cl:SB} with $Z=\bar Q$, $\lambda_q\le w(\bar Q)+f(q)$ for every
$q$. Let $W_t$ be a shortest walk in $A_1$ from $p_t^0$ to $p_t^1$. Then
$B^0[i\to t]\cdot W_t\cdot B^1[t\to i]$ runs from $p_i^0$ to $p_i^1$, so $\lambda_i\le\lambda_t+2d(i,t)$,
and by~(a),
\[
  \lambda_t\ \ge\ w(\bar Q)+f(i)-2d(i,t).
\]

Now let $i_1<t<i_2$.

\emph{Case~1: $t\le m_1$ or $t\ge m_2$.} Say $t\le m_1$; the other case is symmetric. Then
$i_1<t\le m_1$ forces $i_1=i$, so $i<t\le m_1$, and $i_2=m_2$. Since $t$ lies between $i$ and $J$,
$f(t)=f(i)-2d(i,t)$, and~(c) gives $\lambda_t\ge w(\bar Q)+f(t)$. For every $q\in[t,m_2]$,
$f(q)\le f(t)$, so $\lambda_q\le w(\bar Q)+f(q)\le\lambda_t$. This is the interval $[t,i_2]$. If
$t\ge m_2$, the same argument gives $[i_1,t]$.

\emph{Case~2: $m_1<t<m_2$.} By Claim~\ref{cl:Ssep}, $W_t$ meets $\bar Q$ at a node $z$. Split
$W_t=X\cdot Y$ and $\bar Q=Q_1\cdot Q_2$ at visits to $z$, with $X$ from $p_t^0$ and $Q_1$ from $p_x^0$.
The walks $B^0[i\to t]\cdot X\cdot Q_2\cdot V$ and $U\cdot Q_1\cdot Y\cdot B^1[t\to i]$ run from $p_i^0$
to $p_i^1$, so by~(a)
\[
  w(X)\ge w(Q_1)+d(i,x)-d(i,t),\qquad w(Y)\ge w(Q_2)+d(i,y)-d(i,t).
\]
Let $e:=m_2$ if $i\le t$ and $e:=m_1$ if $i>t$. Then $t$ lies between $i$ and $e$, so
$d(i,e)=d(i,t)+d(t,e)$, and the interval between $t$ and $e$ is $[t,i_2]$ if $i\le t$ and $[i_1,t]$ if
$i>t$.
\begin{itemize}
\item If $e=y$, let $W':=X\cdot Q_2\cdot B^1[y\to t]$, a walk from $p_t^0$ to $p_t^1$. Then
  \[
    w(W')=w(X)+w(Q_2)+d(y,t)\le w(X)+w(Y)-d(i,y)+d(i,t)+d(t,y)=w(W_t)=\lambda_t .
  \]
  $W'$ visits $p_t^0$ and $p_y^1$, so by Claim~\ref{cl:SW}, $\lambda_q\le\lambda_t$ for every $q$ between
  $t$ and $y$.
\item If $e=x$, let $W'':=B^0[t\to x]\cdot Q_1\cdot Y$. In the same way $w(W'')\le\lambda_t$, and $W''$
  visits $p_x^0$ and $p_t^1$, so $\lambda_q\le\lambda_t$ for every $q$ between $t$ and $x$.
\end{itemize}
This proves the \LemWit{}. Only (P), (B), (W) and (Sep) in $A_1$ are used.

\subsection{Proof of the Repaired Recursion Lemma}

\emph{Soundness.} The lengths that the recursion compares are of two kinds: the length of a path
computed in a live annulus between its pieces of $p_t^0$ and $p_t^1$, which are unique by
part~(c) of the \LemInc{} (the paths $P_i$, $P_{i^-}$, $P_{i^+}$ and the paths of the leaves); and the
length of a walk of the overlap rule from $g_t^0$ through a node $z$ to $g_t^1$. Either maps to a walk
in $A_1$ from $p_t^0$ to $p_t^1$ of the same length, so its length is at least $\lambda_t\ge\Lambda$,
and by part~(c) of the \LemAnn{} its $c'$-length is at least $\lambda'$.

\emph{Good annuli.} Call a live annulus $A$ \emph{good} if $\lambda_t=\Lambda$ for some
$t\in[j_A,k_A]$. The annulus $A_1$, with the range $[0,m]$, is good. We show that if $A$ is good, then
a path or walk of length $\Lambda$ is computed in $A$, or a called child of $A$ is good. The ranges of
the children are shorter than that of $A$, since $j_A<i<k_A$ when $A$ is not a leaf, so the recursion
is finite, and a path or walk of length $\Lambda$ is computed somewhere. Fix
$t\in[j_A,k_A]$ with $\lambda_t=\Lambda$.
\begin{itemize}
\item \emph{$A$ is a leaf.} It computes $P_{j_A}$ and $P_{k_A}$, of lengths $\lambda_{j_A}$ and
  $\lambda_{k_A}$ by (INV) of the \LemInc{}, and $t$ is one of these two indices.
\item \emph{$i_1<t<i_2$.} By (INV), $P_i$ has length $\lambda_i$, so its image in $A_1$ is a walk as in
  the \LemWit{}, with the same $x$, $y$, $i_1$ and $i_2$. By the \LemWit{},
  $\lambda_q\le\lambda_t=\Lambda$ for all $q\in[i_1,t]$ or for all $q\in[t,i_2]$. Since
  $j_A\le t\le k_A$, we have $i^-=\max(i_1,j_A)\in[i_1,t]$ and $i^+=\min(i_2,k_A)\in[t,i_2]$, and both
  lie in $[j_A,k_A]$. By (INV), the repair computes paths of lengths $\lambda_{i^-}$ and
  $\lambda_{i^+}$ (or $P_i$ itself, if one of these indices is $i$), and one of these lengths is at
  most $\Lambda$, hence equal to it.
\item \emph{$t\le i_1$.} Then $t\in[j_A,i_1]\subseteq[j_A,i]$. If $P_i$ meets $P_a$, the overlap rule
  computes, from a common node $z$, a walk from $g_t^0$ through $z$ to $g_t^1$ of length
  $\dist_A(g_t^0,z)+\dist_A(z,g_t^1)$, which is $\lambda_t$ by (C23) of the \LemInc{}. Otherwise
  the $j$-side is called with the range $[j_A,i_1]$, which contains $t$; by part~(b) of the \LemInc{} it is
  one subgraph without bad edges, so it is a live annulus of the next level, and it is good.
\item \emph{$t\ge i_2$.} This is symmetric, with $P_b$, the range $[i_2,k_A]$ and the second part of
  (C23).
\end{itemize}

\emph{The $c'$-length.} A computed path or walk of $w$-length $\Lambda$ has $c'$-length at least
$\lambda'$ by soundness, and at most $\lambda'$ since $K$ times its $c'$-length is less than its
$w$-length $\Lambda<K(\lambda'+1)$ (Section~\ref{sec:thmD}). Every other computed path or walk has
$c'$-length at least $\lambda'$ by soundness. The tree paths of the \LemTrees{} have fewer than
$K$ edges, so (R2) reads their $c'$-lengths exactly, and a walk of the overlap rule is the union of two
of them. So the least $c'$-length that the recursion reads is $\lambda'=\lfloor\Lambda/K\rfloor$. This
proves the \LemRec{}.

\end{document}